%% file: main.tex
\documentclass[acmsmall,screen, nonacm]{acmart}
\AtBeginDocument{%
  }

\setcopyright{acmlicensed}
\copyrightyear{2026}
\acmYear{2026}
\acmDOI{XXXXXXX.XXXXXXX}

\acmJournal{JACM}
\acmVolume{37}
\acmNumber{4}
\acmArticle{111}
\acmMonth{8}

\usepackage{math-setup}
\begin{document}

\title {Assessing LLM Program Reasoning through Equivalence Proofs}

\author{Haozhi Wang}
\email{haozhi5@ualberta.ca}
\affiliation{%
  \institution{University of Alberta}
  \city{Edmonton}
  \state{Alberta}
  \country{Canada}
}

\author{Anders Miltner}
\email{miltner@cs.sfu.ca}
\affiliation{%
  \institution{Simon Fraser University}
  \city{Burnaby}
  \state{British Columbia}
  \country{Canada}
}

\author{Jocelyn Qiaochu Chen}
\additionalaffiliation{%
  \institution{Alberta Machine Intelligence Institute (Amii)}
}
\email{jocelyn.chen@ualberta.ca}
\affiliation{%
  \institution{University of Alberta}
  \city{Edmonton}
  \state{Alberta}
  \country{Canada}
}

\newcommand{\States}[0]{\ensuremath{\mathcal{S}}}

\input{abstract}
\maketitle
\input{intro}
\input{overview}
\input{evaluation-framework}
\input{problem-formulation}
\input{proof-framework}
\input{language-design}
\input{evaluation}
\input{related}
\input{conclu}
\input{acknowledgment}

\bibliography{references}
\bibliographystyle{ACM-Reference-Format}

\pagebreak
\appendix
\input{proof-framework-appendix}
\end{document}

%% file: abstract.tex
\begin{abstract}
  Language models increasingly optimize and migrate code, and
  trusting these rewrites requires independently checkable arguments that they
  preserve program behavior. We ask whether models can construct such arguments
  as equivalence proofs. We propose a lightweight textual proof format based on
  relational Hoare logic, with automatic checks of intermediate claims and their
  compositions. Models may leave claims underived as explicit assumptions,
  keeping partial proofs checkable while recording how much of each argument
  they derive. We evaluate five models on 200 equivalent program pairs in two
  imperative languages with unfamiliar syntax and guard semantics. In free-form
  reasoning, every model reaches the correct verdict on at least
  {66}\% of tasks, yet proof acceptance without feedback ranges
  from 36\% to 94\%. A verifier-assisted workflow uses the checks to guide
  proof refinement, raising overall acceptance from 56.4\% to 80\%, with the
  two strongest models exceeding 97\%. Proof records also reveal
  what acceptance rates alone obscure: among
  accepted proofs, some models derive nearly every step, while others leave
  most of the argument as explicit assumptions. How these models fail to construct a valid proof gives a profile on their relative strengths and weaknesses.
\end{abstract}

%% file: intro.tex
\section{Introduction}
\label{sec:intro}

Large language models (LLMs) increasingly transform programs, as in optimization and migration~\cite{shypula2024performance,khatry2025crust}.
Transformations implicitly claim the new program behaves like the original. Unless that claim is checked independently, it is only as reliable as the reasoning of the model that made it.

Program equivalence offers a direct test of this reasoning: give a model two programs and ask whether they are equivalent \cite{wei-etal-2025-equibench}. However, a correct answer does not mean that the model reasoned correctly about the programs. For example, a model may correctly judge two programs equivalent by analyzing their behavior or by relying on superficial similarities. Carefully designed benchmarks can reveal such shortcuts across many tasks, but a correct answer alone does not provide an argument that can be independently checked.


This limitation matters as LLMs take on more software engineering work. A
model that transforms a program should be able to accompany it with a checkable argument that behavior is preserved. Yet prior
work does not systematically compare models on constructing such arguments.
Equivalence benchmarks score verdicts
\cite{maveli-etal-2025-large,wei-etal-2025-equibench,oza-etal-2026-llms},
while proof-generation benchmarks concern a single program
\cite{lin2024fvel,loughridge2024dafnybench,ye2025verina}; neither checks a
model's proof that two programs behave the same.

In this paper, we ask whether an LLM can \emph{construct a checkable
equivalence proof}. We propose a lightweight textual proof format in which the
model makes intermediate claims and their dependencies explicit for
independent checking. We represent this proof as a derivation of relational
Hoare judgments over pairs of program fragments
\cite{nick2004rhl}. Given a whole-program relational precondition and
postcondition, the model chooses how to decompose the programs and proposes
the intermediate relations needed to connect the resulting fragment judgments
to the whole-program goal.

Checking such a derivation requires verifying both its individual claims and how they combine to establish the whole-program goal. A language-specific checker validates each relational judgment against the language semantics, while a shared composition checker verifies that the judgments connect correctly. This separation distinguishes invalid local claims from errors in composing valid claims. Each language supplies its own local proof rules and semantic checks; our instantiations use a general-purpose solver to discharge the semantic obligations. We prove that composition is sound for deterministic languages with standard sequential composition, provided that the local proof rules and entailment checks are sound and any retained assumptions are valid.


A model may not be able to justify every step of an equivalence proof. Following work on reasoning under hypotheses~\cite{banerjee2016hypotheses}, we allow underived relational judgments to remain explicit assumptions. Each assumption becomes a proof obligation: the model can derive the judgment or retain it as an assumption for validation by the model or a solver. This distinction lets us measure \emph{derivational coverage}: how much of an accepted argument the model derives rather than assumes. This structure also supports verifier-assisted construction: the verifier identifies failed judgments, entailments, or composition steps so the model can revise its argument without starting over.



To evaluate models' ability to construct such proofs, we instantiate the framework for two imperative languages designed to emphasize explicitly defined semantics and reduce reliance on familiar programming conventions:  {\proclangobs} expresses standard operations using unfamiliar symbols, while {\proclangalt} additionally inverts the guards of conditionals and loops. Our benchmark contains 100 equivalent program pairs per language, with corresponding pairs adapted to each language's guard semantics.

We evaluate five models from four families (GPT-6.1-Sol, GPT-6-Luna, Opus-5.5, GLM-5.3 and Qwen3.8-27B), comparing proof generation with and without verifier assistance. With feedback, models can identify and correct errors in their proofs, as well as attempt to justify claims initially left as assumptions. We assess each proof using the evidence criteria in Section~\ref{sec:evidence-assessment}. To interpret these results, we also evaluate concrete execution, which tests whether models can apply the supplied semantics to individual runs, and free-form equivalence reasoning, which measures their ability to reach correct equivalence verdicts.

Our results show that models with similar accuracy on equivalence judgments can differ substantially in their constructed proofs. Reasoning in free form, every model decides correctly on at least
{66}\% of tasks, yet proof acceptance without feedback ranges from 36.0\% to
94\% across models. Execution accuracy does not predict this ranking either:
Qwen3.8-27B predicts both outputs correctly on over 90\% of pairs but
produces the fewest accepted proofs. Models also accept 27 of 411 local claims
that cvc5 refutes (6.6\%), showing that semantic self-checking misses some
proof errors.


Second, the framework provides useful verifier feedbacks for models to revise their proofs. Under the verifier-assisted workflow, overall proof acceptance rises from 56.4\% to 80\%, with improvements for every model in both languages. The improvements are particularly large for models with relatively lower initial acceptance rates: GLM-5.3 rises from 53.0\% to 79.5\%, and GPT-6-Luna from 40.5\% to 73.5\%. These results show that the framework can help models construct checked equivalence proofs, in addition to assessing their reasoning.

Third, the proof records show how models reason. On the 146 tasks for which
both GPT-6.1-Sol and GPT-6-Luna have accepted selected-final proofs, assumed
judgments cover 6.61\% and 76.62\% of program statement occurrences,
respectively: proof correctness cannot tell the two models apart, but the
records show that one derives almost all of its argument and the other
leaves most of it to the solver. The records also locate failures. Among
GPT-6-Luna's first responses that contain assumptions and have every assumed
claim validated by cvc5, 14.3\% still contain a refuted connecting entailment.

In summary, our contributions are as follows:
\vspace{-2.5pt}
\begingroup
\settowidth{\leftmargini}{\labelitemi}
\addtolength{\leftmargini}{\labelsep}
\begin{itemize}
  \item We introduce a lightweight compositional framework in which models
    write relational equivalence proofs as plain text, without a proof
    assistant, checked by a shared composition checker and solver. We
    prove it sound assuming soundness of the underlying theory checkers.

  \item We make underived judgments explicit proof obligations, enabling
    partial proofs, verifier-assisted revision, and measurement of derivational
    coverage: how much of an argument a model derives.

  \item We evaluate five LLMs on two small imperative languages, with and
    without verifier assistance. We show that checked proofs separate
    models with similar verdicts, that assistance raises acceptance
    sharply, and that proof records reveal how models fail: how much of an
    argument they derive, whether their claims compose, and whether they
    can check their own claims.
\end{itemize}
\endgroup

%% file: overview.tex
\section{Overview}
\label{sec:overview}
\newcommand{\ovfrag}[2]{\ensuremath{{\color{blue!55!black}\mathsf{#1}_{#2}}}}
\colorlet{ovR}{orange!22}
\DeclareRobustCommand{\ovn}[1]{\smash{\tikz[baseline=(n.base)]\node[circle,draw=red!65!black,line width=0.45pt,inner sep=0pt,minimum size=8.4pt,font=\sffamily\scriptsize\bfseries,text=red!65!black](n){#1};}}

\begin{figure}[!tbp]
  \centering
  \newcommand{\ovhl}[2]{{\setlength{\fboxsep}{0.7pt}\colorbox{#1}{#2}}}%
  \newcommand{\ovpt}[1]{\ \ {\normalfont\color{blue!55!black}$\blacktriangleleft$\,$#1$}}%
  \newcommand{\ovlab}[2]{\llap{\makebox[1.7em][l]{\normalfont\small\ovfrag{#1}{#2}}}}%
  \lstset{basicstyle=\ttfamily\footnotesize, columns=fullflexible, keepspaces=true,
    escapechar=|, aboveskip=2pt, belowskip=0pt, showstringspaces=false,
  xleftmargin=1.7em}%
  \begin{minipage}[t]{0.44\linewidth}
    \hspace*{1.7em}\textbf{\footnotesize Source}
\begin{lstlisting}
|\ovlab{A}{1}|x, y, m := 0, 0, 0
px, total := 0, 0
while px < width:|\ovpt{H}|
|\ovlab{B}{1}|  it := 0
|\ovlab{C}{1}|  while it < maxit and m < bound:
    x, y := x*x - y*y + cx,
            2*x*y + cy
    m := x*x + y*y
    it += 1
|\ovlab{D}{1}|  total += it
  cx += 1;  px += 1
|\ovlab{R}{1}|  |\ovhl{ovR}{if px < width:\ \ \# next pixel}|
  |\ovhl{ovR}{\ \ x, y, m := 0, 0, 0}|
\end{lstlisting}
  \end{minipage}\hfill
  \begin{minipage}[t]{0.54\linewidth}
    \hspace*{1.7em}\textbf{\footnotesize Target}
\begin{lstlisting}
|\ovlab{A}{2}|x, y, xx, yy := 0, 0, 0, 0
first := 1;  px, total := 0, 0
while px < width:|\ovpt{H}|
|\ovlab{R}{2}|  |\ovhl{ovR}{if first = 0:\ \ \# this pixel}|
  |\ovhl{ovR}{\ \ x, y, xx, yy := 0, 0, 0, 0}|
|\ovlab{B}{2}|  it := 0;  mag := xx + yy
|\ovlab{C}{2}|  while it < maxit and mag < bound:
    y := (x+y)*(x+y) - xx - yy + cy
    x := xx - yy + cx
    xx, yy := x*x, y*y
    mag := xx + yy
    it += 1
|\ovlab{D}{2}|  total += it
  cx += 1;  px += 1
|\ovlab{F}{2}|  first := 0
\end{lstlisting}
  \end{minipage}
  \caption{The escape-time loop over one row of pixels and its optimized
    version. Letters name corresponding fragments (subscript~1: source,
    2: target); for instance, $\ovfrag{C}{1}$ and $\ovfrag{C}{2}$ are the two inner
    loops. The source resets at the end of a pixel ($\ovfrag{R}{1}$), the target
  at the start of the next ($\ovfrag{R}{2}$), on either side of the loop head $H$.}
  \Description{Two programs side by side, with fragment labels in the left
    margin. Source: A1 initialization, then the outer loop with head H containing
    B1 (it := 0), C1 the inner loop, D1 the bookkeeping, and R1, which resets x, y
    and m if another pixel follows. Target: A2 initialization, then the outer loop
    with head H containing R2, which resets x, y, xx and yy at the start of every
    pixel after the first, B2, C2 the inner loop with cached squares, D2 the
  bookkeeping, and F2, which sets first to 0. The resets are highlighted.}
  \label{fig:overview-pair}
\end{figure}

Suppose we ask a model whether an optimized program preserves the behavior of
the original, and to justify its answer in its own words.
Figure~\ref{fig:overview-pair} shows a escape-time loop of a Mandelbrot renderer next to its
standard optimized form. The loop walks one row of pixels,
with $\mathit{cx}$ and $\mathit{cy}$ the current pixel's coordinates. For each
pixel, the source starts $x$ and $y$ at~0 and keeps updating them until
\texttt{x*x~+~y*y} reaches $\mathit{bound}$ or $\mathit{maxit}$ iterations have
run, then adds the iteration count to $\mathit{total}$. The optimized form keeps
\texttt{x*x} and \texttt{y*y} in $\mathit{xx}$ and $\mathit{yy}$ for reuse, and
computes \texttt{2*x*y} as \texttt{(x+y)*(x+y)~-~xx~-~yy}. If we are to
trust the optimization on the strength of the model's explanation, we must be able to tell whether that explanation is
correct.

\paperparagraph{Free-form arguments.}
We ask each model whether the two programs are equivalent
and to justify its answer step by step in natural language, without prescribing
a proof format. We provide both programs, the language grammar and semantics,
and the relational objective.
Both GPT-6.1-Sol and Opus-5.5 correctly conclude that the programs are
equivalent with reasoning traces
(Figure~\ref{fig:overview-freeform}).\footnote{We collected three answers
from each model; all six give the verdict \texttt{EQUIVALENT}. We quote one answer from each model.}
Both arguments rely on a coupling: $x$ and $y$ agree across the programs,
and the target's $xx$ and $yy$ hold their squares. They check the inner-loop
update using $(x+y)^2-x^2-y^2=2xy$ and appeal to the resets at the start
of each pixel. Both arguments look plausible. But where does each claim
hold, how do the claims depend on one another, and do they establish
the whole-program goal?


\begin{figure}[tp]
  \centering
  \newcommand{\ovrd}[1]{\textcolor{red!65!black}{#1}}%
  \newcommand{\ovs}[2]{\textbf{[#1]}~#2}%
  \footnotesize
  \setlength{\tabcolsep}{0pt}%
  \renewcommand{\arraystretch}{1.05}%
  \begin{tabularx}{\linewidth}{@{}>{\raggedright\arraybackslash}p{0.735\linewidth}@{\hspace{7pt}}!{\color{red!35}\vrule width 0.8pt}@{\hspace{7pt}}>{\raggedright\arraybackslash}X@{}}
    \multicolumn{2}{@{}l}{\small\textbf{(a) Opus-5.5}}\\
    \ovs{4}{``Coupling invariant for Program~2 at the head of each inner loop:
      $xx = x^2$ and $yy = y^2$. \ldots{} So Program~2's new $y = 2xy + cy$ and new
    $x = x^2 - y^2 + cx$, exactly matching Program~1. \ldots''} & \\
    \ovs{5}{``Relational invariant \ovrd{at each outer-loop head: $x$, $y$, $px$,
      $\mathit{total}$ and $cx$ are equal in both programs} ($cy$, $\mathit{width}$,
      $\mathit{maxit}$ and $\mathit{bound}$ are equal and never modified). In
    addition, Program~2 satisfies $xx = x^2$ and $yy = y^2$. \ldots''} &
    \ovn{1}~\ovrd{Can this hold together with [6], which puts $\ovfrag{R}{1}$ before
    $H$ and $\ovfrag{R}{2}$ after it?} \\
    \ovs{6}{``Reset timing: Program~1 resets $x$, $y$, $m$ to 0 \ovrd{at the end of
      an outer iteration} exactly when $px < \mathit{width}$ after the increment,
      which is exactly when another outer iteration will run. Program~2 resets $x$,
      $y$, $xx$, $yy$ \ovrd{at the start of every outer iteration} except the first,
    because $\mathit{first} = 0$ after the first iteration. \ldots''} & \\
    \multicolumn{2}{@{}l}{\rule{0pt}{13pt}\small\textbf{(b) GPT-6.1-Sol}}\\
    \ovs{3}{``\ovrd{The resets correspond for every pixel that is processed}. \ldots{}
      Consequently, \ovrd{each processed pixel starts with $x = y = 0$ in both
    programs}, $m = 0$ in Program~1, and $xx = yy = 0$ in Program~2.''} &
    \ovn{2}~\ovrd{The resets run at different points. What holds between them, at $H$?} \\
    \ovs{4}{``Immediately before each inner loop, both programs set $\mathit{it}$
      to zero, and Program~2 sets $\mathit{mag}$ to $xx + yy$. Use the inner-loop
      invariant that $x$, $y$, $\mathit{it}$, and $cx$ agree across programs; \ldots{}
    \ovrd{Step~3 establishes this invariant initially}. \ldots''} &
    \ovn{3}~\ovrd{Rests on [3].} \\
    \ovs{6}{``\ldots{} If the new $px$ is below $\mathit{width}$, Program~1 resets
      immediately and Program~2 resets at the next pixel's beginning, \ovrd{as
    established in step~3}. \ldots''} &
    \ovn{3}~\ovrd{Rests on [3] again.} \\
    \ovs{7}{``\ovrd{Induction over processed pixels} therefore gives equal
      accumulated totals and $cx$ throughout corresponding execution, and equal $x$
    and $y$ when the outer loops finish. \ldots''} &
    \ovn{2}~\ovrd{The induction needs that relation at $H$.} \\
  \end{tabularx}
  \caption{Excerpts from one free-form answer by each model, verbatim (both
    \texttt{EQUIVALENT}). We add, in the right margin, the questions a reader
    must answer to check these steps.}
  \Description{Quoted argument steps from the two models, each with a margin
    question in red. Opus: step 4, the inner-loop coupling, with no question;
    step 5, equality of x and y at each outer-loop head, with question 1, whether
    it can hold together with step 6, which puts the source's reset before H and
    the target's after it; step 6, the reset timing. Sol: step 3, each pixel starts
    with zeros in both programs, with question 2, starts where; steps 4 and 6,
    which rest on step 3, with question 3, under which reading; step 7, induction
  over pixels, with question 4, which relation at H, never stated.}
  \label{fig:overview-freeform}
\end{figure}

\paperparagraph{Gaps in the arguments.}
The programs reset $x$ and $y$ at different points in the outer loop:
the source at the end of a pixel ($\ovfrag{R}{1}$), and the target at
the start of the next ($\ovfrag{R}{2}$). Between pixels, their states
may therefore differ at the loop head $H$, where the source has reset
these variables but the target has not. Both arguments leave this
difference unresolved at the steps marked in
Figure~\ref{fig:overview-freeform}. Opus claims that $x$ and $y$ agree
across the programs at $H$ (step~5), but does not justify this equality
given the reset locations it identifies in step~6~\ovn{1}. Sol claims
that the resets ``correspond for every pixel,'' so that each pixel
``starts with $x = y = 0$ in both programs'' (step~3). It never states
how the programs' states relate between the resets~\ovn{2}, although
its induction over pixels needs this relation; steps~4 and~6 inherit
the gap~\ovn{3}. To resolve these points, a reader must align the
programs, establish what holds at $H$, and check each step against
it---in effect, reconstruct the proof.


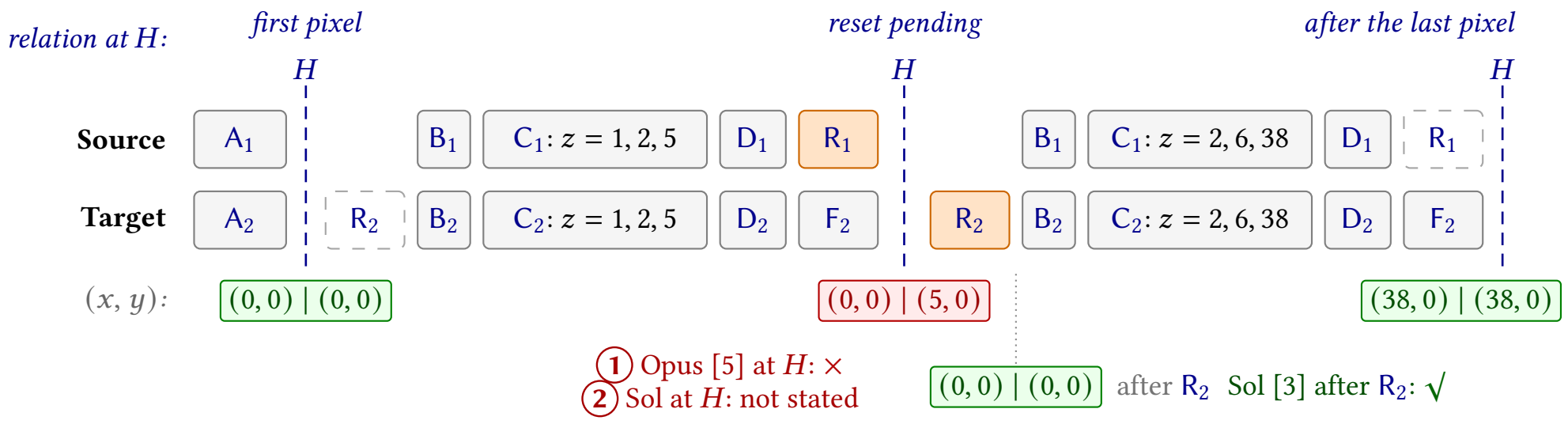
\begin{figure}[!b]
  \centering
  \begin{tikzpicture}[x=1.1cm,y=0.9cm,
      seg/.style={draw=black!50, fill=black!4, rounded corners=1.5pt, minimum height=0.48cm,
      inner sep=0.5pt, font=\scriptsize},
      rst/.style={seg, draw=orange!80!black, fill=ovR},
      skp/.style={seg, draw=black!35, dashed, fill=white, text=black!45},
      hline/.style={densely dashed, blue!55!black, line width=0.5pt},
      ok/.style={draw=green!50!black, fill=green!8, text=green!30!black, rounded corners=1pt,
      font=\scriptsize, inner sep=1.5pt},
      bad/.style={draw=red!70!black, fill=red!8, text=red!60!black, rounded corners=1pt,
      font=\scriptsize, inner sep=1.5pt},
      case/.style={font=\scriptsize\itshape, text=blue!55!black, align=center},
      note/.style={font=\scriptsize, align=center, text=red!65!black},
    ]
    \def\ys{0}\def\yt{-0.75}
    \newcommand{\ovsg}[5][seg]{\node[#1, minimum width={(#3-#2)*1.1cm}] at ({(#2+#3)/2},#4) {#5};}
    \node[font=\scriptsize\bfseries, anchor=east] at (-0.1,\ys) {Source};
    \node[font=\scriptsize\bfseries, anchor=east] at (-0.1,\yt) {Target};
    \ovsg{0}{0.7}{\ys}{$\ovfrag{A}{1}$}
    \ovsg{1.7}{2.1}{\ys}{$\ovfrag{B}{1}$}
    \ovsg{2.2}{3.9}{\ys}{$\ovfrag{C}{1}$: $z = 1, 2, 5$}
    \ovsg{4.0}{4.5}{\ys}{$\ovfrag{D}{1}$}
    \ovsg[rst]{4.6}{5.2}{\ys}{$\ovfrag{R}{1}$}
    \ovsg{6.3}{6.7}{\ys}{$\ovfrag{B}{1}$}
    \ovsg{6.8}{8.5}{\ys}{$\ovfrag{C}{1}$: $z = 2, 6, 38$}
    \ovsg{8.6}{9.1}{\ys}{$\ovfrag{D}{1}$}
    \ovsg[skp]{9.2}{9.8}{\ys}{$\ovfrag{R}{1}$}
    \ovsg{0}{0.7}{\yt}{$\ovfrag{A}{2}$}
    \ovsg[skp]{1.0}{1.6}{\yt}{$\ovfrag{R}{2}$}
    \ovsg{1.7}{2.1}{\yt}{$\ovfrag{B}{2}$}
    \ovsg{2.2}{3.9}{\yt}{$\ovfrag{C}{2}$: $z = 1, 2, 5$}
    \ovsg{4.0}{4.5}{\yt}{$\ovfrag{D}{2}$}
    \ovsg{4.6}{5.2}{\yt}{$\ovfrag{F}{2}$}
    \ovsg[rst]{5.6}{6.2}{\yt}{$\ovfrag{R}{2}$}
    \ovsg{6.3}{6.7}{\yt}{$\ovfrag{B}{2}$}
    \ovsg{6.8}{8.5}{\yt}{$\ovfrag{C}{2}$: $z = 2, 6, 38$}
    \ovsg{8.6}{9.1}{\yt}{$\ovfrag{D}{2}$}
    \ovsg{9.2}{9.8}{\yt}{$\ovfrag{F}{2}$}
    \foreach \x in {0.85, 5.4, 9.95} {
      \draw[hline] (\x,0.5) -- (\x,-1.25);
      \node[font=\scriptsize, text=blue!55!black] at (\x,0.65) {$H$};
    }
    \node[case, anchor=south] at (0.85,0.8)  {first pixel};
    \node[case, anchor=south] at (5.4,0.8)   {reset pending};
    \node[case, anchor=south east] at (10.15,0.8)  {after the last pixel};
    \node[font=\scriptsize\itshape, text=blue!55!black, anchor=east] at (-0.1,0.95) {relation at $H$:};
    \node[ok]  (s0) at (0.85,-1.5)  {$(0,0)\,|\,(0,0)$};
    \node[bad] (s1) at (5.4,-1.5)   {$(0,0)\,|\,(5,0)$};
    \node[ok]  (s2) at (6.25,-2.3)  {$(0,0)\,|\,(0,0)$};
    \node[ok, anchor=east]  (s3) at (10.4,-1.5)  {$(38,0)\,|\,(38,0)$};
    \draw[densely dotted, black!45] (6.25,-1.25) -- (s2.north);
    \node[font=\scriptsize, text=black!55, anchor=west] at (s2.east) {after $\ovfrag{R}{2}$};
    \node[font=\scriptsize\itshape, text=black!60, anchor=east] at (-0.1,-1.5) {$(x,y)$:};
    \node[note, anchor=north] at (4.0,-1.85) {\ovn{1} Opus [5] at $H$: \textcolor{red!70!black}{$\times$}\\ \ovn{2} Sol at $H$: not stated};
    \node[note, anchor=west, text=green!30!black] at (7.75,-2.3) {Sol [3] after $\ovfrag{R}{2}$: \textcolor{green!40!black}{$\surd$}};
  \end{tikzpicture}
  \caption{A reconstructed run ($\mathit{cx}=1$, $\mathit{cy}=0$,
    $\mathit{width}=2$, $\mathit{maxit}=3$, $\mathit{bound}=100$). The programs are
    aligned at visits to $H$; badges give $(x,y)$ as source\,|\,target. Boxes name
    fragments of Figure~\ref{fig:overview-pair}; dashed boxes mark resets skipped
    by their tests, and inner-loop boxes list $z$, which remains real because
    $\mathit{cy}=0$. The second visit exposes the between-pixel case behind
  questions~\ovn{1} and~\ovn{2}.}
  \Description{A two-row timeline of one run of the source and target, aligned at
    three visits to the outer-loop head H. Source: A1, then B1, C1 (z goes to 1, 2,
    5), D1, R1; then B1, C1 (z goes to 2, 6, 38), D1, and a skipped R1. Target: A2, a
    skipped R2, B2, C2, D2, F2; then R2, B2, C2, D2, F2. Above the visits, the case of
    the relation at H: first pixel, reset pending, after the last pixel. Below
    them, (x, y) in source and target: (0,0) and (0,0); (0,0) and (5,0), marked
    red, where Opus's step 5 fails and Sol states nothing; (38,0) in both. After
  the target's R2 in the second pixel, both are (0,0), where Sol's step 3 holds.}
  \label{fig:overview-run}
\end{figure}

Figure~\ref{fig:overview-run} exposes the three cases needed at $H$. Before the
first pixel, both programs are zeroed. Between pixels, the source has run
$\ovfrag{R}{1}$ but the target has not yet run $\ovfrag{R}{2}$, so only the
source is known to be zeroed. After the last pixel, $x$ and $y$ agree. In the
middle case, Opus's step~5 can fail~\ovn{1}: at the second visit to $H$ in the
run, the source holds $(0,0)$ while the target holds $(5,0)$. Sol's
conclusion in step~3 holds after $\ovfrag{R}{2}$, but connecting the two resets
requires the three-case relation at $H$, which Sol never states~\ovn{2}; once
supplied, it supports steps~4 and~6~\ovn{3}.

A reader must repeat this reconstruction for every argument, and even
then it only goes so far: one execution can refute a claim like Opus's step~5,
but cannot certify Sol's argument, which needs the relation at $H$ to hold
in every state. Both models have the right idea of why the
programs are equivalent; their answers lack a structure that lets
anyone else confirm it. Without one, a nearly correct proof and a correct one
look the same, and telling them apart means redoing the proof.

\paperparagraph{Automatically checked derivations.}
To avoid manually reconstructing free-form arguments, we introduce a framework
for automatically checking model-generated equivalence proofs. Given two
programs and an equivalence goal, we ask the model to construct a derivation
of relational judgments. Each judgment relates a pair of program fragments
under an explicit precondition and postcondition. The framework checks whether
these judgments are valid and whether they compose to establish the
whole-program goal. If a check fails, the framework identifies the problematic
step so the model can revise its proof (Section~\ref{sec:technical}).
Figure~\ref{fig:overview-derivation} shows Opus's derivation for this pair.
Three features of the framework turn the questions raised in
Figure~\ref{fig:overview-freeform} into checks.


\begin{figure}[!tbp]
  \centering
  \newcommand{\ovok}{\textcolor{green!40!black}{$\surd$}}%
  \newcommand{\ovxy}{\rhlTagOne{x}=\rhlTagTwo{x}\land\rhlTagOne{y}=\rhlTagTwo{y}}%
  \newcommand{\ovin}{\hspace*{1em}}%
  \footnotesize
  \setlength{\tabcolsep}{4pt}%
  \renewcommand{\arraystretch}{1.1}%
  \begin{tabularx}{\linewidth}{@{}>{\raggedright\arraybackslash}p{2.2cm}>{\raggedright\arraybackslash}p{4.75cm}c>{\raggedright\arraybackslash}X@{}}
    \toprule
    \textbf{Fragments} & \textbf{Precondition $\Rightarrow$ postcondition} & & \textbf{Question of Figure~\ref{fig:overview-freeform} it answers} \\
    \midrule
    $\ovfrag{A}{1}\sim\ovfrag{A}{2}$ & $\Phi_{\mathrm{in}}\Rightarrow E\land Z\land\rhlTagTwo{\mathit{first}}=1$ & \ovok & \\
    \rowcolor{ovR}
    outer loops & $I\Rightarrow I\land\neg G$\quad(while rule) & \ovok &
    \ovn{2}~The relation at $H$ is $I$, below. With Opus's step~5 in its place, this check fails~\ovn{1}. \\
    \rowcolor{ovR}
    \ovin$\mathtt{skip}\sim\ovfrag{R}{2}$ & $I\land G\Rightarrow E\land Z$ & \ovok &
    \ovn{2}~Sol's ``each pixel starts with $x=y=0$'' is this postcondition, after $\ovfrag{R}{2}$. \\
    \ovin$\ovfrag{B}{1}\sim\ovfrag{B}{2}$ & $E\land Z\Rightarrow K$ & \ovok &
    \ovn{3}~Sol's step~4 rests on this link. \\
    \ovin$\ovfrag{C}{1}\sim\ovfrag{C}{2}$ & $K\Rightarrow K\land\neg G_{\mathrm{in}}$\quad(while rule) & \ovok & \\
    \ovin$\ovfrag{D}{1}\sim\ovfrag{D}{2}$ & $K\Rightarrow E\land\ovxy$ & \ovok & \\
    \rowcolor{ovR}
    \ovin$\ovfrag{R}{1}\sim\ovfrag{F}{2}$ & $E\land\ovxy\Rightarrow I$ & \ovok &
    \ovn{2}~How the resets correspond: $\ovfrag{R}{1}$ here, then $\mathtt{skip}\sim\ovfrag{R}{2}$ at the next pixel, through $I$. \\
    loop exit & $I\land\neg G\Rightarrow\Phi_{\mathrm{eq}}$ & \ovok & \\
    \bottomrule
  \end{tabularx}\\[4pt]
  \begin{minipage}{\linewidth}
    \footnotesize
    \setlength{\tabcolsep}{3pt}%
    \renewcommand{\arraystretch}{1.1}%
    \begin{tabular}{@{}r@{\quad}l@{}}
      $I$, the invariant at $H$: & $E$ and one of three cases:\\
      \emph{first pixel} & $\rhlTagTwo{\mathit{first}}=1\land Z$\\
      \colorbox{ovR}{\emph{reset pending}} & $\rhlTagTwo{\mathit{first}}=0\land\rhlTagOne{(\mathit{px}<\mathit{width})}\land\rhlTagOne{x}=\rhlTagOne{y}=\rhlTagOne{m}=0$\\
      \emph{after the last pixel} & $\rhlTagTwo{\mathit{first}}=0\land\neg\rhlTagOne{(\mathit{px}<\mathit{width})}\land\ovxy$\\[2pt]
      $K$, the inner-loop invariant: & $E$, equal $x$, $y$, $\mathit{it}$, and $\rhlTagTwo{xx}=\rhlTagTwo{x}^2$, $\rhlTagTwo{yy}=\rhlTagTwo{y}^2$, $\rhlTagOne{m}=\rhlTagTwo{\mathit{mag}}$\\
      & $E$: equal $\mathit{px}$, $\mathit{total}$, $\mathit{cx}$, inputs.\quad $Z$: $x$, $y$, $m$, $xx$, $yy$ all $0$.\quad $G$, $G_{\mathrm{in}}$: the loop tests.
    \end{tabular}
  \end{minipage}
  \caption{Opus's derivation for the pair (outline of 87 steps). Each row relates
    two fragments of Figure~\ref{fig:overview-pair} and is checked automatically
  ($\surd$); indented rows make up one iteration of the outer loops.}
  \Description{A table of the judgments in Opus's derivation, with columns for
    the fragments, the precondition and postcondition, a check mark, and the
    question of Figure 2 that the row answers. Rows: A1 with A2; the outer loops by
    the while rule with invariant I, answering question 4, the relation at H, and
    noting that with Opus's step 5 in its place the check fails; indented, skip
    with R2, answering question 2, since starting with zeros is its postcondition
    after R2; B1 with B2, the link Sol's step 4 rests on; C1 with C2 by the while
    rule; D1 with D2; R1 with F2, the reset timing of step 6; then the loop exit.
    Below, the invariant I is defined with three cases (first pixel, reset pending,
  after the last pixel), followed by K, E, Z and G.}
  \label{fig:overview-derivation}
\end{figure}

\begin{list}{\textbullet}{\setlength{\leftmargin}{0pt}\setlength{\labelwidth}{0.6em}%
    \setlength{\labelsep}{0.5em}\setlength{\itemindent}{1.1em}%
    \setlength{\itemsep}{2pt}\setlength{\parsep}{0pt}\setlength{\topsep}{3pt}%
  \renewcommand{\makelabel}[1]{#1\hfil}}
  \item \textbf{Making claims precise.} Free-form arguments may leave unclear
where a claim holds. Our framework makes this explicit through relational
judgments over named fragment pairs, with explicit preconditions and
postconditions. In the example above, Sol's claim that the resets correspond~\ovn{2} becomes two
judgments on either side of $H$:
$\ovfrag{R}{1}\sim\ovfrag{F}{2}$ and
$\mathtt{skip}\sim\ovfrag{R}{2}$. The second judgment's postcondition states
that each pixel starts with $x = y = 0$ in both programs. These judgments
are checked against the language semantics.
\item \textbf{Checking dependencies.} Even if individual claims are correct, they may not form a valid proof when combined. Our framework requires judgments to be composed
using explicit proof rules, checking their premises and entailment conditions. As highlighted in Figure~\ref{fig:overview-freeform}, 
Sol's step~4 depends on step~3~\ovn{3}; the sequencing rule makes this
dependency explicit by joining $\mathtt{skip}\sim\ovfrag{R}{2}$ to
$\ovfrag{B}{1}\sim\ovfrag{B}{2}$. For the outer loops, the while rule requires
a relation at $H$ preserved by every iteration~\ovn{2}. Opus supplies the
three-case invariant $I$. Replacing it with the relation from its free-form
step~5 fails this check because it omits the between-pixel case~\ovn{1}.
\item \textbf{Handling incomplete arguments.} A model may be unable to justify every step of its argument. Rather than leaving such gaps implicit, our framework allows the model to record underived judgments as explicit assumptions, which can subsequently be checked or discharged through further derivation. A failed check will be reported back at the step where it occurs, so the model can revise its derivation or supply a sub-proof.
\end{list}

\paperparagraph{Checking instead of reconstructing.}
Free-form explanations make a model's reasoning visible, but not checkable. A
model can reach the right verdict with a largely correct argument, yet leave its claims
unlocated, its dependencies implicit, and its gaps unmarked, so checking
the argument means rebuilding the proof. Here, showing that Opus's step~5
can fail and supplying the relation at $H$ that Sol leaves out took a
hand reconstruction (Figure~\ref{fig:overview-run}). In our framework, the
model instead writes a derivation: claims are tied to
fragments, dependencies are checked links, and gaps are explicit assumptions,
so the same questions are answered by automatic checks
(Figure~\ref{fig:overview-derivation}). To produce that derivation, Opus
received the same programs, language grammar, and relational goal as in the
free-form setting, together with the proof rules, and chose the
invariants $I$ and $K$ itself.

\paperparagraph{From checked arguments to model comparison.}
Even when two models ultimately produce accepted, assumption-free proofs,
their proof records can reveal substantial differences in how much of the
argument they derive without assistance. For this pair, Sol's initial
response derives every judgment and passes every check. Opus initially
derives only the initialization, leaving the relation between the outer
loops, which covers 80.33\% of the two programs' statement occurrences, as one
assumption. Its first recursive refinement derives this relation using
the invariant $I$ of Figure~\ref{fig:overview-derivation}, but leaves the
inner-loop bodies assumed; the second derives the bodies and closes the
proof.

We therefore compare models using both their proof evidence and their
refinement records: whether the argument is valid, how much the model
derives unaided, and which open claims it can derive when asked.
Section~\ref{sec:evidence-assessment} formalizes the assessment of proof
evidence; Section~\ref{sec:evaluation} applies it to five models using
task-level measures and diagnostics for individual checks.



%% file: evaluation-framework.tex
\section{Semantic Preliminaries}\label{sec:semantic-setting}

We define language semantics and the execution and relational
judgments that state and justify claims about program behavior.
Section~\ref{sec:problem-formulation} formulates
the reasoning tasks and how to assess them.

A language $\mathcal{L} = (\mathrm{Prog},\States,O,\llbracket \cdot \rrbracket)$ is a 4-tuple consisting of: (1) a set of programs $\mathrm{Prog}$ (2) a set of states $\States$; (3) a set of outputs $O$; and (4) a semantics function $\llbracket \cdot \rrbracket : \mathrm{Prog}
\rightarrow (\States \rightharpoonup O)$ from programs to partial maps
between initial states and outputs.

Throughout this section, we fix $\mathcal L$ and take programs, initial
states, and outputs from its sets $\mathrm{Prog}$, $\States$, and $O$,
respectively. For the languages studied here, each execution either
terminates with an output or diverges; an undefined denotation represents
divergence.

\subsection{Semantic Evidence}
\label{sec:execution-task}

Two kinds of derivations justify claims about behavior in
$\mathcal L$:
execution evidence for one initial state and relational evidence for
related ones. Both derivations must be sound 
to the semantics $\llbracket\cdot\rrbracket$.

\paperparagraph{Execution outcomes.}
To justify an execution outcome, the model supplies a derivation showing
that program $p$ produces output $o$ from state $s$ under the supplied
rules. These rules inductively define judgments $R_{\rightarrow}(p,s,o)$
and must satisfy the adequacy condition
\[
R_{\rightarrow}(p,s,o)
\ \Longleftrightarrow \
\llbracket p\rrbracket(s)=o.
\]

Small-step traces establish these judgments. Each configuration
pairs a state with the remaining computation. Let $\rho_i$ denote the
execution rule claimed to justify the transition from $c_i$ to $c_{i+1}$.
A trace links configurations by these claimed rule applications:
\[
c_0\xrightarrow{\rho_0}c_1\;\cdots\;
\xrightarrow{\rho_{n-1}}c_n.
\]
The checker verifies that the trace starts with $p$ in state $s$,
that each transition follows the named rule $\rho_i$, including its
premises and side conditions, and that the final configuration yields $o$.

\paperparagraph{Relational satisfaction.} A relational claim specifies how two programs transform related initial
states into related outputs. It covers every pair of initial
states satisfying the input relation.

The model justifies such a claim with an inspectable derivation of
a judgment $R_r(p,q,\Phi,\Phi')$, where precondition $\Phi$ relates
inputs and postcondition $\Phi'$ relates outputs. Soundness
requires that, if this judgment holds, every pair
$s_1,s_2\in\States$ with $\Phi(s_1,s_2)$
either jointly divergences or outputs $o_1,o_2\in O$
such that
{
  \small
  \[
    \llbracket p\rrbracket(s_1)=o_1
    \quad\land\quad
    \llbracket q\rrbracket(s_2)=o_2
    \quad\land\quad
    \Phi'(o_1,o_2).
  \]
  \normalsize
}

\subsection{Program Equivalence}\label{sec:equivalence-task}

Program equivalence is defined by the same semantics of
$\mathcal L$. Two programs are
equivalent on a set of admissible inputs $I\subseteq\States$ when they
agree on termination and produce equal outputs:
{
  \small
  \[
    p\equiv_{\mathcal L,I}q
    \quad\Longleftrightarrow\quad
    \forall{}s\in I.\;
    \llbracket p\rrbracket(s)=\llbracket q\rrbracket(s).
  \]
  \normalsize
}

For a single input $s$, executions evidence can
established that the programs agree. Derivations of
$R_\rightarrow(p,s,o)$ and $R_\rightarrow(q,s,o)$ show
that both programs output $o$, and hence that
$p\equiv_{\mathcal L,\{s\}}q$. Derivations yielding different
outputs establish a counterexample to equivalence on any input set
containing $s$.

To establish equivalence over general sets $I$, relational evidence must cover the
admissible inputs and ensure equality of outputs. A derivation of
$R_r(p,q,\Phi,\Phi')$ suffices when $\Phi(s,s)$ holds for
every $s\in I$ and $\Phi'(o_1,o_2)$ implies $o_1=o_2$.
Soundness then ensures that, on every admissible input, the programs
either both diverge or both terminate with equal outputs.

The derivation records intermediate claims and their dependencies.
Section~\ref{sec:technical} presents our compositional relational Hoare
proof framework, which assembles judgments about program fragments into a
whole-program argument.

%% file: problem-formulation.tex
\section{Problem Formulation}
\label{sec:problem-formulation}

A correct execution output or equivalence verdict can be accompanied by
invalid or incomplete evidence. To assess the justification as well as the
answer, we formulate two tasks with explicit evidence requirements. We use
the semantics of Section~\ref{sec:semantic-setting} to define answer
correctness, then specify how to assess the submitted evidence.

\subsection{Reasoning Tasks}
\label{sec:reasoning-tasks}

For both tasks, the model receives the language specification as part of
the instance and returns an answer together with evidence supporting it.

\paperparagraph{Execution.}
An execution task $\tau_\rightarrow=(p_1,p_2,s)$ is a triple consisting of: two programs $p_1,p_2\in\mathrm{Prog}$ and an initial state $s \in \States$ that
$p_1$ and $p_2$ terminate on.
The goal of the task is to compute a pair of valid execution traces $t_1,t_2$,
concluding $R_\rightarrow(p_1,s,o_1)$ and $R_\rightarrow(p_2,s,o_2)$.
The task succeeds when $o_1 = o_2$ exactly when the semantics compute the
same output. Namely,
{
  \small
  \[
    \textsc{Success}_\rightarrow(\tau_\rightarrow,t_1,t_2) := (o_1 = o_2) \Leftrightarrow
    (\llbracket p_1 \rrbracket(s) = \llbracket p_2 \rrbracket(s)) \wedge (t_1,t_2) \text{ valid}.
  \]
  \normalsize
}



\paperparagraph{Relational Reasoning.}
A relational reasoning task $\tau_r=(p_1,p_2,I)$ is a triple consisting of two programs $p_1,p_2\in\mathrm{Prog}$ and a set of admissible input states $I \subseteq \States$. The goal of this task is to compute a relational derivation $d$ concluding
that the two outputs are equal exactly when the two programs have the
same semantics. When the programs are semantically distinct, the model should
respond \textsc{None}. Namely, if $d$ is the response, then:
{
  \vspace{-5pt}
  \small
  \[
    \textsc{Success}_r(p_1,p_2,I) :=
    \begin{cases}
      \top & \text{if } R_r(p_1,p_2,s_1\in I \wedge s_2\in I \wedge
      s_1=s_2,o_1 = o_2) \text{ is the root of
      $d$} \\
      & \wedge \ p_1\equiv_{\mathcal L,I}p_2 \wedge d \text{ valid}\\
      \top & \text{if } d = \textsc{None} \wedge p_1\not\equiv_{\mathcal L,I}p_2\\
      \bot & \text{otherwise}
    \end{cases}
  \]
  \normalsize
}

%

\subsection{Evidence Assessment and Failure Modes}
\label{sec:evidence-assessment}

The success conditions in Section~\ref{sec:reasoning-tasks} require valid
evidence, defined by the criteria below.

\paperparagraph{Syntax and well-formedness.}
A \emph{syntax error} prevents the response from parsing under the required
grammar. A \emph{well-formedness error} occurs when a response parses but
cannot be interpreted as a trace or proof, for example because it applies an
operator to the wrong number of arguments or refers to a proof node that
does not exist.

\paperparagraph{Derivation validity.}
Each step of a proof either applies a rule or assumes a claim. A rule
application must follow the supplied execution or proof rules, including
their premises and side conditions; an assumed claim must hold under the
language semantics, which we check directly. A step that fails its check,
such as an incorrect execution step, a misapplied rule, or a claim or
entailment with a counterexample, is \emph{invalid}. A step whose check is
inconclusive, for example because a solver times out, is \emph{unresolved}.

\paperparagraph{Completeness.}
Every premise that the derivation requires must be justified by a rule
application or an assumption. A proof that leaves a required premise
unjustified, for example by not connecting a rule premise to any derivation,
is \emph{incomplete}.

\paperparagraph{Connection to the task goal.}
Each execution trace must start from its program and the given initial state
and end with the reported output. An equivalence proof must concern the given
program pair and establish, for every admissible input, agreement on
termination and equal outputs whenever both programs terminate. Evidence is \emph{accepted} only if it is well formed, every step is checked
valid, it is complete, and it establishes the task goal.

%% file: proof-framework.tex
\section{A Compositional Proof Framework for Equivalence Verification}\label{sec:technical}

We design the proof framework to check equivalence arguments for the
task of Section~\ref{sec:reasoning-tasks} while letting the model decide
how to construct them. The model chooses which parts of the programs to
compare and what relations to prove between their states.
The framework provides rules for checking whether the resulting argument
establishes the required equivalence. If the argument is incomplete, the
framework records which claims still need justification.

To check the resulting argument, the framework separates the justification
of individual claims from their composition. A \emph{local proof layer}
justifies claims under the supplied language semantics, while a
\emph{composition layer} checks how they connect to the whole-program
goal. A completed argument must establish both the local claims and their
connections.

Our compositional proof system expresses intermediate claims as relational
judgments, which describe how two program fragments transform related
states. We adapt relational Hoare logic (RHL)~\cite{nick2004rhl} to justify
these claims locally, using the semantics of the language under study. This
lets the model focus on the behavior of individual fragments when developing
its argument. The composition layer then combines claims about successive
fragments or different properties of the same programs to establish the
whole-program goal.

We first introduce the relational proof interface
(Section~\ref{sec:proof-foundations}) and local relational reasoning
(Section~\ref{sec:local-reasoning}), then the composition system
(Section~\ref{sec:proof-system}). Finally, we establish soundness
(Section~\ref{sec:proof-guarantees}).

\subsection{Relational Proof Interface}
\label{sec:proof-foundations}
\label{sec:judgments}

We use relational judgments to specify the behavior of pairs of program
fragments. Throughout this section, we fix a deterministic object language
with stores
$\rhlStates=(\rhlVars\rightarrow\rhlValues)$, where $\rhlValues$ is a
nonempty value domain. We use \emph{program fragment} to refer to any
well-formed command of the underlying language, including an entire
program. Each fragment either terminates in a final store or diverges.
Appendix~\ref{sec:object-language} gives the formal language interface.

We express the preconditions and postconditions of relational judgments in
an assertion language with a fixed interpretation $\rhlInterp$. Each
assertion $\rhlPhi$ denotes a relation
$\rhlSem{\rhlPhi}\subseteq\rhlStates\times\rhlStates$: a pair of stores
satisfies $\rhlPhi$ when it belongs to this relation. Within an assertion,
$\rhlTagOne{x}$ and $\rhlTagTwo{x}$ refer to the value of variable $x$ in
the first and second store, respectively. All judgments and implications
below use this interpretation. Appendix~\ref{app:proof-foundations}
defines the assertion language and its connection to program expressions
under the supplied language semantics.

\begin{definition}[Relational judgment]
  Let $\rhlC$ and $\rhlD$ be program fragments, and let $\rhlPhi$ and
  $\rhlPhiPrime$ be assertions in this language. The judgment
  $
  \rhlJudgmentForm{\rhlC}{\rhlD}{\rhlPhi}{\rhlPhiPrime}
  $
  has \emph{precondition} $\rhlPhi$ and \emph{postcondition}
  $\rhlPhiPrime$. It is \emph{semantically valid} if, for every pair of
  initial stores satisfying $\rhlPhi$, the fragments either both diverge
  or both terminate in final stores satisfying $\rhlPhiPrime$.
\end{definition}

\subsection{Local Relational Reasoning}
\label{sec:local-reasoning}

We instantiate the local proof layer with RHL~\cite{nick2004rhl}, adapting its rules to the supplied language
semantics. In this layer, the judgment
$
\rhlJudgment{\rhlC}{\rhlD}{\rhlPhi}{\rhlPhiPrime}
$
records a derivation of the corresponding relational claim using the
adapted RHL rules. We require these rules to be sound for the supplied
language under the relational validity condition of
Section~\ref{sec:proof-foundations}.

RHL lets the model justify one fragment claim at a time, starting from
the assumptions supplied by its precondition and establishing the
relation specified by its postcondition. The derivation records the rule
applications and assertion implications that justify this relation,
allowing the local argument to be checked against the supplied semantics.
These local relations can describe how different intermediate values
correspond across the two executions, while the surrounding argument
uses them to establish equality of the final outputs.

\begin{example}
  Consider the fragments $x\rhlAssign x+1$ and $y\rhlAssign y+2$ under
  ordinary integer arithmetic and assignment semantics. Let
  $\rhlPhi\equiv\rhlTagTwo{y}=2\rhlTagOne{x}$. RHL derives
  \[
  \rhlJudgment
  {x\rhlAssign x+1}
  {y\rhlAssign y+2}
  {\rhlPhi}
  {\rhlPhi}.
  \]
  The assignment and consequence rules reduce this claim to the
  implication
  \[
  \rhlTagTwo{y}=2\rhlTagOne{x}
  \ \Longrightarrow \
  \rhlTagTwo{y}+2=2(\rhlTagOne{x}+1),
  \]
  obtained by substituting the assigned expressions into the
  postcondition. This implication holds by integer arithmetic, so the
  two updates preserve the relation that the second value is twice the
  first.
\end{example}

Once a local derivation is complete, the composition system uses its
conclusion through the \textsc{Embed} rule, described next.

\subsection{\proofSysName}
\label{sec:proof-system}

Our {\proofSysName} (\proofSysAbbrev) lets a model develop an equivalence
argument by proposing fragment claims and determining how they would
establish the target judgment. The key design choice is to keep these
hypotheses separate from the rules governing their use. The model chooses
the claims and their decomposition; fixed rules specify the premises
that must be justified before the argument is accepted.
This separation lets us distinguish a claim awaiting proof from an
invalid attempt to justify it, as required by
Section~\ref{sec:evidence-assessment}.

Composition uses the conclusions of local proofs and assertion checks,
allowing its rules to be shared across language instantiations. We
realize this design using relational sequencing, conjunction, and
consequence~\cite{nick2004rhl,banerjee2016hypotheses}, together with
reasoning under hypotheses~\cite{banerjee2016hypotheses}.

A composition judgment has the form
$
\proofSysContextJudgment{\proofSysAssumptions}{\prog}{\progprime}{\rhlPhi}{\rhlPhiPrime},
$
where {\proofSysAssumptions} is the finite set of depending assumed relational
judgments. The {\proofSysAbbrev} and RHL
subscripts distinguish composition derivations from local proofs; both use the relational meaning defined in
Section~\ref{sec:proof-foundations}.

To develop a derivation of this judgment, the model constructs a
\emph{proof skeleton}: a finite tree of proposed rule applications
rooted at the target judgment, possibly with open leaves for premises
still to be justified. The skeleton records how the model proposes to
connect its fragment claims. The model may revise these claims or supply
missing justifications as it develops the tree; the rules constrain the
completed proof without prescribing a construction order.

Figure~\ref{fig:composition-rules} presents the rules of the system. We
explain them according to three roles: decomposing the goal, connecting
fragment claims, and establishing local claims from hypotheses.

\begin{figure}[t]
  \centering
  \small
  \setlength{\abovedisplayskip}{3pt}
  \setlength{\belowdisplayskip}{3pt}
  \setlength{\abovedisplayshortskip}{3pt}
  \setlength{\belowdisplayshortskip}{3pt}
  \[
    \frac{
      \begin{array}{c}
        \prog \equiv \subprogone \rhlSeq \subprogtwo
        \qquad
        \progprime \equiv
        \subprogoneprime \rhlSeq \subprogtwoprime
        \\[1mm]
        \proofSysContextJudgment
        {{\proofSysAssumptions}_1}
        {\subprogone}
        {\subprogoneprime}
        {\rhlPhiOne}
        {\rhlPhiOnePrime}
        \qquad
        \proofSysContextJudgment
        {{\proofSysAssumptions}_2}
        {\subprogtwo}
        {\subprogtwoprime}
        {\rhlPhiTwo}
        {\rhlPhiTwoPrime}
        \\[1mm]
        \rhlEntailmentDerivation{\rhlPhi}{\rhlPhiOne}
        \qquad
        \rhlEntailmentDerivation{\rhlPhiOnePrime}{\rhlPhiTwo}
        \qquad
        \rhlEntailmentDerivation{\rhlPhiTwoPrime}{\rhlPhiPrime}
      \end{array}
    }{
      \proofSysContextJudgment
      {{\proofSysAssumptions}_1\cup{\proofSysAssumptions}_2}
      {\prog}
      {\progprime}
      {\rhlPhi}
      {\rhlPhiPrime}
    }
    \;[\text{\textsc{Sequential Composition}}]
  \]

  \[
    \frac{
      \begin{array}{c}
        \proofSysContextJudgment
        {{\proofSysAssumptions}_1}
        {\prog}
        {\progprime}
        {\rhlPhiOne}
        {\rhlPhiOnePrime}
        \quad
        \cdots
        \quad
        \proofSysContextJudgment
        {{\proofSysAssumptions}_n}
        {\prog}
        {\progprime}
        {\rhlPhiN}
        {\rhlPhiNPrime}
        \\[1mm]
        \rhlEntailmentDerivation
        {\rhlPhi}
        {\displaystyle\bigwedge_{i=1}^{n}\rhlPhiIndexed{i}}
        \qquad
        \rhlEntailmentDerivation
        {\displaystyle\bigwedge_{i=1}^{n}\rhlPhiIndexedPrime{i}}
        {\rhlPhiPrime}
        \qquad n\geq1
      \end{array}
    }{
      \proofSysContextJudgment
      {\textstyle\bigcup_{i=1}^{n}{\proofSysAssumptions}_i}
      {\prog}
      {\progprime}
      {\rhlPhi}
      {\rhlPhiPrime}
    }
    \;[\text{\textsc{Conjunctive Composition}}]
  \]

  \[
    \frac{
      \rhlContextJudgment
      {\proofSysAssumptions}
      {\prog}
      {\progprime}
      {\rhlPhi}
      {\rhlPhiPrime}
    }{
      \proofSysContextJudgment
      {\proofSysAssumptions}
      {\prog}
      {\progprime}
      {\rhlPhi}
      {\rhlPhiPrime}
    }
    \;[\text{\textsc{Embed}}]
    \qquad\qquad
    \frac{
      \rhlEntailmentCheck{\Psi}{\Theta}=\rhlOracleValid
    }{
      \rhlEntailmentDerivation{\Psi}{\Theta}
    }
    \;[\text{\textsc{Oracle}}]
  \]

  \[
    \frac{
      {}
    }{
      {\proofSysAssumptions},
      \rhlJudgmentForm
      {\prog}
      {\progprime}
      {\rhlPhi}
      {\rhlPhiPrime}
      \rhlDerives_{\mathrm{RHL}}
      \rhlJudgmentForm
      {\prog}
      {\progprime}
      {\rhlPhi}
      {\rhlPhiPrime}
    }
    \;[\text{\textsc{Assumption}}]
  \]

  \[
    \frac{
      \begin{array}{c}
        \rhlJudgment
        {\subprogone}
        {\subprogoneprime}
        {\rhlPhiOne}
        {\rhlPhiOnePrime}
        \\[1mm]
        {\proofSysAssumptions},
        \rhlJudgmentForm
        {\subprogone}
        {\subprogoneprime}
        {\rhlPhiOne}
        {\rhlPhiOnePrime}
        \rhlDerives_{\mathrm{RHL}}
        \rhlJudgmentForm
        {\subprogtwo}
        {\subprogtwoprime}
        {\rhlPhiTwo}
        {\rhlPhiTwoPrime}
      \end{array}
    }{
      \rhlContextJudgment
      {\proofSysAssumptions}
      {\subprogtwo}
      {\subprogtwoprime}
      {\rhlPhiTwo}
      {\rhlPhiTwoPrime}
    }
    \;[\text{\textsc{Obligation Discharge}}]
  \]
  \caption{Rules of the composition proof framework. Sequential and Conjunctive
    Composition combine established claims and collect their assumptions. Embed and Oracle connect composition
    to local proofs and assertion checks. Assumption and Obligation Discharge
  record and resolve hypotheses within local proofs.}
  \Description{Six inference rules in order: Sequential Composition, Conjunctive
    Composition, Embed and Oracle on one line, Assumption, and Obligation Discharge.
  Each rule shows its premises above its conclusion.}
  \vspace{-10pt}
  \label{fig:composition-rules}
\end{figure}

\paperparagraph{Decomposing the goal.}
The model can decompose a goal in two complementary ways.
\textsc{Sequential Composition} splits each program into two successive
fragments and reduces the goal to claims relating the corresponding pairs.
\textsc{Conjunctive Composition} separates the properties to be established
about the same program pair. It combines $n\geq1$ judgments whose
postconditions jointly imply the target
postcondition, provided the target precondition entails every premise
precondition. Each property can use its own decomposition into fragments,
so the model can choose boundaries suited to that property. For $n=1$,
the rule reduces to consequence, adapting the precondition and
postcondition of a single judgment.

\vspace{-4pt}
\begin{example}[Decomposing by property]
  \label{ex:property-decomposition}
  Consider the following extension of the fragments in
  Section~\ref{sec:local-reasoning}, with the same ordinary integer
  arithmetic and assignment semantics:
  {
    \small
    \[
      \begin{aligned}
        \prog &\equiv x\rhlAssign x+1\rhlSeq
        z\rhlAssign 2\times x\rhlSeq w\rhlAssign z+1,\\
        \progprime &\equiv y\rhlAssign y+2\rhlSeq
        w\rhlAssign y+1\rhlSeq z\rhlAssign y.
      \end{aligned}
    \]
    \normalsize
  }
  From the precondition $\rhlTagTwo{y}=2\rhlTagOne{x}$, we wish to
  establish both $\rhlTagOne{z}=\rhlTagTwo{z}$ and
  $\rhlTagOne{w}=\rhlTagTwo{w}$. To prove equality of $z$, we can split
  $\prog$ after its second assignment and $\progprime$ after its first.
  To prove equality of $w$, we can instead split $\prog$ after its first
  assignment and $\progprime$ after its second. Each split separates the
  computation of the relevant output in one program from its computation
  in the other. Sequential Composition combines the fragment claims for
  each output, and Conjunctive Composition combines the resulting claims
  about the complete program pair.
\end{example}
\vspace{-5pt}

\paperparagraph{Connecting fragment claims.}
\label{sec:entailment}
Once the model has chosen a decomposition, its fragment claims must
connect to establish the desired property. The composition rules express
these connections through implications between assertions. We record a
checked implication as
$
\rhlEntailmentDerivation{\Psi}{\Theta}.
$
This judgment establishes that, under the fixed interpretation
$\rhlInterp$, every state pair satisfying $\Psi$ also satisfies
$\Theta$. Here, both assertions concern the same pair of stores.

In \textsc{Sequential Composition}, the first pair's postcondition must
imply the second pair's precondition. This
connecting implication lets the local claims use different assertions,
provided that the first supplies what the second requires. The remaining
entailments connect these claims to the target's precondition and
postcondition.

For the $z$-property in Example~\ref{ex:property-decomposition}, the
chosen prefixes can establish
$\Psi\equiv\rhlTagOne{z}=2\rhlTagOne{x}\land
\rhlTagTwo{y}=2\rhlTagOne{x}$, while the suffixes only require
$\Theta\equiv\rhlTagOne{z}=\rhlTagTwo{y}$ to establish equal final
values of $z$. Checking $\rhlEntailmentDerivation{\Psi}{\Theta}$ connects
these two claims.

We represent implication checking by an entailment oracle
$\rhlEntailmentOracle_{\rhlInterp}$, which validates claims under
$\rhlInterp$. The \textsc{Oracle} rule establishes the entailment judgment
when $\rhlEntailmentCheck{\Psi}{\Theta}=\rhlOracleValid$.
We require this result to certify that $\Psi$ entails $\Theta$, but do
not require the oracle to decide every implication. Any entailment checker satisfying the soundness condition in
Appendix~\ref{app:entailment-validation} can serve as the oracle.

\paperparagraph{Establishing local claims.}
\label{sec:obligations}
An open relational premise in the skeleton identifies what a fragment
pair needs to establish before a local proof is available. The model
can develop the surrounding argument while this premise remains open
and return to its justification later.

A local proof may itself use another fragment claim whose proof is still
missing. We record these dependencies in a finite set
{\proofSysAssumptions} of assumed relational judgments; implication
premises must be justified separately through \textsc{Oracle}. We write
$
\rhlContextJudgment
{\proofSysAssumptions}
{\prog}
{\progprime}
{\rhlPhi}
{\rhlPhiPrime}
$
for a local RHL derivation under these assumptions. Each assumed judgment
is a \emph{proof obligation} on which the local derivation remains
conditional.

\textsc{Assumption}, \textsc{Obligation Discharge}, and \textsc{Embed}
govern how a proposed local claim becomes an established result.
Assumption permits a local derivation to use an unproved relational
judgment and records that dependency in {\proofSysAssumptions}. The
context records which claims are assumed, while the local proof rules
continue to govern every inference. Obligation Discharge removes a
dependency by substituting a completed, assumption-free RHL proof for
uses of the assumed judgment. Once all other premises are justified,
Embed imports the local conclusion into the composition system,
preserving its programs and assertions together with any assumptions that
remain undischarged. The composition rules collect the assumptions of
their premises, so the context of a composition judgment records every
assumed claim on which it still depends.

For example, a local RHL proof of the $z$-property for the complete
program pair can use the prefix claim as a hypothesis. Even after the
suffix reasoning is complete, this proof remains conditional on that
claim. Supplying a completed, assumption-free local proof of the prefix
claim allows Obligation Discharge to remove the dependency; otherwise,
the prefix claim remains an assumption of the $z$-property after Embed. With all
premises justified, Embed makes the $z$-property
available for composition with the $w$-property. The model chooses where
to use \textsc{Embed}: it can justify a large fragment as a unit or decompose it
further.

\paperparagraph{Completing a proof.}
A well-formed skeleton rooted at the requested judgment becomes a
completed composition proof when all its premises are justified and
every proposed rule application is validated. Each composition step
must match a rule in Figure~\ref{fig:composition-rules}, each
\textsc{Embed} premise must be supplied by a completed local RHL proof,
and each entailment premise must be established through \textsc{Oracle}.
A completed proof establishes its root judgment under the assumptions in
its context; it is \emph{assumption-free} when this context is empty.
Appendix~\ref{app:proof-skeletons} defines proof skeletons and the conditions
for a completed composition proof.

\subsection{Soundness}
\label{sec:proof-guarantees}

Composition soundness ensures that a completed composition proof
establishes the relational judgment at its root whenever the assumptions
in its context are valid.

The local RHL rules and entailment checker must be sound for the chosen
language instantiation. Sequential Composition also requires ordinary
sequencing semantics: the second fragment starts in the first fragment's
final store, and the sequence diverges if the first fragment diverges or
the second diverges from that store. Appendix~\ref{sec:object-language}
states this requirement formally.

\vspace{-5pt}
\begin{proposition}[Composition soundness]
  \label{prop:composition-soundness}
  Fix a deterministic language instantiation with interpretation \(\rhlInterp\), ordinary sequencing semantics, sound local RHL rules, and a sound entailment oracle.
  Let $\pi$ be a completed composition proof with root
  $
  \proofSysContextJudgment{\proofSysAssumptions}{\prog}{\progprime}{\rhlPhi}{\rhlPhiPrime},
  $
  and suppose every judgment in {\proofSysAssumptions} is semantically
  valid. Then for every $(s_1,s_2)\in\rhlSem{\rhlPhi}$, the executions
  of $\prog$ from $s_1$ and $\progprime$ from $s_2$
  either both diverge or both terminate in final stores
  $s'_1,s'_2$ satisfying
  $(s'_1,s'_2)\in\rhlSem{\rhlPhiPrime}$.
\end{proposition}
\vspace{-5pt}
\begin{proof}[Proof sketch]
  We argue by induction on $\pi$. Embed supplies
  valid local judgments by contextual soundness of the local rules, since
  every remaining assumption is valid, and Oracle supplies valid
  entailments by the soundness of the entailment oracle. Sequential Composition connects the
  intermediate stores and preserves agreement on termination. Conjunctive
  Composition combines properties of the same deterministic executions, so
  all its postconditions hold of the same final stores. The full proof
  appears in Appendix~\ref{app:composition-soundness}.
\end{proof}

%% file: language-design.tex
\section{Framework Instantiation}
\label{sec:language-design}

We instantiate the semantic setting of Section~\ref{sec:semantic-setting}
and the compositional proof framework of Section~\ref{sec:technical} on
two variants of a imperative language {\proclang}, called
{\proclangobs} and {\proclangalt}. We first
describe the languages and the design choices relevant to our evaluation.
We then instantiate the local proof interface with an assertion
interpretation and checking procedures for these languages.

\subsection{Evaluation Languages}
\label{sec:familiarity-controls}
\label{sec:synthetic-semantics}

\begin{wrapfigure}{R}{0.50\linewidth}
  \centering
  \footnotesize
  \setlength{\jot}{1pt}
  \hrule
  \vspace{2pt}
  \(
    \begin{gathered}
      \rhlVar \rhlIn \rhlVars
      \quad
      \rhlNum \rhlIn \rhlInts
      \quad
      \rhlBool \rhlIn \{\rhlTrue,\rhlFalse\}
      \\[2pt]
      \begin{aligned}
        \rhlE &::= \rhlNum \mid \rhlVar
        \mid \rhlE + \rhlE
        \mid \rhlE \times \rhlE
        \mid \rhlE - \rhlE
        \\
        \rhlB &::= \rhlBool
        \mid \rhlE < \rhlE
        \mid \rhlE = \rhlE
        \mid \rhlNeg\,\rhlB
        \mid \rhlB \rhlAnd \rhlB
        \\[2pt]
        \rhlC &::= \rhlSkip
        \mid \rhlVar \rhlAssign \rhlE
        \mid \rhlC \rhlSeq \rhlC
        \\[-1pt]
        &\phantom{::=}\mathrel{}\mid
        \rhlIf\ \rhlB\ \rhlThen\ \rhlC\ \rhlElse\ \rhlC
        \\[-1pt]
        &\phantom{::=}\mathrel{}\mid
        \rhlWhile\ \rhlB\ \rhlDo\ \rhlC
      \end{aligned}
      \\[3pt]
      \text{\proclangobs\ substitutions (throughout):}
      \\[1pt]
      {+}\mapsto\proclangAdd
      \quad {\times}\mapsto\proclangMul
      \quad {-}\mapsto\proclangSub
      \\
      {<}\mapsto\proclangLt
      \quad {=}\mapsto\proclangEq
      \quad {\rhlNeg}\mapsto\proclangNot
      \quad {\rhlAnd}\mapsto\proclangAnd
    \end{gathered}
  \)\par
  \vspace{1pt}
  \hrule
  \caption{{\proclang} syntax and Obs mappings.}
  \Description{Conventional ProcLang grammar for integer expressions, Boolean
    expressions, and commands. The ProcLang-Obs annotation replaces addition
    by a square, multiplication by a triangle, subtraction by a circle,
    less-than by a star, equality by a diamond, negation by a hash, and
  conjunction by a boxed plus. These substitutions apply throughout.}
  \label{fig:proclang-syntax}
\end{wrapfigure}

We instantiate the framework in an imperative setting that supports
execution and equivalence reasoning about arithmetic, branching, and
iteration. Keeping the languages compact lets us supply their syntax and
semantics with each task in the LLM prompt, while defining them ourselves lets us vary the
programming conventions that models encounter. This control matters because a model may interpret a familiar language using prior knowledge in addition to the supplied specification. Following prior work~\cite{wu-etal-2024-reasoning,sharma2026esolangbench},
we use unfamiliar operator notation and inverted control-flow guards
to reduce the influence of familiar programming conventions.

\noindent{\proclang} operates on integer-valued stores and includes arithmetic and
Boolean expressions, assignment, sequential composition, ordinary
if-else statements, and while loops. It uses conventional operator symbols.
The {\proclangobs} variant (``Obs'' for ``obscure'') replaces the arithmetic,
comparison, and Boolean operator symbols with uncommon ones. For example, $\proclangAdd$
denotes addition and $\proclangMul$ denotes multiplication. Thus the
language retains familiar operations while changing their notation.

\noindent{\proclangalt} shares {\proclangobs}'s stores, expressions, assignment, and
sequencing, and changes its control-flow semantics. Its grammar replaces
the ordinary conditional and loop keywords throughout with
$\proclangAltIf$ and $\proclangAltWhile$. The $\proclangAltIf$ conditional executes its second
branch when the guard is true and its first branch otherwise.
The $\proclangAltWhile$ loop repeats its body while the guard is false
and stops when the guard is true.
Figure~\ref{fig:proclang-syntax} gives the grammar of {\proclang} and the
operator substitutions defining {\proclangobs}.

For these languages, an execution's observable output consists of the final
values of all variables still in scope at program termination. Equivalence
requires agreement on termination and equal observable outputs whenever
the programs terminate, following Section~\ref{sec:equivalence-task}.

\par\WFclear
\subsection{Proof Checking}
\label{sec:instantiation}
\label{sec:checker-implementation}

The framework of Section~\ref{sec:technical} checks how relational claims
compose, while their meaning and local justification depend on the
language. For {\proclangobs} and {\proclangalt}, we instantiate this local
interface by fixing the value domain, assertion interpretation, command
rules, and procedures for assessing local claims and entailments. This
connects the abstract proof judgments to the program behavior defined in
Section~\ref{sec:familiarity-controls}.

\paperparagraph{Assertion interpretation.}
We take $\rhlValues=\rhlInts$, so stores assign integer values to
variables. Relational assertions compare two such stores:
$\rhlTagOne{x}$ denotes the value of $x$ in the first store, and
$\rhlTagTwo{x}$ its value in the second, as in
Section~\ref{sec:proof-foundations}. The interpretation $\rhlInterp$
assigns the arithmetic and comparison symbols the integer operations
and relations specified in Section~\ref{sec:familiarity-controls}.
Logical connectives in assertions retain their usual meanings. Both
languages use this interpretation; Appendix~\ref{app:language-instantiation}
gives the full expression-interface construction.

\paperparagraph{Local rule checking.}
Our verifier checks relational derivations against the local RHL rules,
including the program fragments, assertions, and premises of each
application. The command rules must reflect the selected language's
control flow. For {\proclangobs}, conditionals and loops use their
ordinary guard conditions. For {\proclangalt}, premises for executing
the first conditional branch require a false guard, and those for the
second require a true guard; loop-body premises likewise use a false
guard.

\paperparagraph{Semantic checks.}
We instantiate the entailment oracle of Section~\ref{sec:entailment} with
cvc5, and use cvc5 to validate the assumptions that remain in a completed
proof. Values and arithmetic operations are encoded as bit-vectors, and
loops are handled by bounded unrolling. Appendix~\ref{app:implementation}
gives the implementation details.

\paperparagraph{Composition and the task goal.}
The common composition checker uses the local judgments and entailments
under this interpretation. Its rules are shared across both languages;
acceptance requires their premises to be established, the remaining
assumptions to be validated, and the root judgment to concern the task's
program pair, cover its admissible inputs,
and establish termination agreement and equality of the final values
still in scope.

%% file: evaluation.tex
\section{Evaluation}
\label{sec:evaluation}

We evaluate models at three levels: executing programs under supplied
semantics, constructing equivalence proofs without feedback, and
revising those proofs with verifier assistance. We then use the resulting
proof records to analyze the validity, composition, and derivational coverage
of the submitted arguments.

\begingroup
\settowidth{\leftmargini}{\labelitemi}
\addtolength{\leftmargini}{\labelsep}
\begin{itemize}
  \item RQ1:
    How often do LLMs construct valid equivalence proofs without verifier
    feedback, and how accurately can LLMs judge proofs' local claims?

  \item RQ2:
    How much does verifier feedback improve proof correctness?

  \item RQ3:
    What do equivalence-proof records reveal about LLM program reasoning?
\end{itemize}
\endgroup

\subsection{Experimental Setup}
\label{sec:experimental-setup}

\paperparagraph{Benchmarks.}
\label{sec:data-generation}
For each of {\proclangobs} and {\proclangalt}, we evaluate all five models on
the same 100 program pairs. Each pair is used in three tasks. In the
\emph{execution} task, the model executes both programs on the same concrete input using the supplied semantic rules. In the \emph{free-form} task, it decides whether the programs satisfy the relational goal and justifies its decision in natural language. In the \emph{proof} task, it constructs a proof of the relational goal using the supplied proof rules. Each pair includes a relational precondition and postcondition specifying the required input and output relations. The pairs correspond across the two languages, with control-flow constructs rewritten to preserve behavior under {\proclangalt}'s inverted
guards.

We use Gemini-3.8-Flash~\cite{googledeepmind2026gemini38flash} to generate program
templates and propose structurally different rewrites. A sampler inserts
computations from a library of \emph{local transformations} into
these generated contexts; each entry pairs a direct implementation with a more
involved alternative (Appendix~\ref{app:base-programs}). We use the number of
insertions (\emph{insertion count}) as a heuristic control on proof difficulty.
Each language contains 35, 36, and 29 pairs with zero, one, and two insertions,
respectively. Across both programs, a pair contains between 25 and 86 statement
occurrences (median 42).

We retain only pairs for which our cvc5-based checker validates the required relational goal. In this case, every retained pair is labeled \texttt{EQUIVALENT}, so the benchmark evaluates the ability to establish equivalence rather than to discriminate between equivalent and inequivalent pairs.

\paperparagraph{Free-form reasoning.}
\label{sec:free-form-protocol}
The free-form task follows the setting of Section~\ref{sec:overview}. The
model receives the language grammar and semantic rules, the two programs, and
the relational precondition and postcondition, and is asked whether every pair
of initial states satisfying the precondition leads, whenever both programs
terminate, to final states satisfying the postcondition. It returns a
decision, \texttt{EQUIVALENT} or \texttt{NOT\_EQUIVALENT}, with numbered
reasoning steps in natural language.

\paperparagraph{Proof-generation protocols.}
\label{sec:proof-generation}
Every proof-construction prompt contains the grammar, RHL and
composition rules, two programs, and relational pre- and
post-condition, but not the language's semantic rules. We compare two protocols: \emph{without feedback} and \emph{verifier feedback}. Without feedback, the model returns one proof. If the proof's JSON does not parse, a deterministic procedure repairs it; nothing is returned to the model.

Verifier feedback adds solver checks and feedback around simple generation. With verifier feedback, a proof that does not parse or is malformed
is returned to the model for up to three parsing corrections. Afterwards, cvc5 checks each assumption the model
left underived. Assumptions that cvc5 refutes are recorded and kept as they
are. The remaining assumptions are \emph{refined}: each is posed to the model
as a sub-goal, and simple generation creates a sub-proof for it, whose own
assumptions are refined in turn, up to a recursive depth of 3. A sub-proof
that parses replaces its assumption, raising derivational coverage; if the
model declines or no sub-proof parses, the assumption stays. Because
sub-proofs are checked only after they are inserted, refinement can also
introduce errors, and cvc5's verdicts are not shown to the model at this
stage. Finally, the step verifier and cvc5 check the refined proof. The proof
and its problems (failed proof steps, unreadable assertions, and assumptions
or connecting entailments that cvc5 refutes)
are given to the model as feedback for another full proof-generation step, in
which it revises its proof. This loop runs for at most two rounds of
feedback, so a task receives at most three proofs; it stops early once the
verifier reports no problems or a round yields no parsable proof. We report
the best of these proofs.

\paperparagraph{Semantic judgment.}
To test whether a model can check its own proof, we extract the leaves of the
proofs it produced, its assumed judgments and its connecting entailments, skipping
trivial assumptions. A separate call to the same model
receives one leaf together with the language's semantic rules and returns a
verdict (\texttt{true}, \texttt{false}, or \texttt{indeterminte}) with a
justification. This is the only proof related call that includes the semantic
rules. We compare each verdict with cvc5's; verdicts never affect acceptance
and are never returned to proof construction.

\paperparagraph{Models and generation settings.}
We evaluate GPT-6.1-Sol~\cite{openai2026gpt61sol},
Opus-5.5~\cite{anthropic2026opus55}, GLM-5.3~\cite{zai2026glm53},
GPT-6-Luna~\cite{openai2026gpt6luna}, and
Qwen3.8-27B~\cite{qwen2026qwen3827b}. The set spans four model families and includes both proprietary and open-weight models. Sol, Luna, and Qwen use the \texttt{xhigh} reasoning-effort setting; GLM and Opus use \texttt{high}.\footnote{GLM does not support the \texttt{xhigh} setting. For Opus, we use \texttt{high} because \texttt{xhigh} leaves insufficient capacity for the configured output-token budget.} The output-token budget is 128,000 per call, except for Qwen's proof-generation calls, which allow 262,144; Qwen's execution calls use 128,000. Temperature is 0.7 for GLM and 1.0 for Qwen; requests to the other models omit temperature. We report one retained proof-generation run per model--task combination.
Appendix~\ref{app:setup} gives serving details, token usage, and cost.

\paperparagraph{Use of AI in the research.}
LLMs play two roles in the study: they help construct the benchmark and are the systems evaluated on program reasoning. For benchmark construction, Gemini-3.8-Flash generates program templates and proposes transformations; a sampler instantiates the templates, and the cvc5-based checker filters the resulting pairs as described above. In the evaluation, the five models generate execution predictions, free-form decisions, and proofs, including parsing repairs, recursive refinements, and revisions under the verifier-assisted protocol. We assess these outputs against reference executions and the formal proof checks described in this paper. The semantic judgment stage uses the same generation settings as proof construction (Appendix~\ref{app:checking-procedures}).

\paperparagraph{Execution metrics.}
\emph{Pair execution accuracy} is the percentage of pairs
for which both predicted outputs match their reference outputs on the
declared observable variables. We use \emph{execution rule F1} as an
approximate measure of trace agreement. It compares the multisets of rule
applications in the predicted and reference traces,
pooling both sides of each pair. Each token records the program side and
rule label; repeated applications count separately, while AST identifiers
and rule order are ignored. We report the mean pair-level F1, giving
partial credit for overlap in rule usage.

\paperparagraph{Decision and proof metrics.}
\emph{Decision correctness} is the percentage of free-form responses whose
decision is \texttt{EQUIVALENT}, the correct decision for every benchmark pair.
\emph{Proof correctness} is the percentage of tasks whose final proof is
a completed composition proof of the requested judgment
(Section~\ref{sec:proof-system}) and whose remaining assumptions cvc5
validates (Section~\ref{sec:checker-implementation}).
Because accepted proofs can leave parts of the argument to cvc5 in this way,
we also measure \emph{derivational coverage}: the number of accepted proofs
with no remaining assumptions, and the fraction of statement occurrences
in the two programs that remaining assumptions cover.
For the semantic-judgment stage, \emph{cvc5 agreement} is the percentage of
judged claims on which the model's verdict matches cvc5's, counting both
joint acceptance and joint rejection; we exclude claims lacking a Boolean
model verdict or a decided cvc5 result.
Finally, we classify each proof's diagnostics by the assessment criteria of
Section~\ref{sec:evidence-assessment}; RQ3 introduces the finer categories
it reports.


\subsection{RQ1: Proof Construction and Semantic Judgment}
\label{sec:proof-results}

RQ1 asks two halves of one question: can a model construct an accepted proof
without verifier assistance, and can it judge the leaves of its proofs?
Unassisted proof acceptance (Table~\ref{tab:one-shot-results}) answers the
first; comparing leaf verdicts with cvc5
(Table~\ref{tab:first-iteration-semantic-judgments}) answers the second.

\begin{table*}[t]
  \centering
  \footnotesize
  \setlength{\tabcolsep}{1.5pt}
  \caption{Execution, free-form decisions, and proof construction without
    feedback on 100 program pairs per model and language. Outputs is pair
    execution accuracy; Rule F1 measures agreement in rule-application counts.
    Free-form is decision correctness in free-form reasoning. Proof scores
    the first response after deterministic JSON cleanup, before model feedback
    or refinement.
  All numeric entries are percentages.}
  \label{tab:one-shot-results}
  \begin{tabular}{@{}lrrrr@{\hspace{6pt}}rrrr@{}}
    \toprule
    & \multicolumn{4}{c}{{\proclangobs}}
    & \multicolumn{4}{c}{{\proclangalt}} \\
    \cmidrule(lr){2-5}\cmidrule(l){6-9}
    & \multicolumn{2}{c}{Execution} & \multicolumn{2}{c}{Equivalence}
    & \multicolumn{2}{c}{Execution} & \multicolumn{2}{c}{Equivalence} \\
    \cmidrule(lr){2-3}\cmidrule(lr){4-5}
    \cmidrule(lr){6-7}\cmidrule(l){8-9}
    Model & Outputs & Rule F1 & Free-form & Proof
    & Outputs & Rule F1 & Free-form & Proof \\
    \midrule
    GPT-6.1-Sol & {100} & {100.00} & {100} & {94}
    & {99} & {97.88} & {100} & {94} \\
    Opus-5.5 & 96 & 99.79 & {99} & 59 & 93 & 76.86 & {83} & 50 \\
    GLM-5.3 & 72 & 93.79 & {93} & 53 & 55 & 58.76 & 80 & 53 \\
    GPT-6-Luna & 83 & 95.41 & 91 & 45 & 73 & 87.45 & 66 & 36 \\
    Qwen3.8-27B & 96 & 97.76 & 98 & 39 & 91 & 70.41 & 96 & 41 \\
    \midrule
    Overall & 89.4 & 97.35 & 96.2 & 58 & 82.2 & 78.27 & 85.0 & 54.8 \\
    \bottomrule
  \end{tabular}
  \vspace{-6pt}
\end{table*}

\paperparagraph{Models generally succeed on the preliminary execution task.}
We first ask whether models can apply the supplied semantic rules to a single
concrete input. On {\proclangobs} and
{\proclangalt}, both outputs are correct for
858/1,000 program pairs: 447/500
(89.4\%) on {\proclangobs} and 411/500 (82.2\%) on {\proclangalt}. Output correctness, however, does not capture every
error in the reported reasoning: mean pair-level rule F1 falls from 97.35\%
on {\proclangobs} to 78.27\% on {\proclangalt}. Every model has lower F1 on
{\proclangalt}, and the largest reductions occur for GLM (93.79\% to 58.76\%) and
Qwen (97.76\% to 70.41\%). Together, these results show that models can
generally reason about individual executions, while inverted control flow
makes their rule-level reasoning less reliable.

\paperparagraph{Proof correctness varies far more than decision correctness.}
Reasoning in free form, the models decide correctly on {906}/1,000 tasks
({90.6}\%), and every model does so on at least {157} of its 200 tasks.
Proof correctness is much more variable: Sol's proofs are accepted on 188/200 tasks (94\%) and Opus's on
109 (54.5\%), whereas GLM, Luna, and Qwen obtain 106, 81, and 80 accepted
proofs. Overall, 564/1,000 proofs (56.4\%) are accepted. On 377 tasks, a
model decides correctly in free form yet does not produce an accepted proof.
Execution accuracy does not explain this ranking: Qwen predicts both outputs
correctly on 96\% and 91\% of pairs but produces the fewest accepted proofs.
The cross-language gap is also smaller for proof construction: inverted
guards lower rule F1 by 19 points, but proof correctness by fewer than nine.

\begin{wraptable}{r}{0.32\linewidth}
  \vspace{-10pt}
  \centering
  \footnotesize
  \setlength{\tabcolsep}{3pt}
  \caption{LLM--cvc5 agreement on the first composed proof (\%).}
  \label{tab:first-iteration-semantic-judgments}
  \begin{tabular}{@{}lr@{}}
    \toprule
    Model & Agreement \\
    \midrule
    GPT-6.1-Sol & 99.98\% \\
    Opus-5.5 & 99.97\% \\
    GLM-5.3 & 97.93\% \\
    GPT-6-Luna & 98.23\% \\
    Qwen3.8-27B & 96.21\% \\
    \midrule
    Overall & 98.82\% \\
    \bottomrule
  \end{tabular}
\end{wraptable}

\paperparagraph{Models usually agree with cvc5 on local claims.}
We compare Boolean model verdicts with decided cvc5 results for assumed
judgments and nontrivial connecting entailments from the first composed
iteration, after refinement and before whole-proof revision. The models
agree with cvc5 on 98.82\% of the usable checks, ranging from 96.21\%
for Qwen to 99.98\% for Sol
(Table~\ref{tab:first-iteration-semantic-judgments}).
This high agreement is unsurprising because semantic judgment is a much more
constrained task than constructing a complete proof: the model receives a
single candidate claim together with the language semantics and only needs to
assess whether that claim holds. Proof construction additionally requires
choosing intermediate relations, applying and composing proof rules, and
assembling the resulting claims into a valid derivation. Nevertheless, the models still incorrectly accept
27 of 411 refuted claims (6.6\%), indicating that a check based solely on the model's
verdict is not sufficient for correctness.

\rqtakeaway{1}{Equivalence reasoning must be assessed through the evidence
models supply.}{%
  Checked proofs expose differences that answer accuracy misses. Independent
validation also catches false claims accepted by the models' own judgments.}

\subsection{RQ2: Proof Generation with Verifier Feedback}
\label{sec:feedback-results}

\begin{figure}[!htbp]
  \centering
  {\small $\circ$ Unassisted \qquad $\bullet$ Verifier assisted\par}
  \smallskip
  \begin{minipage}[t]{0.49\linewidth}
    \centering
    \begin{tikzpicture}
      \begin{axis}[
          width=0.97\linewidth,
          height=4.2cm,
          title={{\proclangobs}},
          title style={font=\small},
          xmin=0, xmax=100,
          ymin=0.5, ymax=5.5,
          xtick={0,20,40,60,80,100},
          ytick={1,2,3,4,5},
          yticklabels={Qwen,Luna,GLM,Opus,Sol},
          ytick style={draw=none},
          xlabel={Proof correctness (\%)},
          axis y line*=left,
          y axis line style={draw=none},
          axis x line*=bottom,
          xmajorgrids,
          grid style={black!12},
          tick align=outside,
          tick label style={font=\scriptsize},
          label style={font=\small},
          clip=false,
        ]
        \draw[black!40,line width=0.7pt]
        (axis cs:94,5) -- (axis cs:99,5);
        \draw[black!40,line width=0.7pt]
        (axis cs:59,4) -- (axis cs:99,4);
        \draw[black!40,line width=0.7pt]
        (axis cs:53,3) -- (axis cs:79,3);
        \draw[black!40,line width=0.7pt]
        (axis cs:45,2) -- (axis cs:73,2);
        \draw[black!40,line width=0.7pt]
        (axis cs:39,1) -- (axis cs:46,1);

        \addplot[
          only marks, mark=o,
          mark size=2.2pt,
          mark options={solid,draw=black,fill=white,line width=0.7pt},
          point meta=x, nodes near coords,
          every node near coord/.append style={
          font=\scriptsize,anchor=east,xshift=-3pt}
        ] coordinates {
          (94,5) (59,4) (53,3) (45,2) (39,1)
        };

        \addplot[
          only marks, mark=*,
          mark size=2.2pt,
          mark options={solid,draw=black,fill=black},
          point meta=x, nodes near coords,
          every node near coord/.append style={
          font=\scriptsize,anchor=west,xshift=3pt}
        ] coordinates {
          (99,5) (99,4) (79,3) (73,2) (46,1)
        };
      \end{axis}
    \end{tikzpicture}
  \end{minipage}\hfill
  \begin{minipage}[t]{0.49\linewidth}
    \centering
    \begin{tikzpicture}
      \begin{axis}[
          width=0.97\linewidth,
          height=4.2cm,
          title={{\proclangalt}},
          title style={font=\small},
          xmin=0, xmax=100,
          ymin=0.5, ymax=5.5,
          xtick={0,20,40,60,80,100},
          ytick={1,2,3,4,5},
          yticklabels={Qwen,Luna,GLM,Opus,Sol},
          ytick style={draw=none},
          xlabel={Proof correctness (\%)},
          axis y line*=left,
          y axis line style={draw=none},
          axis x line*=bottom,
          xmajorgrids,
          grid style={black!12},
          tick align=outside,
          tick label style={font=\scriptsize},
          label style={font=\small},
          clip=false,
        ]
        \draw[black!40,line width=0.7pt]
        (axis cs:94,5) -- (axis cs:100,5);
        \draw[black!40,line width=0.7pt]
        (axis cs:50,4) -- (axis cs:96,4);
        \draw[black!40,line width=0.7pt]
        (axis cs:53,3) -- (axis cs:80,3);
        \draw[black!40,line width=0.7pt]
        (axis cs:36,2) -- (axis cs:74,2);
        \draw[black!40,line width=0.7pt]
        (axis cs:41,1) -- (axis cs:51,1);

        \addplot[
          only marks, mark=o,
          mark size=2.2pt,
          mark options={solid,draw=black,fill=white,line width=0.7pt},
          point meta=x, nodes near coords,
          every node near coord/.append style={
          font=\scriptsize,anchor=east,xshift=-3pt}
        ] coordinates {
          (94,5) (50,4) (53,3) (36,2) (41,1)
        };

        \addplot[
          only marks, mark=*,
          mark size=2.2pt,
          mark options={solid,draw=black,fill=black},
          point meta=x, nodes near coords,
          every node near coord/.append style={
          font=\scriptsize,anchor=west,xshift=3pt}
        ] coordinates {
          (100,5) (96,4) (80,3) (74,2) (51,1)
        };
      \end{axis}
    \end{tikzpicture}
  \end{minipage}
  \caption{Proof correctness with and without verifier feedback on
    100 tasks per model and language. Unassisted results score the first
  response; assisted results score the selected final proof.}
  \Description{Two panels compare first-response and selected-final proof
    correctness. On ProcLang-Obs, Sol improves from 94 to 99 percent,
    Opus from 59 to 99, GLM from 53 to 79, Luna from 45 to 73,
    and Qwen from 39 to 46. On ProcLang-Alt, the respective changes
  are 94 to 100, 50 to 96, 53 to 80, 36 to 74, and 41 to 51 percent.}
  \label{fig:assisted-proof-success}
  \vspace{-6pt}
\end{figure}
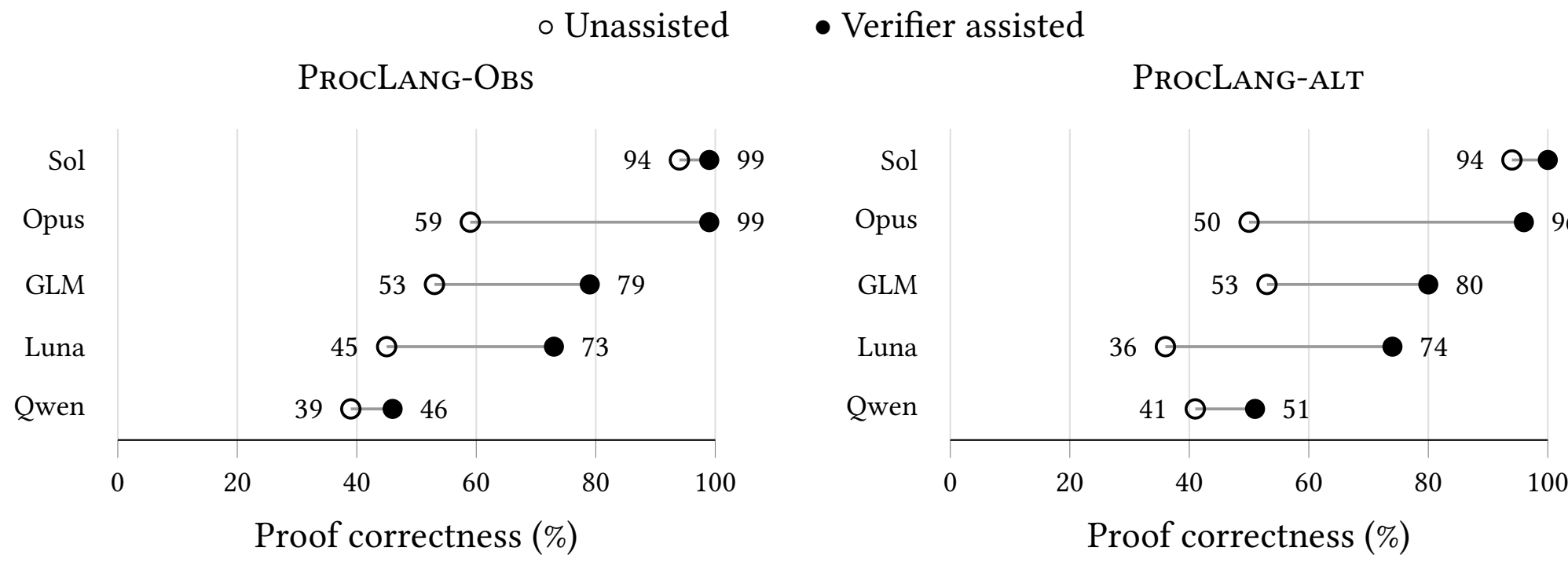

RQ2 measures the effect of verifier and cvc5 diagnostics on
whole-proof revision. Figure~\ref{fig:assisted-proof-success} compares proof
correctness with and without verifier feedback on the
model--task evaluations.

Across the 1,000 model--task evaluations, proof correctness rises
from 56.4\% without feedback to 79.7\% with verifier feedback.
Every model improves in both languages. Pooling the two languages,
the largest increase is for Opus, from 54.5\% to 97.5\%, followed
by Luna, from 40.5\% to 73.5\%, and GLM, from 53.0\% to 79.5\%. These gains show that the assisted workflow can
repair many initially unsuccessful proofs across models.

\paperparagraph{Recovery varies by the initial failure.}
We classify each initially rejected proof by the first acceptance check it
fails. Recovery is highest for proofs that parse and conclude the required
goal but contain a rejected derivation step: 125 of 159 (78.6\%) have an
accepted selected-final proof. These step failures include incorrect rule
applications and premises that do not match earlier conclusions. Initially
unparsed responses recover on 126 of 170 tasks (74.1\%), compared with
35 of 61 (57.4\%) for failed local checks and 17 of 46 (37.0\%) for
goal mismatches. Rejected derivation steps also remain the most common
failure after assistance, in 127 of the 203 tasks without an accepted
proof, so step-level proof construction is both the main opportunity for
revision and the main remaining bottleneck.

\rqtakeaway{2}{The proof interface supports construction as well as
assessment.}{%
  The same interface that exposes reasoning errors also supports
  verifier-assisted construction of checked arguments for behavior
preservation on these tasks.}

\subsection{RQ3: What Equivalence Proofs Reveal about Program Reasoning}
\label{sec:reasoning-results}
\label{sec:qualitative-results}

RQ3 compares how models reason about program behavior through their proofs. Table~\ref{tab:proof-diagnostics} summarizes the types and frequencies of errors in model-generated proofs, following the criteria in Section~\ref{sec:evidence-assessment}.
We then examine two findings acceptance rates hide:
individually valid claims can compose into an invalid argument, and accepted
proofs differ in how much of the argument the model derives.

\begin{table}[!htbp]
  \vspace{-4pt}
  \centering
  \caption{Proof diagnostics per model over 200 tasks (100 per language):
    tasks whose first response has each outcome, grouped by
    the criteria of Section~\ref{sec:evidence-assessment}. A proof can have
  several outcomes.}
  \label{tab:proof-diagnostics}
  \footnotesize
  \setlength{\tabcolsep}{4pt}
  \begin{tabular}{@{}lrrrrrr@{}}
    \toprule
    & Syntax & \multicolumn{3}{c}{Derivation validity}
    & Completeness & Task goal \\
    \cmidrule(lr){3-5}
    Model & Malformed & Rule & Claim & Connection & Unlinked & Mismatch \\
    \midrule
    GPT-6.1-Sol & 1  & 6  & 2  & 3  & 1  & 0  \\
    Opus-5.5    & 58 & 2  & 7  & 5  & 23 & 5  \\
    GLM-5.3     & 51 & 11 & 32 & 3  & 2  & 10 \\
    GPT-6-Luna  & 44 & 77 & 31 & 29 & 57 & 21 \\
    Qwen3.8-27B & 76 & 25 & 16 & 3  & 4  & 10 \\
    \bottomrule
  \end{tabular}
  \vspace{-6pt}
\end{table}

\paperparagraph{The assessment criteria give each model a failure profile.} Table~\ref{tab:proof-diagnostics} shows different obstacles to
first-response proof construction.
Beyond malformed output, the models fail in different parts of the
derivation. Luna's errors are mostly
structural, with rule errors in 77 tasks and unlinked premises in 57,
whereas GLM's are mostly refuted assumed claims (32 tasks); Sol and Opus
have at most 7 tasks in any derivation-validity category.
Goal mismatches are rare (at most 21 tasks per model), so the models
differ in the evidence they supply rather than in the claim they target.

\definecolor{rq3Sol}{HTML}{0072B2}
\definecolor{rq3Opus}{HTML}{E69F00}
\definecolor{rq3GLM}{HTML}{009E73}
\definecolor{rq3Luna}{HTML}{D55E00}
\definecolor{rq3Qwen}{HTML}{CC79A7}
\begin{figure}[!htbp]
  \vspace{-4pt}
  \centering
  {\sffamily\footnotesize\color{black!80} \tikz[baseline=-0.6ex]\fill[rq3Sol] (0,0) circle (1.2mm);\,Sol\enspace \tikz[baseline=-0.6ex]\fill[rq3Opus] (0,0) circle (1.2mm);\,Opus\enspace \tikz[baseline=-0.6ex]\fill[rq3GLM] (0,0) circle (1.2mm);\,GLM\enspace \tikz[baseline=-0.6ex]\fill[rq3Luna] (0,0) circle (1.2mm);\,Luna\enspace \tikz[baseline=-0.6ex]\fill[rq3Qwen] (0,0) circle (1.2mm);\,Qwen\qquad
  \tikz[baseline=-0.6ex]\node[circle, minimum width=2.4mm, minimum height=2.4mm, inner sep=0pt, fill=black!45, font=\sffamily\bfseries\tiny, text=white] {0};\,\tikz[baseline=-0.6ex]\node[circle, minimum width=2.4mm, minimum height=2.4mm, inner sep=0pt, fill=black!45, font=\sffamily\bfseries\tiny, text=white] {1};\,\tikz[baseline=-0.6ex]\node[circle, minimum width=2.4mm, minimum height=2.4mm, inner sep=0pt, fill=black!45, font=\sffamily\bfseries\tiny, text=white] {2};\; insertions\enspace \tikz[baseline=-0.6ex]\node[rounded rectangle, minimum width=4.2mm, minimum height=2.4mm, inner sep=0pt, fill=black!45, font=\sffamily\bfseries\tiny, text=white] {1--2};\; merged\par}
  \smallskip
  \begin{tikzpicture}
    \begin{axis}[scale only axis, height=3.0cm, xmin=0, xmax=104, ymin=-5, ymax=104,
        xtick={0,25,50,75,100}, ytick={0,25,50,75,100}, xticklabels={0,25,50,75,100},
        tick label style={font=\sffamily\scriptsize, text=black!55},
        label style={font=\sffamily\scriptsize, text=black!75},
        title style={font=\sffamily\footnotesize, text=black!85, at={(0,1)}, anchor=south west, xshift=-2pt, yshift=1pt},
        grid style={black!9, line width=0.45pt}, xmajorgrids, ymajorgrids,
      axis line style={draw=none}, major tick length=0pt, clip=false, width=3.45cm, at={(0cm,0cm)}, anchor=north west, title={\textbf{(a)} Unassisted}, ylabel={Assumed statements (\%)}, yticklabels={0,25,50,75,100}, xticklabels={}]
      \draw[black!28, line width=0.7pt, -{Stealth[length=4pt, width=3.5pt]}] (axis cs:84,50) -- (axis cs:100,32);
      \node[font=\sffamily\itshape\scriptsize, text=black!45, anchor=south east, inner sep=1pt] at (axis cs:86,50) {better};
      \node[circle, minimum width=2.4mm, minimum height=2.4mm, fill=rq3Qwen, draw=white, line width=0.6pt, inner sep=0pt, font=\sffamily\bfseries\tiny, text=black!85] at (axis cs:40.00,94.10) {};
      \node[font=\sffamily\scriptsize, text=black!75, anchor=west, inner sep=1.2pt] at (axis cs:43.00,91.10) {Qwen};
      \node[circle, minimum width=2.4mm, minimum height=2.4mm, fill=rq3Luna, draw=white, line width=0.6pt, inner sep=0pt, font=\sffamily\bfseries\tiny, text=white] at (axis cs:40.50,93.91) {};
      \node[font=\sffamily\scriptsize, text=black!75, anchor=west, inner sep=1.2pt] at (axis cs:24.50,98.00) {Luna};
      \node[circle, minimum width=2.4mm, minimum height=2.4mm, fill=rq3GLM, draw=white, line width=0.6pt, inner sep=0pt, font=\sffamily\bfseries\tiny, text=white] at (axis cs:53.00,77.41) {};
      \node[font=\sffamily\scriptsize, text=black!75, anchor=west, inner sep=1.2pt] at (axis cs:56.00,77.41) {GLM};
      \node[circle, minimum width=2.4mm, minimum height=2.4mm, fill=rq3Opus, draw=white, line width=0.6pt, inner sep=0pt, font=\sffamily\bfseries\tiny, text=black!85] at (axis cs:54.50,29.86) {};
      \node[font=\sffamily\scriptsize, text=black!75, anchor=south, inner sep=1.2pt] at (axis cs:54.50,34.66) {Opus};
      \node[circle, minimum width=2.4mm, minimum height=2.4mm, fill=rq3Sol, draw=white, line width=0.6pt, inner sep=0pt, font=\sffamily\bfseries\tiny, text=white] at (axis cs:94.00,16.29) {};
      \node[font=\sffamily\scriptsize, text=black!75, anchor=west, inner sep=1.2pt] at (axis cs:97.00,16.29) {Sol};
    \end{axis}
    \begin{axis}[scale only axis, height=3.0cm, xmin=0, xmax=104, ymin=-5, ymax=104,
        xtick={0,25,50,75,100}, ytick={0,25,50,75,100}, xticklabels={0,25,50,75,100},
        tick label style={font=\sffamily\scriptsize, text=black!55},
        label style={font=\sffamily\scriptsize, text=black!75},
        title style={font=\sffamily\footnotesize, text=black!85, at={(0,1)}, anchor=south west, xshift=-2pt, yshift=1pt},
        grid style={black!9, line width=0.45pt}, xmajorgrids, ymajorgrids,
      axis line style={draw=none}, major tick length=0pt, clip=false, width=8.4cm, at={(3.9000000000000004cm,0cm)}, anchor=north west, title={\textbf{(b)} Unassisted {\color{black!50}\,\textperiodcentered\, by insertion count}}, yticklabels={}, xticklabels={}]
      \draw[rq3Luna!45, line width=1.3pt, line cap=round, line join=round] (axis cs:28.57,83.27) -- (axis cs:44.44,95.65) -- (axis cs:50.00,97.60);
      \draw[rq3Qwen!45, line width=1.3pt, line cap=round, line join=round] (axis cs:37.14,85.74) -- (axis cs:41.54,97.24);
      \draw[rq3GLM!45, line width=1.3pt, line cap=round, line join=round] (axis cs:65.71,79.20) -- (axis cs:54.17,73.86) -- (axis cs:36.21,79.26);
      \draw[rq3Opus!45, line width=1.3pt, line cap=round, line join=round] (axis cs:72.86,23.73) -- (axis cs:58.33,28.76) -- (axis cs:27.59,48.07);
      \draw[rq3Sol!45, line width=1.3pt, line cap=round, line join=round] (axis cs:95.07,12.68) -- (axis cs:91.38,23.22);
      \node[circle, minimum width=2.4mm, minimum height=2.4mm, fill=rq3Luna, draw=white, line width=0.6pt, inner sep=0pt, font=\sffamily\bfseries\tiny, text=white] at (axis cs:50.00,97.60) {2};
      \node[circle, minimum width=2.4mm, minimum height=2.4mm, fill=rq3Luna, draw=white, line width=0.6pt, inner sep=0pt, font=\sffamily\bfseries\tiny, text=white] at (axis cs:44.44,95.65) {1};
      \node[circle, minimum width=2.4mm, minimum height=2.4mm, fill=rq3Luna, draw=white, line width=0.6pt, inner sep=0pt, font=\sffamily\bfseries\tiny, text=white] at (axis cs:28.57,83.27) {0};
      \node[font=\sffamily\scriptsize, text=black!75, anchor=west, inner sep=1.2pt] at (axis cs:53.00,97.60) {Luna};
      \node[rounded rectangle, minimum width=4.2mm, minimum height=2.4mm, fill=rq3Qwen, draw=white, line width=0.6pt, inner sep=0pt, font=\sffamily\bfseries\tiny, text=black!85] at (axis cs:41.54,97.24) {1--2};
      \node[circle, minimum width=2.4mm, minimum height=2.4mm, fill=rq3Qwen, draw=white, line width=0.6pt, inner sep=0pt, font=\sffamily\bfseries\tiny, text=black!85] at (axis cs:37.14,85.74) {0};
      \node[font=\sffamily\scriptsize, text=black!75, anchor=east, inner sep=1.2pt] at (axis cs:37.34,96.94) {Qwen};
      \node[circle, minimum width=2.4mm, minimum height=2.4mm, fill=rq3GLM, draw=white, line width=0.6pt, inner sep=0pt, font=\sffamily\bfseries\tiny, text=white] at (axis cs:36.21,79.26) {2};
      \node[circle, minimum width=2.4mm, minimum height=2.4mm, fill=rq3GLM, draw=white, line width=0.6pt, inner sep=0pt, font=\sffamily\bfseries\tiny, text=white] at (axis cs:54.17,73.86) {1};
      \node[circle, minimum width=2.4mm, minimum height=2.4mm, fill=rq3GLM, draw=white, line width=0.6pt, inner sep=0pt, font=\sffamily\bfseries\tiny, text=white] at (axis cs:65.71,79.20) {0};
      \node[font=\sffamily\scriptsize, text=black!75, anchor=west, inner sep=1.2pt] at (axis cs:68.61,79.20) {GLM};
      \node[circle, minimum width=2.4mm, minimum height=2.4mm, fill=rq3Opus, draw=white, line width=0.6pt, inner sep=0pt, font=\sffamily\bfseries\tiny, text=black!85] at (axis cs:27.59,48.07) {2};
      \node[circle, minimum width=2.4mm, minimum height=2.4mm, fill=rq3Opus, draw=white, line width=0.6pt, inner sep=0pt, font=\sffamily\bfseries\tiny, text=black!85] at (axis cs:58.33,28.76) {1};
      \node[circle, minimum width=2.4mm, minimum height=2.4mm, fill=rq3Opus, draw=white, line width=0.6pt, inner sep=0pt, font=\sffamily\bfseries\tiny, text=black!85] at (axis cs:72.86,23.73) {0};
      \node[font=\sffamily\scriptsize, text=black!75, anchor=east, inner sep=1.2pt] at (axis cs:24.39,48.07) {Opus};
      \node[circle, minimum width=2.4mm, minimum height=2.4mm, fill=rq3Sol, draw=white, line width=0.6pt, inner sep=0pt, font=\sffamily\bfseries\tiny, text=white] at (axis cs:91.38,23.22) {2};
      \node[rounded rectangle, minimum width=4.2mm, minimum height=2.4mm, fill=rq3Sol, draw=white, line width=0.6pt, inner sep=0pt, font=\sffamily\bfseries\tiny, text=white] at (axis cs:95.07,12.68) {0--1};
      \node[font=\sffamily\scriptsize, text=black!75, anchor=north, inner sep=1.2pt] at (axis cs:92.27,7.48) {Sol};
    \end{axis}
    \begin{axis}[scale only axis, height=3.0cm, xmin=0, xmax=104, ymin=-5, ymax=104,
        xtick={0,25,50,75,100}, ytick={0,25,50,75,100}, xticklabels={0,25,50,75,100},
        tick label style={font=\sffamily\scriptsize, text=black!55},
        label style={font=\sffamily\scriptsize, text=black!75},
        title style={font=\sffamily\footnotesize, text=black!85, at={(0,1)}, anchor=south west, xshift=-2pt, yshift=1pt},
        grid style={black!9, line width=0.45pt}, xmajorgrids, ymajorgrids,
      axis line style={draw=none}, major tick length=0pt, clip=false, width=3.45cm, at={(0cm,-3.65cm)}, anchor=north west, title={\textbf{(c)} Verifier-assisted}, ylabel={Assumed statements (\%)}, yticklabels={0,25,50,75,100}, xlabel={Proof acceptance (\%)}]
      \node[circle, minimum width=2.4mm, minimum height=2.4mm, fill=rq3Qwen, draw=white, line width=0.6pt, inner sep=0pt, font=\sffamily\bfseries\tiny, text=black!85] at (axis cs:50.00,74.61) {};
      \node[font=\sffamily\scriptsize, text=black!75, anchor=north, inner sep=1.2pt] at (axis cs:50.00,69.71) {Qwen};
      \node[circle, minimum width=2.4mm, minimum height=2.4mm, fill=rq3Luna, draw=white, line width=0.6pt, inner sep=0pt, font=\sffamily\bfseries\tiny, text=white] at (axis cs:74.00,76.44) {};
      \node[font=\sffamily\scriptsize, text=black!75, anchor=west, inner sep=1.2pt] at (axis cs:77.00,76.44) {Luna};
      \node[circle, minimum width=2.4mm, minimum height=2.4mm, fill=rq3GLM, draw=white, line width=0.6pt, inner sep=0pt, font=\sffamily\bfseries\tiny, text=white] at (axis cs:79.50,23.18) {};
      \node[font=\sffamily\scriptsize, text=black!75, anchor=west, inner sep=1.2pt] at (axis cs:82.50,23.18) {GLM};
      \node[circle, minimum width=2.4mm, minimum height=2.4mm, fill=rq3Opus, draw=white, line width=0.6pt, inner sep=0pt, font=\sffamily\bfseries\tiny, text=black!85] at (axis cs:97.50,10.84) {};
      \node[font=\sffamily\scriptsize, text=black!75, anchor=east, inner sep=1.2pt] at (axis cs:94.50,16.04) {Opus};
      \node[circle, minimum width=2.4mm, minimum height=2.4mm, fill=rq3Sol, draw=white, line width=0.6pt, inner sep=0pt, font=\sffamily\bfseries\tiny, text=white] at (axis cs:99.50,6.39) {};
      \node[font=\sffamily\scriptsize, text=black!75, anchor=east, inner sep=1.2pt] at (axis cs:96.50,0.99) {Sol};
    \end{axis}
    \begin{axis}[scale only axis, height=3.0cm, xmin=0, xmax=104, ymin=-5, ymax=104,
        xtick={0,25,50,75,100}, ytick={0,25,50,75,100}, xticklabels={0,25,50,75,100},
        tick label style={font=\sffamily\scriptsize, text=black!55},
        label style={font=\sffamily\scriptsize, text=black!75},
        title style={font=\sffamily\footnotesize, text=black!85, at={(0,1)}, anchor=south west, xshift=-2pt, yshift=1pt},
        grid style={black!9, line width=0.45pt}, xmajorgrids, ymajorgrids,
      axis line style={draw=none}, major tick length=0pt, clip=false, width=8.4cm, at={(3.9000000000000004cm,-3.65cm)}, anchor=north west, title={\textbf{(d)} Verifier-assisted {\color{black!50}\,\textperiodcentered\, by insertion count}}, yticklabels={}, xlabel={Proof acceptance (\%)}, xticklabels={,25,50,75,100}]
      \draw[rq3Luna!45, line width=1.3pt, line cap=round, line join=round] (axis cs:78.57,75.95) -- (axis cs:76.39,67.45) -- (axis cs:65.52,87.12);
      \draw[rq3Qwen!45, line width=1.3pt, line cap=round, line join=round] (axis cs:62.86,66.15) -- (axis cs:41.67,71.67) -- (axis cs:44.83,87.91);
      \draw[rq3GLM!45, line width=1.3pt, line cap=round, line join=round] (axis cs:92.86,25.46) -- (axis cs:81.94,23.97) -- (axis cs:60.34,18.74);
      \node[circle, minimum width=2.4mm, minimum height=2.4mm, fill=rq3Luna, draw=white, line width=0.6pt, inner sep=0pt, font=\sffamily\bfseries\tiny, text=white] at (axis cs:65.52,87.12) {2};
      \node[circle, minimum width=2.4mm, minimum height=2.4mm, fill=rq3Luna, draw=white, line width=0.6pt, inner sep=0pt, font=\sffamily\bfseries\tiny, text=white] at (axis cs:76.39,67.45) {1};
      \node[circle, minimum width=2.4mm, minimum height=2.4mm, fill=rq3Luna, draw=white, line width=0.6pt, inner sep=0pt, font=\sffamily\bfseries\tiny, text=white] at (axis cs:78.57,75.95) {0};
      \node[font=\sffamily\scriptsize, text=black!75, anchor=west, inner sep=1.2pt] at (axis cs:80.97,75.95) {Luna};
      \node[circle, minimum width=2.4mm, minimum height=2.4mm, fill=rq3Qwen, draw=white, line width=0.6pt, inner sep=0pt, font=\sffamily\bfseries\tiny, text=black!85] at (axis cs:44.83,87.91) {2};
      \node[circle, minimum width=2.4mm, minimum height=2.4mm, fill=rq3Qwen, draw=white, line width=0.6pt, inner sep=0pt, font=\sffamily\bfseries\tiny, text=black!85] at (axis cs:41.67,71.67) {1};
      \node[circle, minimum width=2.4mm, minimum height=2.4mm, fill=rq3Qwen, draw=white, line width=0.6pt, inner sep=0pt, font=\sffamily\bfseries\tiny, text=black!85] at (axis cs:62.86,66.15) {0};
      \node[font=\sffamily\scriptsize, text=black!75, anchor=north, inner sep=1.2pt] at (axis cs:62.86,62.05) {Qwen};
      \node[circle, minimum width=2.4mm, minimum height=2.4mm, fill=rq3GLM, draw=white, line width=0.6pt, inner sep=0pt, font=\sffamily\bfseries\tiny, text=white] at (axis cs:60.34,18.74) {2};
      \node[circle, minimum width=2.4mm, minimum height=2.4mm, fill=rq3GLM, draw=white, line width=0.6pt, inner sep=0pt, font=\sffamily\bfseries\tiny, text=white] at (axis cs:81.94,23.97) {1};
      \node[circle, minimum width=2.4mm, minimum height=2.4mm, fill=rq3GLM, draw=white, line width=0.6pt, inner sep=0pt, font=\sffamily\bfseries\tiny, text=white] at (axis cs:92.86,25.46) {0};
      \node[font=\sffamily\scriptsize, text=black!75, anchor=south, inner sep=1.2pt] at (axis cs:92.86,28.86) {GLM};
      \node[circle, minimum width=2.4mm, minimum height=2.4mm, fill=rq3Opus, draw=white, line width=0.6pt, inner sep=0pt, font=\sffamily\bfseries\tiny, text=black!85] at (axis cs:97.50,10.84) {};
      \node[font=\sffamily\scriptsize, text=black!75, anchor=east, inner sep=1.2pt] at (axis cs:94.00,16.04) {Opus};
      \node[circle, minimum width=2.4mm, minimum height=2.4mm, fill=rq3Sol, draw=white, line width=0.6pt, inner sep=0pt, font=\sffamily\bfseries\tiny, text=white] at (axis cs:99.50,6.39) {};
      \node[font=\sffamily\scriptsize, text=black!75, anchor=east, inner sep=1.2pt] at (axis cs:96.50,0.99) {Sol};
    \end{axis}
  \end{tikzpicture}
  \caption{Proof acceptance versus the share of statement occurrences
    covered by assumptions in accepted proofs, pooled over both languages.
    Panels (a,b) score first responses. Panels (c,d) count a task as
    accepted if any composed iteration passes and measure coverage in
    its first accepted iteration. Left: overall results over 200 tasks
    per model; right: results by insertion count. Ranges pool the
    indicated insertion counts; unlabeled markers in (d) pool all three.
  Lower right indicates higher acceptance and a smaller assumed share.}
  \Description{Four scatter plots compare proof acceptance and assumed
    statement share. Assistance increases acceptance for every model.
    GLM's assumed share falls substantially, from 77.4 to 23.2 percent.
    Luna and Qwen also improve, but their accepted assisted proofs
    still leave 76.4 and 74.6 percent of statements to assumptions.
    Sol and Opus have the smallest assumed shares under both protocols.
    By insertion count, unassisted Opus acceptance falls from 72.9
    to 27.6 percent, while Luna acceptance rises from 28.6 to 50.0
  percent as its assumed share rises from 83.3 to 97.6 percent.}
  \label{fig:rq3-success-assumptions}
  \vspace{-6pt}
\end{figure}
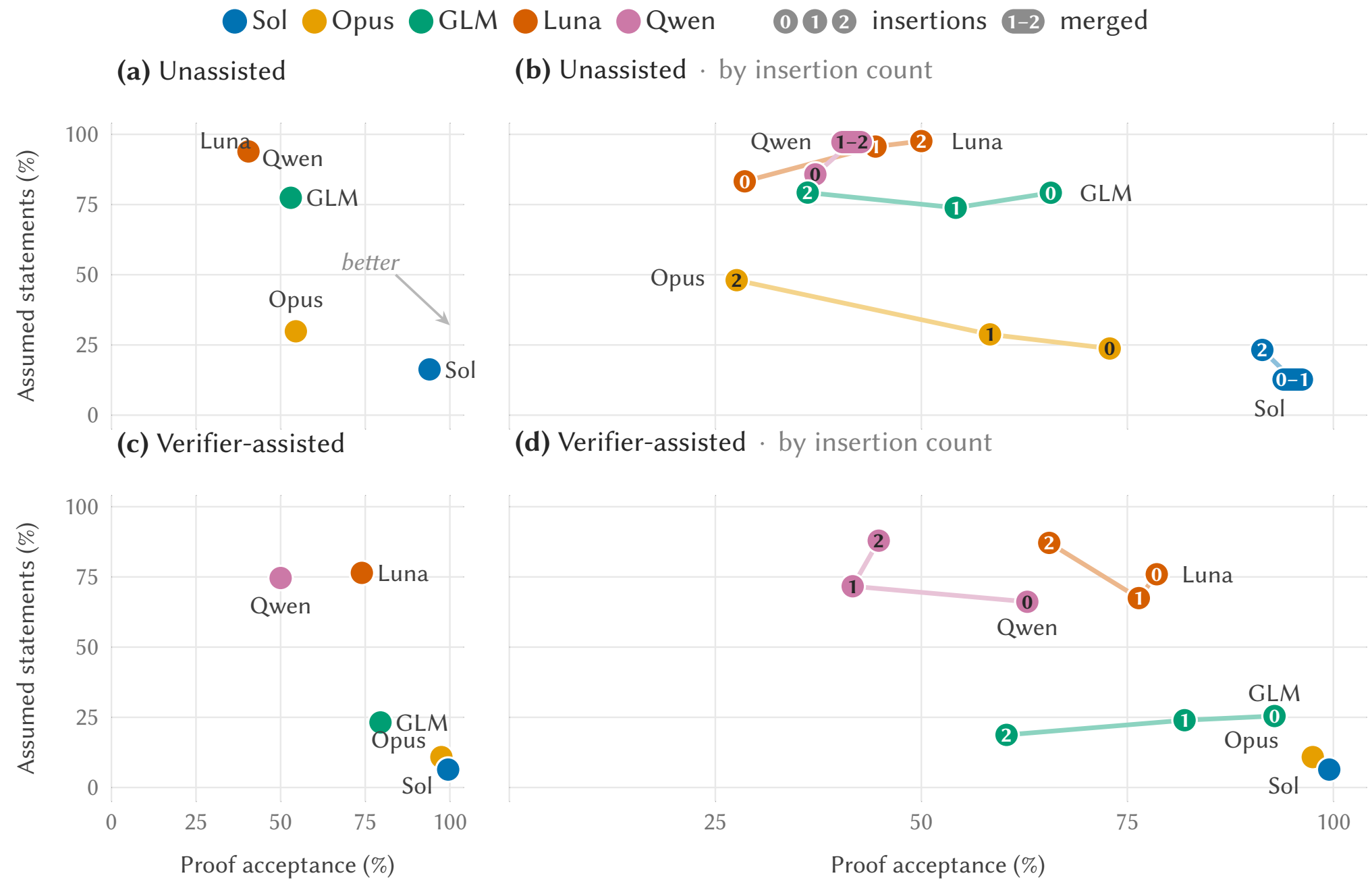

\paperparagraph{Models differ in where their arguments fail.}
Luna's arguments most often fail where valid claims are connected, and
GLM's in an assumed claim itself. Of the 542 first responses in which
cvc5 validates every assumed claim, 32 still contain a refuted connecting
entailment, and 23 of these are Luna's (14.3\% of its 161 such responses).
In 19 of the 23, the refuted entailment is the connecting implication of
\textsc{Sequential Composition}: the claims for two successive fragments
hold, but the first does not supply what the second requires.
GLM instead assumes a false claim most often, in 32 of its 145 responses
with assumptions (22.1\%), compared with 16.0\% for Luna and 12.3\% for Qwen.
Across models, assumed claims about loops are refuted more than twice as
often as loop-free ones (11.8\% versus 5.3\%);
Appendix~\ref{app:evaluation-examples} shows one failure of each kind.

\paperparagraph{Accepted proofs differ in how much the model derives.}
An accepted proof may retain assumptions that cvc5 validates.
Figure~\ref{fig:rq3-success-assumptions} relates proof acceptance
to the share of statement occurrences covered by these assumptions. Without feedback, assumptions
cover 16.3\% and 29.9\% of statement occurrences in Sol's and Opus's
accepted proofs, but 93.9\% and 94.1\% in Luna's and Qwen's, and 38 of
Luna's 81 and 50 of Qwen's 80 accepted proofs derive no statement at all.
Loops separate the models: Sol applies the while rule, which requires a
relational loop invariant, in 77.1\% of these proofs, whereas Luna and Qwen
do so in 17.3\% and 17.5\% and otherwise leave every loop inside an
assumption. Verifier assistance closes the gap only for GLM: its use of the
while rule rises from 31.1\% to 69.8\%, and its assumed share falls from
77.4\% to 23.2\%, whereas Luna and Qwen still assume 76.4\% and 74.6\%.
Because these shares are computed over the proofs accepted under each
protocol, they also reflect which tasks are accepted.

Insertion count shows that higher acceptance need not mean more
derivation. Without
feedback, Luna's acceptance rises from 28.6\% with no insertions to 50.0\%
with two, but the rise comes from proofs that derive nothing: 20 of its 29
accepted proofs with two insertions derive no statement, and without them
its acceptance falls from 27.1\% to 15.5\%.
GLM instead keeps deriving: with assistance, its assumptions stay below
26\% at every insertion count, while its acceptance falls from 92.9\% to
60.3\%. Even Sol derives less as insertions increase: assumption-free proofs
account for 60 of its 68 accepted first responses with no insertions, but
only 16 of 53 with two.

Proof correctness alone cannot separate Sol and Luna: on the 146 tasks
that both pass with their selected-final proofs, Sol produces 93
assumption-free proofs and Luna two, and assumptions cover 6.61\% versus
76.62\% of statement occurrences (Appendix~\ref{app:evaluation-examples}
shows one such task).

\rqtakeaway{3}{Proof success and the model's contribution must be measured
separately.}{%
  Higher acceptance need not mean more model-derived reasoning. Derivational
  coverage measures how much of the argument is left to solver-validated
assumptions.}

%% file: related.tex
\section{Related Work}\label{sec:related}

\paperparagraph{LLM reasoning on program semantics.}
Prior work has evaluated LLMs on execution and runtime
prediction~\cite{gu2024cruxeval,jain2025livecodebench,liu2026codemind,10.1109/ICSE55347.2025.00012,11028290},
formal specification inference~\cite{pmlr-v202-pei23a,10.1145/3660791,le-cong-etal-2025-llms},
and equivalence
checking~\cite{maveli-etal-2025-large,wei-etal-2025-equibench,oza-etal-2026-llms}.
Execution traces have also been used to train models to reason about
program behavior and repair code~\cite{ni2024next}.
EquiBench~\cite{wei-etal-2025-equibench} and
CETBench~\cite{oza-etal-2026-llms} ask models to decide whether two
programs are equivalent. We study the ability to construct a checkable
proof of a supplied equivalence claim.
REFUTE~\cite{sinha2025can} studies the complementary problem of generating
counterexamples: given an incorrect program and a reference solution,
the model produces an input on which they disagree. Executing the
programs on this input provides checkable evidence of inequivalence.

Several studies also examine intermediate execution states, dependency
traces, and explanations, or probe models' sensitivity to misleading
cues~\cite{roy2026codesense,lam2025codecrash,10.1145/3744916.3773272,10.52202/085713-3192,li-etal-2026-beyond-output-correctness}.
Liu et al.~\cite{10.1145/3744916.3773272} check predicted states and their
coherence along concrete program executions; our task asks for a symbolic
proof covering all admissible inputs. Li et
al.~\cite{li-etal-2026-beyond-output-correctness} evaluate explanations
against human annotations and propose VERA, an LLM judge that retrieves
evidence for technical claims. Our checker validates the submitted
derivation using explicit proof rules and solver-checked entailments.
For sequential composition, it checks that a fragment's postcondition
implies the precondition required by the next fragment: individually valid
fragment claims may still fail to form a valid equivalence proof.

\paperparagraph{LLM-guided proof construction.}
There is a growing body of work on using LLMs to generate and repair proofs
in interactive
provers~\cite{10.1145/3611643.3616243,10.1145/3808164,zhang-etal-2024-selene,thakur2024an}
and verification-aware
languages~\cite{10.1145/3763174,aggarwal2025alphaverus}, and to synthesize
program invariants~\cite{wei2026quokka,10.1145/3691620.3695014}.
Other approaches combine language models with automated theorem provers,
using learned prover calls~\cite{jiang2022thor} or formal sketches derived
from informal proofs~\cite{jiang2023draft}.
Related benchmarks evaluate proof generation and verifiable code generation
within existing verification
toolchains~\cite{lin2024fvel,loughridge2024dafnybench,ye2025verina,zeng2026veriequivbench}.

Among these efforts, LLMLift~\cite{bhatia2024verified} is particularly
related to our work. It generates a summary of a source program in a
target DSL together with loop invariants in a fixed proof template, and
uses an SMT solver to check equivalence. Our proof interface lets the
model choose the program decomposition and intermediate relations.
Lemur~\cite{wu2024lemur} similarly combines LLM-proposed invariants with
a verifier, treating the invariants as hypotheses to be discharged.
We use explicit hypotheses about program fragments to assess partial
arguments as well as completed proofs.

LLMs have also been combined with symbolic solvers for natural-language
reasoning~\cite{ye2023satlm,pan2023logiclm,olausson2023linc}.
SatLM~\cite{ye2023satlm} translates problems into declarative
specifications and uses an automated prover to solve them. In our setting,
the model constructs a derivation, and the proof records distinguish
model-derived claims from assumptions validated by the solver.

\paperparagraph{Automated relational verification.}
Program equivalence and transformation correctness are relational verification
problems, with automated approaches based on product
programs~\cite{barthe2011product}, regression
verification~\cite{godlin2009regression}, data-driven
invariant inference~\cite{sharma2013ddec}, execution
alignment~\cite{churchill2019alignment}, learned proof
search~\cite{chen2019coeus}, and translation
validation~\cite{pnueli1998tv,necula2000tv,lopes2021alive2}.
Product programs reduce relational properties to safety properties of a
combined program, while regression verification exploits correspondences
between successive program versions. Execution traces guide
simulation-relation inference in DDEC~\cite{sharma2013ddec} and loop
alignment in semantic program alignment~\cite{churchill2019alignment}.
Coeus~\cite{chen2019coeus} learns strategies for relational proof search.
Our evaluation asks the model to supply the fragment
correspondences and intermediate relations, and checks how they compose.
Translation validation checks individual compiler runs. In particular,
Alive2 validates LLVM optimizations using bounded loop unrolling; our
solver checks of entailments and assumed fragment claims likewise have a
bounded interpretation.

\paperparagraph{Relational program logics.}
Relational program logics prove relationships between executions
compositionally~\cite{nick2004rhl,sousa2016chl}. Banerjee et al. introduce explicit
procedure hypotheses that implementation proofs can
discharge~\cite{banerjee2016hypotheses}, while Maillard et al. separate
generic relational rules from effect-specific
reasoning~\cite{10.1145/3371072}. Our framework adapts these ideas to
model-proposed fragment relations: shared rules compose the relations, and
language-specific checkers validate them.

A key difference is termination. Banerjee et al., Beringer, and Maillard et
al. use partial-correctness interpretations that do not require agreement
on termination~\cite{banerjee2016hypotheses,2033939.2033947,10.1145/3371072}.
Our judgments instead follow Benton's termination-sensitive relational
lifting~\cite{nick2004rhl}; each language instantiation must justify its
local rules accordingly. These proof records let us assess whether the
model's intermediate claims hold, whether they establish the task goal,
and how much of the argument the model derives.

%% file: conclu.tex
\section{Conclusion}
\label{sec:conclusion}
In this paper, we presented a framework that evaluates LLM program
reasoning through relational Hoare logic proof skeletons rather than
verdicts: checkers validate each step, and a solver checks whatever the
model leaves as an assumption. These proofs reveal differences that simple
verdicts hide. Models that usually reach the right verdict differ widely in
whether their proofs check, and accepted proofs range from almost fully
derived to almost fully delegated to the solver. With verifier and solver
feedback, models repair many failed proofs, yet as judges of individual
steps they accept some claims the solver refutes, so their judgment cannot
replace an external check.

The framework also opens questions that simpler benchmark cannot
answer. Language variants can change one semantic feature at a time, as
\proclangalt\ inverts the guards of \proclangobs, to find which features
models fail to reason about once familiar conventions no longer apply. The
checkers accept any relational pre- and postcondition, so the evaluation
could extend beyond equivalence, for instance to noninterference. And
because assumptions mark where each derivation stops, repeated evaluations
across model generations can track the share of each proof that models
derive themselves, even after acceptance saturates. As models improve, the
informative question will be less whether they reach the right answer than
where their checkable reasoning stops.

%% file: acknowledgment.tex
\section*{Acknowledgment}

This work was supported in part by a Canada CIFAR AI Chair award and an NSERC Discovery Grant to JQC, as well as a resource allocation panel grant from Amii. Computational support was provided in part through a Gemini Academic Program awarded to JQC. This research was also enabled in part by support provided by the Digital Research Alliance of Canada.

%% file: proof-framework-appendix.tex
\section{Relational Interface and Composition Soundness}
\label{app:proof-foundations}

This appendix gives the full definitions underlying the relational proof
interface of Section~\ref{sec:proof-foundations}. It defines the parameterized
assertion language (Appendix~\ref{sec:assertions}), its connection to
object-language expressions (Appendices~\ref{sec:object-language}
and~\ref{app:language-instantiation}), and the semantic and validation
judgments used by the composition system
(Appendices~\ref{app:relational-judgments}--\ref{app:proof-skeletons}). It
then proves composition soundness
(Appendix~\ref{app:composition-soundness}) and connects completed proofs to
the equivalence task (Appendix~\ref{app:task-encoding}).

\subsection{Assertion Syntax and Interpretation}
\label{sec:assertions}

Benton's RHL~\cite{nick2004rhl} constructs relational assertions using the object
language's arithmetic, comparison, and Boolean operators over tagged
variables. We generalize this presentation by separating the symbols used in
assertions from their interpretation: a signature determines the available
operation and predicate symbols, and an interpretation assigns their meanings.
The logical connectives and equality have fixed meanings independent of the
object language.

An instantiation connects this assertion language to an object language through
a signature, an interpretation, and translations of program expressions.
Thus, the same assertion syntax can be interpreted differently across language
instantiations, while the logical structure is shared. We first define the
assertion language and then state the conditions on these translations.

Fix a set $\rhlVars$ of program variables and a nonempty value domain
$\rhlValues$. We specialize the state space of
Section~\ref{sec:semantic-setting} to stores
$\rhlStates=(\rhlVars\rightarrow\rhlValues)$, leaving the choice of values to
the language instantiation. We let $s,s_1,s_2$ range over stores and write
$\rhlBools=\{\rhlTrue,\rhlFalse\}$ for the truth values.

\begin{definition}[Signature]
  A \emph{signature} $\rhlSig = (\rhlOps,\rhlPreds)$ consists of a set $\rhlOps$
  of operation symbols and a set $\rhlPreds$ of predicate symbols, together with
  an arity $\rhlArity(\cdot) \rhlIn \rhlNats$ for each symbol. Constant symbols
  are operation symbols of arity zero.
\end{definition}

\begin{definition}[Generalized expressions and relational assertions]
  Fix a signature $\rhlSig$, and let $\rhlConst$ range over its constant symbols
  and $\rhlVar \rhlIn \rhlVars$ over variables. Generalized expressions and
  relational assertions have the syntax
  \[
    \begin{aligned}
      \rhlGExp \ni \rhlGE
      &::= \rhlConst
      \mid \rhlTagOne{\rhlVar}
      \mid \rhlTagTwo{\rhlVar}
      \mid \rhlOpSym(\rhlGE_1,\ldots,\rhlGE_k),\\
      \rhlRelExp \ni \rhlPhi
      &::= \rhlTop
      \mid \rhlBot
      \mid \rhlPredSym(\rhlGE_1,\ldots,\rhlGE_k)
      \mid \rhlGE = \rhlGE\\
      &\quad\ \mid \rhlNeg \rhlPhi
      \mid \rhlPhi \rhlAnd \rhlPhi
      \mid \rhlPhi \rhlOr \rhlPhi
      \mid \rhlPhi \relIff \rhlPhi,
    \end{aligned}
  \]
  where $\rhlOpSym \rhlIn \rhlOps$ and $\rhlPredSym \rhlIn \rhlPreds$, and $k$ is
  the arity of the symbol concerned; operation applications in the last
  expression clause have $k\geq 1$. Binary symbols are written infix. The tags
  $\langle 1\rangle$ and $\langle 2\rangle$ select values from the first and the
  second state of a pair, which is what allows a single assertion to compare the
  two executions.
\end{definition}

The symbols of $\rhlSig$ supply the atoms of an assertion; the connectives
$\rhlNeg$, $\rhlAnd$, $\rhlOr$ and $\relIff$ are fixed and standard, and are
never drawn from the object language. Equality is a \emph{logical} symbol
rather than a member of $\rhlPreds$: it is available for every signature and is
always interpreted as identity on $\rhlValues$. This makes equality of observable
variable values expressible even when the object language supplies no equality
operator.

\begin{definition}[Interpretation]
  An \emph{interpretation} $\rhlInterp$ of $\rhlSig$ over $\rhlValues$ assigns
  to each constant symbol $\rhlConst$ a value
  $\rhlInterp(\rhlConst)\in\rhlValues$, to each
  $\rhlOpSym \rhlIn \rhlOps$ of positive arity a function
  $\rhlInterp(\rhlOpSym) : \rhlValues^{\rhlArity(\rhlOpSym)} \rightarrow \rhlValues$,
  and to each $\rhlPredSym \rhlIn \rhlPreds$ a relation
  $\rhlInterp(\rhlPredSym) \rhlSubseteq \rhlValues^{\rhlArity(\rhlPredSym)}$.
\end{definition}

For the fixed state space, a signature determines which assertions are
well-formed, and an interpretation determines their denotations. In particular,
changing a symbol's interpretation can change the truth of an assertion
without changing its syntax.

\begin{definition}[Denotation]
  \setlength{\jot}{1.5pt}
  Fix an interpretation $\rhlInterp$. Over a state pair
  $(\rhlStateOne,\rhlStateTwo)$, a generalized expression denotes a value in
  $\rhlValues$:
  \[
    \begin{aligned}
      \rhlSem{\rhlConst}(\rhlStateOne,\rhlStateTwo)
      &= \rhlInterp(\rhlConst),\\
      \rhlSem{\rhlTagOne{\rhlVar}}(\rhlStateOne,\rhlStateTwo)
      &= \rhlStateOne(\rhlVar),\\
      \rhlSem{\rhlTagTwo{\rhlVar}}(\rhlStateOne,\rhlStateTwo)
      &= \rhlStateTwo(\rhlVar),\\
      \rhlSem{\rhlOpSym(\rhlGE_1,\ldots,\rhlGE_k)}(\rhlStateOne,\rhlStateTwo)
      &= \rhlInterp(\rhlOpSym)\bigl(
        \rhlSem{\rhlGE_1}(\rhlStateOne,\rhlStateTwo),\ldots,
      \rhlSem{\rhlGE_k}(\rhlStateOne,\rhlStateTwo)\bigr).
    \end{aligned}
  \]
  A relational assertion $\rhlPhi$ denotes a binary relation on states. Its truth
  function $\rhlChi{\rhlPhi}^{\rhlInterp} : \rhlStates \times \rhlStates
  \rightarrow \rhlBools$ is defined recursively by
  \[
    \begin{aligned}
      \rhlChi{\rhlTop}(\rhlStateOne,\rhlStateTwo)
      &= \rhlTrue,\\
      \rhlChi{\rhlBot}(\rhlStateOne,\rhlStateTwo)
      &= \rhlFalse,\\
      \rhlChi{\rhlPredSym(\rhlGE_1,\ldots,\rhlGE_k)}(\rhlStateOne,\rhlStateTwo)
      &= \bigl(
        (\rhlSem{\rhlGE_1}(\rhlStateOne,\rhlStateTwo),\ldots,
        \rhlSem{\rhlGE_k}(\rhlStateOne,\rhlStateTwo))
      \rhlIn \rhlInterp(\rhlPredSym)\bigr),\\
      \rhlChi{\rhlGE_1 = \rhlGE_2}(\rhlStateOne,\rhlStateTwo)
      &= \bigl(\rhlSem{\rhlGE_1}(\rhlStateOne,\rhlStateTwo)
      = \rhlSem{\rhlGE_2}(\rhlStateOne,\rhlStateTwo)\bigr),\\
      \rhlChi{\rhlNeg \rhlPhi}(\rhlStateOne,\rhlStateTwo)
      &= \rhlNeg\, \rhlChi{\rhlPhi}(\rhlStateOne,\rhlStateTwo),\\
      \rhlChi{\rhlPhi \rhlAnd \rhlPhiPrime}(\rhlStateOne,\rhlStateTwo)
      &= \rhlChi{\rhlPhi}(\rhlStateOne,\rhlStateTwo)
      \rhlAnd
      \rhlChi{\rhlPhiPrime}(\rhlStateOne,\rhlStateTwo),\\
      \rhlChi{\rhlPhi \rhlOr \rhlPhiPrime}(\rhlStateOne,\rhlStateTwo)
      &= \rhlChi{\rhlPhi}(\rhlStateOne,\rhlStateTwo)
      \rhlOr
      \rhlChi{\rhlPhiPrime}(\rhlStateOne,\rhlStateTwo),\\
      \rhlChi{\rhlPhi \relIff \rhlPhiPrime}(\rhlStateOne,\rhlStateTwo)
      &= \rhlChi{\rhlPhi}(\rhlStateOne,\rhlStateTwo)
      \relIff
      \rhlChi{\rhlPhiPrime}(\rhlStateOne,\rhlStateTwo).
    \end{aligned}
  \]
  On the right-hand sides, a parenthesized condition stands for its truth
  value, and $\rhlNeg$, $\rhlAnd$, $\rhlOr$, and $\relIff$ are the usual
  operations on $\rhlBools$. The relation denoted by $\rhlPhi$ is
  \[
    \rhlSem{\rhlPhi}
    =
    \rhlSet{
      (\rhlStateOne,\rhlStateTwo)
      \rhlIn \rhlStates \times \rhlStates
      \rhlSuchThat
      \rhlChi{\rhlPhi}(\rhlStateOne,\rhlStateTwo)=\rhlTrue
    }.
  \]
\end{definition}

An assertion can therefore be read in either of two ways: as a formula over two
states, or as the set of state pairs that satisfy it. Both $\rhlSem{\cdot}$ and
$\rhlChi{\cdot}$ depend on the interpretation $\rhlInterp$. We write
$\rhlSem{\rhlPhi}^{\rhlInterp}$ and $\rhlChi{\rhlPhi}^{\rhlInterp}$ when this
dependence matters and suppress the index whenever the interpretation is
fixed by the context.

\begin{definition}[Relational entailment]
  Relational entailment under an interpretation $\rhlInterp$ is defined by
  \[
    \rhlEntailment{\rhlPhi}{\rhlPhiPrime}
    \rhlIff
    \rhlSem{\rhlPhi}^{\rhlInterp}
    \rhlSubseteq
    \rhlSem{\rhlPhiPrime}^{\rhlInterp}.
  \]
  Equivalently, every pair of states satisfying $\rhlPhi$ also satisfies
  $\rhlPhiPrime$ under $\rhlInterp$. The expression
  $\rhlEntailmentClaim{\rhlPhi}{\rhlPhiPrime}$ is a claim about two assertions;
  it is not a constructor of the assertion language.
\end{definition}

\subsection{Object Languages and Tagging}
\label{sec:object-language}

The proof framework uses deterministic object languages whose commands
either terminate in a final store or diverge. We represent these outcomes
in $\rhlLift{\rhlStates}=\rhlStates\cup\{\rhlBottom\}$, where
$\rhlBottom\notin\rhlStates$ denotes divergence (it is unrelated to the
assertion $\rhlBot$ of Appendix~\ref{sec:assertions}). Command denotations
are total functions into this lifted space. This specialization excludes runtime errors and
stuck configurations; they are not identified with divergence.

\begin{definition}[Object language]
  \label{def:object-language}
  An \emph{object language} $\rhlObjLang$ over $\rhlValues$ consists of value expressions,
  Boolean expressions and commands together with denotation functions
  \[
    \begin{aligned}
      \rhlSemObj{\cdot} &: \rhlValExp_{\rhlObjLang} \rightarrow
      (\rhlStates \rightarrow \rhlValues),\\
      \rhlSemObj{\cdot} &: \rhlBoolExp_{\rhlObjLang} \rightarrow
      (\rhlStates \rightarrow \rhlBools),\\
      \rhlSemObj{\cdot} &: \rhlCom_{\rhlObjLang} \rightarrow
      (\rhlStates \rightarrow \rhlLift{\rhlStates}),
    \end{aligned}
  \]
  where the commands include $\rhlSkip$, assignment, sequential composition,
  conditionals and loops (or corresponding constructs).
\end{definition}

To connect this object language to a task language
$\mathcal L=(\mathrm{Prog},\States,O,\llbracket\cdot\rrbracket)$ of
Section~\ref{sec:semantic-setting}, we take $\States=\rhlStates$ and
identify each task program with a command in $\rhlCom_{\rhlObjLang}$.
The object-language denotation returns final stores, while the task
denotation returns outputs in $O$. An observation function
$\operatorname{obs}_{\mathcal L}:\rhlStates\rightarrow O$ connects them.
For every task program $p$ and initial store $s$, we require
\begin{equation}
  \label{eq:store-output-adequacy}
  \begin{aligned}
    \rhlSemObj{p}(s)=\rhlBottom
    &\quad\Longleftrightarrow\quad
    \llbracket p\rrbracket(s)\text{ is undefined},\\
    \rhlSemObj{p}(s)=s'\in\rhlStates
    &\quad\Longrightarrow\quad
    \llbracket p\rrbracket(s)
    =\operatorname{obs}_{\mathcal L}(s').
  \end{aligned}
\end{equation}
Thus the two semantics agree on termination, and observations of final
stores give exactly the task outputs. Fragment composition uses the full
stores; the correspondence above connects the resulting whole-program
claim to the task semantics, as used in
Appendix~\ref{app:task-encoding}.

To instantiate the proof framework, an object language supplies a signature,
an interpretation, and translations of its expressions into the assertion
language.

\begin{definition}[Instantiation]
  For the fixed domain $\rhlValues$, an \emph{instantiation} of the framework
  at an object language $\rhlObjLang$
  is a triple $(\rhlSig,\rhlInterp,\tau)$ consisting of a signature $\rhlSig$, an
  interpretation $\rhlInterp$ of it, and a pair of \emph{tagging} maps
  $\tau = (\tau_1,\tau_2)$,
  \[
    \tau_i : \rhlValExp_{\rhlObjLang} \rightarrow \rhlGExp,
    \qquad
    \tau_i : \rhlBoolExp_{\rhlObjLang} \rightarrow \rhlRelExp,
  \]
  such that for every $\rhlE \rhlIn \rhlValExp_{\rhlObjLang}$, every
  $\rhlB \rhlIn \rhlBoolExp_{\rhlObjLang}$, every $i \rhlIn \{1,2\}$ and every
  state pair $(\rhlStateOne,\rhlStateTwo)$,
  \begin{equation}
    \label{eq:adequacy}
    \rhlSem{\tau_i(\rhlE)}(\rhlStateOne,\rhlStateTwo) = \rhlSemObj{\rhlE}(\rhlS_i),
    \qquad
    \rhlChi{\tau_i(\rhlB)}(\rhlStateOne,\rhlStateTwo) = \rhlSemObj{\rhlB}(\rhlS_i).
  \end{equation}
  We write $\rhlTag{\rhlE}{i}$ for $\tau_i(\rhlE)$ and $\rhlTag{\rhlB}{i}$ for
  $\tau_i(\rhlB)$.
\end{definition}

Agreement between the tagged guards of the two programs is written
$\rhlTag{\rhlB}{1} \relIff \rhlTag{\rhlB'}{2}$; this is the assertion that
appears in the premises of the RHL rules for conditionals and loops. Benton
writes $\rhlTag{\rhlB}{1} = \rhlTag{\rhlB'}{2}$ for it. We prefer $\relIff$
because the two symbols belong to different sorts: $=$ relates generalized
expressions and $\relIff$ relates relational assertions, so a sort error is
visible in the notation rather than hidden behind an overloaded symbol.

Condition~\eqref{eq:adequacy} is checked once per language rather than once per
proof; Appendix~\ref{app:language-instantiation} gives the tagging construction
for the two languages studied here. Reusing the proof framework also
requires that the adopted command rules be sound for that language's
semantics under the relational validity condition in
Appendix~\ref{app:relational-judgments}. Expression-tagging adequacy alone
does not justify reusing rules for altered control-flow constructs.

Sequential Composition additionally requires ordinary semantic sequencing:
for all commands $\subprogone,\subprogtwo$ and every store $s$,
\begin{equation}
  \label{eq:sequencing}
  \rhlSemObj{\subprogone\rhlSeq\subprogtwo}(s)=
  \begin{cases}
    \rhlBottom, & \text{if } \rhlSemObj{\subprogone}(s)=\rhlBottom,\\
    \rhlSemObj{\subprogtwo}(s'), & \text{if } \rhlSemObj{\subprogone}(s)=s'\in\rhlStates.
  \end{cases}
\end{equation}
The second case includes divergence of $\subprogtwo$ from $s'$. Internal
constructs of either fragment may vary between instantiations; this
equation specifies the composition behavior required at their boundary.

\subsection{Expression Interface for the Evaluation Languages}
\label{app:language-instantiation}

This subsection gives the expression-interface construction summarized
in Section~\ref{sec:checker-implementation}. The languages
{\proclangobs} and {\proclangalt} share the same expressions, so they use
the same signature, interpretation, and tagging maps. Their different
command semantics additionally require appropriate local proof rules.

\paperparagraph{Signature and interpretation.}
Take $\rhlValues=\rhlInts$, so that states are integer-valued stores and
the value expressions $\rhlValExp$ are obtained from the integer-expression
grammar in Figure~\ref{fig:proclang-syntax} by applying the
{\proclangobs} operator substitutions. The operation symbols $\rhlOps$ comprise one numeral
constant for each integer, together with the binary symbols
$\proclangAdd$, $\proclangMul$, and $\proclangSub$.
The predicate symbols are $\rhlPreds=\{\proclangEq,\proclangLt\}$, both binary.
Each numeral is interpreted as its integer value,
$\rhlInterp(\rhlNum)=\rhlNum$; the binary symbols are interpreted by
\[
  \begin{gathered}
    \rhlInterp(\proclangAdd) = {+},
    \qquad
    \rhlInterp(\proclangMul) = {\times},
    \qquad
    \rhlInterp(\proclangSub) = {-},
    \\[2pt]
    \rhlInterp(\proclangEq)
    = \rhlSet{(m,n) \rhlIn \rhlInts \times \rhlInts \rhlSuchThat m = n},
    \qquad
    \rhlInterp(\proclangLt)
    = \rhlSet{(m,n) \rhlIn \rhlInts \times \rhlInts \rhlSuchThat m < n}.
  \end{gathered}
\]

\paperparagraph{Tagging.}
For $i \rhlIn \{1,2\}$ the maps $\tau_i$ are defined by structural recursion.
On integer expressions,
\[
  \tau_i(\rhlNum) = \rhlNum,
  \qquad
  \tau_i(\rhlVar) = \rhlTag{\rhlVar}{i},
  \qquad
  \tau_i(\rhlE_1\ \rhlOpSym\ \rhlE_2)
  = \tau_i(\rhlE_1)\ \rhlOpSym\ \tau_i(\rhlE_2)
  \quad (\rhlOpSym \rhlIn \rhlOps,\ \rhlArity(\rhlOpSym)=2);
\]
and on Boolean expressions,
\[
  \begin{gathered}
    \tau_i(\rhlTrue) = \rhlTop,
    \qquad
    \tau_i(\rhlFalse) = \rhlBot,
    \qquad
    \tau_i(\rhlE_1\ \rhlPredSym\ \rhlE_2)
    = \tau_i(\rhlE_1)\ \rhlPredSym\ \tau_i(\rhlE_2)
    \quad (\rhlPredSym \rhlIn \rhlPreds),
    \\[3pt]
    \tau_i(\proclangNot\ \rhlB) = \rhlNeg\, \tau_i(\rhlB),
    \qquad
    \tau_i(\rhlB_1 \proclangAnd \rhlB_2)
    = \tau_i(\rhlB_1) \rhlAnd \tau_i(\rhlB_2).
  \end{gathered}
\]
Tagging is a homomorphism on integer expressions and on the comparison atoms of
Boolean expressions. On {\proclangobs}'s Boolean connectives it is a
\emph{translation}: $\proclangNot$ and $\proclangAnd$ are replaced by $\rhlNeg$
and $\rhlAnd$. This is the only point at which the concrete operators of
{\proclangobs} meet the assertion language.

With $\rhlValues=\rhlInts$, the triple $(\rhlSig,\rhlInterp,\tau)$ defined above
instantiates the expression interface at {\proclangobs}, and likewise at
{\proclangalt}, whose expressions and expression semantics are the same.
Condition~\eqref{eq:adequacy} follows by structural induction, first on
integer and then on Boolean expressions, for fixed $i$ and
$(\rhlStateOne,\rhlStateTwo)$. Numerals and variables are immediate from the
denotation clauses:
$\rhlSem{\tau_i(\rhlNum)}(\rhlStateOne,\rhlStateTwo)=\rhlInterp(\rhlNum)=\rhlNum$
and
$\rhlSem{\tau_i(\rhlVar)}(\rhlStateOne,\rhlStateTwo)=\rhlS_i(\rhlVar)$.
For a binary operation,
\[
  \begin{aligned}
    \rhlSem{\tau_i(\rhlE_1\ \rhlOpSym\ \rhlE_2)}(\rhlStateOne,\rhlStateTwo)
    &= \rhlInterp(\rhlOpSym)\bigl(
      \rhlSem{\tau_i(\rhlE_1)}(\rhlStateOne,\rhlStateTwo),
    \rhlSem{\tau_i(\rhlE_2)}(\rhlStateOne,\rhlStateTwo)\bigr)\\
    &= \rhlInterp(\rhlOpSym)\bigl(
    \rhlSemObj{\rhlE_1}(\rhlS_i),\rhlSemObj{\rhlE_2}(\rhlS_i)\bigr)
    = \rhlSemObj{\rhlE_1\ \rhlOpSym\ \rhlE_2}(\rhlS_i),
  \end{aligned}
\]
by the induction hypothesis and because $\rhlInterp$ interprets each symbol
by the operation the language assigns it. Comparison atoms are analogous,
the Boolean constants are immediate, and the cases for $\proclangNot$ and
$\proclangAnd$ hold because those operators denote negation and conjunction.

Note that $\rhlInterp(\proclangEq)$ is the identity relation, so
$\rhlGE_1 \proclangEq \rhlGE_2$ and $\rhlGE_1 = \rhlGE_2$ denote the same
relation for all generalized expressions. We therefore write the logical
symbol $=$ in relational assertions; $\proclangEq$ enters an assertion only
through a tagged program guard.

\paperparagraph{Command semantics.}
Expression adequacy connects the generated assertions to program
expressions; the command rules must additionally match each language's
command semantics. Both languages interpret $\rhlSkip$, assignment, and
sequencing in the usual way: $\rhlSkip$ returns its store,
$\rhlVar\rhlAssign\rhlE$ updates $\rhlVar$ to the value of $\rhlE$ in the
current store, and sequencing is defined by~\eqref{eq:sequencing}, so the
sequencing requirement of Appendix~\ref{sec:object-language} holds by
definition. Expressions are total, so loops are the only source of
divergence and every command denotes a total function into
$\rhlLift{\rhlStates}$. {\proclangobs} gives conditionals and loops their
ordinary meanings. {\proclangalt} reads guards with the opposite polarity
(Section~\ref{sec:familiarity-controls}). Writing $\rhlC^{\dagger}$ for the
{\proclangobs} command obtained from a {\proclangalt} command $\rhlC$ by
replacing every $\proclangAltIf\ \rhlB$ with $\rhlIf\ \proclangNot\rhlB$ and
every $\proclangAltWhile\ \rhlB$ with $\rhlWhile\ \proclangNot\rhlB$, the
commands $\rhlC$ and $\rhlC^{\dagger}$ have the same denotation.

\paperparagraph{Local proof rules.}
Local derivations must use command rules that are sound for the chosen
language under the relational validity condition of
Appendix~\ref{app:relational-judgments}. For {\proclangobs}, we instantiate
Benton's rules~\cite{nick2004rhl} with the assertion language above. The
rule set comprises Benton's core rules for $\rhlSkip$, assignment,
sequencing, conditionals, loops, consequence, symmetry, and transitivity;
structural rules that remove $\rhlSkip$ or a command related to it, regroup
sequences, distribute a command over the branches of a conditional, and
unroll loops; one-sided rules, in left- and right-handed versions, for dead
assignments, dead loops, and conditionals whose guard is known or analyzed
by cases; and a rule for an unsatisfiable precondition. The core rules are
sound for this validity condition by Benton's soundness theorem, which uses
the same lifting. Each remaining rule preserves validity by a direct
argument from the command semantics; in particular, each right-handed rule
is the mirror image of a left-handed one, and validity is preserved by
exchanging the two programs and the tags $\langle 1\rangle$ and
$\langle 2\rangle$.

For {\proclangalt}, the rules for $\rhlSkip$, assignment, sequencing, and
consequence are unchanged, and each rule that mentions a guard is stated
with the opposite polarity. For example, the conditional rule relates the
first branches under
$\rhlNeg(\rhlTag{\rhlB}{1}\rhlOr\rhlTag{\rhlB'}{2})$ and the second
branches under $\rhlTag{\rhlB}{1}\rhlAnd\rhlTag{\rhlB'}{2}$, and the loop
rule assumes $\rhlNeg(\rhlTag{\rhlB}{1}\rhlOr\rhlTag{\rhlB'}{2})$ for the
bodies and concludes $\rhlTag{\rhlB}{1}\rhlAnd\rhlTag{\rhlB'}{2}$ on exit.
Since $\tau_i(\proclangNot\rhlB)=\rhlNeg\,\tau_i(\rhlB)$, and guard agreement
$\rhlTag{\rhlB}{1}\relIff\rhlTag{\rhlB'}{2}$ is unchanged by negating both
guards, these are the {\proclangobs} rules applied to $\rhlC^{\dagger}$, up
to logical equivalence of their assertions. An instance of a {\proclangalt}
rule is therefore sound whenever the corresponding {\proclangobs} instance
is. The rules for the task's language are supplied to the model in the
proof-construction prompt (Appendix~\ref{app:proof-generation-prompts}), and
the verifier checks every local step against the same rules.

\subsection{Relational Judgments}
\label{app:relational-judgments}

\begin{definition}[Judgment]
  A judgment has the form
  \[
    \rhlJudgmentForm{\rhlC}{\rhlD}{\rhlPhi}{\rhlPhiPrime}.
  \]
  The assertions $\rhlPhi$ and $\rhlPhiPrime$ are the relational precondition
  and postcondition, respectively. The precondition $\rhlPhi$ relates the
  two initial stores, and the postcondition $\rhlPhiPrime$ relates the final
  stores when both commands terminate. Semantic validity below also
  requires agreement on termination.
\end{definition}

A judgment can hold in two senses, and we keep them apart throughout.
It may be \emph{derivable}, that is, obtained by the rules of a given proof
system, which we write
\[
  \derivableJudgment{\rhlC}{\rhlD}{\rhlPhi}{\rhlPhiPrime},
\]
annotating $\rhlDerives$ with the name of the system whenever more than one is
in play. Alternatively it may be \emph{semantically valid}, a property of the
denotational semantics of the two programs that refers to no proof system at
all.

To preserve the termination-sensitive comparison of
Section~\ref{sec:equivalence-task}, lift a relation
$R\subseteq\rhlStates\times\rhlStates$ to execution results by
\[
  \rhlLift{R}=R\cup\{(\rhlBottom,\rhlBottom)\}.
\]
This lifting permits joint divergence and excludes pairs in which only one
execution terminates. For terminating executions, the final states must
satisfy $R$.

\begin{definition}[Semantic validity]
  A judgment is \emph{semantically valid} when the relationship it requires
  between initial and final states holds according to the denotational semantics
  of the programs. Semantic validity is written
  \[
    \validJudgment{\rhlC}{\rhlD}{\rhlPhi}{\rhlPhiPrime}
  \]
  and holds exactly when
  \[
    \rhlForall
    (\rhlStateOne,\rhlStateTwo)
    \rhlIn
    \rhlSem{\rhlPhi}.\;
    \bigl(
      \rhlSemObj{\rhlC}(\rhlStateOne),
      \rhlSemObj{\rhlD}(\rhlStateTwo)
    \bigr)
    \rhlIn
    \rhlLift{\rhlSem{\rhlPhiPrime}}.
  \]
\end{definition}

Unfolding the lifting, this is the validity condition of
Section~\ref{sec:proof-foundations}: from every pair of initial stores
satisfying $\rhlPhi$, the two commands either both diverge or both
terminate in final stores satisfying $\rhlPhiPrime$. Soundness is the
statement that connects the two notions: every derivable judgment is
semantically valid.

\paperparagraph{Relational Hoare logic.}
\label{sec:rhl}
Relational Hoare logic is Benton's proof system for reasoning about two program
executions simultaneously, and the proof system of
Section~\ref{sec:proof-system} builds on it. Its judgments are those defined
above: a relational precondition and postcondition
together specify how a relation between the initial states of two programs is
transformed into a relation between their resulting states. When
\(\rhlJudgmentForm{\rhlC}{\rhlD}{\rhlPhi}{\rhlPhiPrime}\) is derivable using
the proof rules of RHL, we write
\[
  \rhlJudgment{\rhlC}{\rhlD}{\rhlPhi}{\rhlPhiPrime},
\]
the subscript identifying the system in which the derivation is carried out.

The conclusion of an RHL rule associates a pair of programs $\rhlC$ and $\rhlD$
with a relational precondition $\rhlPhi$ and postcondition $\rhlPhiPrime$.
Applying the rule matches its conclusion against the programs under
consideration and generates the premises that must in turn be established;
repeated application builds a derivation tree whose root is the target
judgment. We use Benton's rule schemata for language instantiations that satisfy the
command-semantics requirements of Appendix~\ref{sec:object-language};
we do not reproduce those rules here. Each instantiated local rule must
preserve semantic validity: whenever its relational premises are
semantically valid, its entailment premises hold under $\rhlInterp$ in the
sense of Appendix~\ref{sec:assertions}, and its other side conditions hold,
its conclusion is semantically valid under the same language semantics and
interpretation. We call this requirement \emph{local rule soundness}. It
applies to the adapted rule instances, including those used under
hypotheses in Appendix~\ref{app:local-obligations}.

\subsection{Entailment Validation}
\label{app:entailment-validation}

Several rules of Section~\ref{sec:proof-system} require one relational assertion to entail another.
Relational entailment as defined in Appendix~\ref{sec:assertions} is a relation
between denotations, so a premise asserting it directly would place a semantic
statement inside a derivation. We introduce instead a judgment
\[
  \rhlEntailmentDerivation{\rhlPhi}{\rhlPhiPrime},
\]
read as: the entailment $\rhlEntailmentClaim{\rhlPhi}{\rhlPhiPrime}$ has been
established. Every rule that needs an entailment takes a judgment of this form
as a premise, so the semantic check is confined to the one rule that
concludes this judgment, the \textsc{Oracle} rule of
Figure~\ref{fig:composition-rules}:
\[
  \frac{
    \rhlEntailmentCheck{\rhlPhi}{\rhlPhiPrime}=\rhlOracleValid
  }{
    \rhlEntailmentDerivation{\rhlPhi}{\rhlPhiPrime}
  }
  \;[\text{\textsc{Oracle}}]
\]
Here $\rhlEntailmentOracle_{\rhlInterp}$ is an entailment oracle, and
$\rhlEntailmentCheck{\rhlPhi}{\rhlPhiPrime}=\rhlOracleValid$ records that it
confirms the claim under the fixed interpretation $\rhlInterp$. We require
the oracle to be sound:
\begin{equation}
  \label{eq:oracle-soundness}
  \rhlEntailmentCheck{\rhlPhi}{\rhlPhiPrime}=\rhlOracleValid
  \rhlImplies
  \rhlEntailment{\rhlPhi}{\rhlPhiPrime}.
\end{equation}
Because \textsc{Oracle} is the only rule concluding an entailment judgment,
\eqref{eq:oracle-soundness} makes the judgment itself sound:
$\rhlEntailmentDerivation{\rhlPhi}{\rhlPhiPrime}$ implies
$\rhlSem{\rhlPhi}\rhlSubseteq\rhlSem{\rhlPhiPrime}$.

The oracle receives assertions interpreted under $\rhlInterp$; where a premise
mentions object-language expressions, the tagging maps of
Appendix~\ref{sec:object-language} have already translated them into assertions.
Its implementation is left unspecified, and we do not require it to decide
every implication. Only the verdict $\rhlOracleValid$ licenses
\textsc{Oracle}; we impose no requirements on other answers. As
distinguished in Section~\ref{sec:evidence-assessment}, a disproved
entailment invalidates the proposed proof step, whereas an inconclusive
check leaves the premise unestablished.

\subsection{Local Proof Assumptions and Discharge}
\label{app:local-obligations}

The assumption context {\proofSysAssumptions} in
Section~\ref{sec:proof-system} is a finite set of relational judgments; we
write $\Gamma,J$ for $\Gamma\cup\{J\}$.
Let $J,J_1,\ldots,J_k$ range over relational judgments and
$E_1,\ldots,E_m$ over entailment judgments established through
\textsc{Oracle}. For each local RHL rule with relational premises
$J_1,\ldots,J_k$, entailment premises $E_1,\ldots,E_m$, and conclusion
$J$, we use the contextual rule
\[
  \frac{
    \Gamma_1\rhlDerives_{\mathrm{RHL}}J_1
    \quad\cdots\quad
    \Gamma_k\rhlDerives_{\mathrm{RHL}}J_k
    \qquad E_1\quad\cdots\quad E_m
  }{
    \left(\bigcup_{i=1}^{k}\Gamma_i\right)
    \rhlDerives_{\mathrm{RHL}}J
  }.
\]
Here $k,m\geq0$; the original rule's side conditions remain in force,
and the union is empty when $k=0$. Contextual derivability is the least
relation closed under these lifted rules and \textsc{Assumption} and
\textsc{Obligation Discharge} in Figure~\ref{fig:composition-rules}.
We write $\rhlDerives_{\mathrm{RHL}}J$ for
$\varnothing\rhlDerives_{\mathrm{RHL}}J$. Only relational judgments
may be assumed; the entailment premises remain unconditional checks.

Obligation Discharge removes one of these dependencies. Its first premise
establishes the judgment for ${\subprogone}$ and ${\subprogoneprime}$ without
assumptions. Its second premise derives the judgment for ${\subprogtwo}$ and
${\subprogtwoprime}$ under {\proofSysAssumptions} together with that
established judgment. Substituting the first derivation for uses of the
assumed judgment leaves only the dependencies in {\proofSysAssumptions}.
Repeated discharge removes all assumed judgments once their proofs have
been supplied. Once every other premise is also justified, Embed can
import the completed local proof into {\proofSysAbbrev}, together with any
assumptions that remain.

Local derivations are sound relative to their contexts.

\begin{lemma}[Contextual soundness]
  \label{lem:contextual-soundness}
  Assume local rule soundness (Appendix~\ref{app:relational-judgments}) and
  a sound entailment oracle (Appendix~\ref{app:entailment-validation}). If
  $\Gamma\rhlDerives_{\mathrm{RHL}}J$ and every judgment in $\Gamma$ is
  semantically valid, then $J$ is semantically valid.
\end{lemma}

\begin{proof}
  By induction on the derivation of $\Gamma\rhlDerives_{\mathrm{RHL}}J$.

  \emph{Assumption.} Here $J\in\Gamma$, so $J$ is valid by hypothesis.

  \emph{Lifted local rule.} The context is
  $\Gamma=\bigcup_{i=1}^{k}\Gamma_i$, so every $\Gamma_i$ contains only
  valid judgments, and the induction hypothesis makes each relational
  premise $J_i$ valid. Each entailment premise is concluded by
  \textsc{Oracle} and therefore holds under $\rhlInterp$
  by~\eqref{eq:oracle-soundness}. The side conditions hold because the step
  is an instance of the rule. Local rule soundness then gives the validity
  of $J$.

  \emph{Obligation Discharge.} The premises are
  $\rhlDerives_{\mathrm{RHL}}J_1$ and
  $\Gamma,J_1\rhlDerives_{\mathrm{RHL}}J$. The induction hypothesis for the
  first premise, whose context is empty, shows that $J_1$ is valid. Every
  judgment in $\Gamma,J_1$ is then valid, and the induction hypothesis for
  the second premise shows that $J$ is valid.
\end{proof}

In particular, an empty-context derivation establishes its conclusion
without hypotheses.

\subsection{Proof Skeletons and Completion}
\label{app:proof-skeletons}

A proof skeleton records the argument proposed by the model; validation
determines whether that argument constitutes a proof. Fix a language
instantiation with interpretation $\rhlInterp$. Let $\mathcal R$ consist of
the rules of Figure~\ref{fig:composition-rules} and the contextual local
RHL rules of Appendix~\ref{app:local-obligations}. We write $K$ for a
well-formed claim: a contextual composition judgment, a contextual local RHL judgment,
an entailment judgment, or a check required by a rule. Checks include
syntactic conditions, such as the decompositions
$\prog\equiv\subprogone\rhlSeq\subprogtwo$ and
$\progprime\equiv\subprogoneprime\rhlSeq\subprogtwoprime$ required by
\textsc{Sequential Composition}, and oracle premises of the form
$\rhlEntailmentCheck{\rhlPhi}{\rhlPhiPrime}=\rhlOracleValid$.

\begin{definition}[Proof skeleton]
  \label{def:proof-skeleton}
  A \emph{proof skeleton} $\pi$ is a finite term generated by
  \[
    t ::= \square_K
    \mid \operatorname{evidence}(K,e)
    \mid \operatorname{step}(K,\rho;t_1,\ldots,t_n),
    \qquad \rho\in\mathcal R,\ n\geq0,
  \]
  whose root is a composition judgment. Here $\square_K$ is an open
  claim, $e$ is supplied evidence for $K$, and a step proposes deriving
  $K$ by rule $\rho$ from the roots of its ordered children
  $t_1,\ldots,t_n$. The root function is
  \[
    \begin{aligned}
      \operatorname{root}(\square_K)&=K,\\
      \operatorname{root}(\operatorname{evidence}(K,e))&=K,\\
      \operatorname{root}(\operatorname{step}(K,\rho;t_1,\ldots,t_n))&=K.
    \end{aligned}
  \]
\end{definition}

The grammar records proposed inferences without asserting their validity.
In particular, a step with no children differs from an open claim:
$\operatorname{step}(K,\rho;)$ proposes a rule with no premises, whereas
$\square_K$ supplies no justification.

To validate a step, write
$\operatorname{Inst}_{\rho}(K_1,\ldots,K_n;K)$ when the ordered premises
$K_1,\ldots,K_n$ and conclusion $K$ form an instance of $\rho$, with all
side conditions satisfied. To validate an evidence leaf, let
$\operatorname{Evidence}_{\rhlInterp}(e,K)$ hold precisely in the following
cases:
\begin{enumerate}
  \item $K=(\Gamma\rhlDerives_{\mathrm{RHL}}J)$ and $e$ is a finite
    derivation of exactly this judgment under the contextual rules of
    Appendix~\ref{app:local-obligations}, with every entailment and check
    premise justified.
  \item $K=(\rhlEntailmentCheck{\rhlPhi}{\rhlPhiPrime}=\rhlOracleValid)$ and $e$
    records the result $\rhlOracleValid$ from that oracle check under
    $\rhlInterp$.
  \item $K$ is a syntactic condition and $e$ is a check establishing $K$.
\end{enumerate}
Thus evidence leaves cannot directly justify composition or entailment
judgments; those judgments require the corresponding rules.

\begin{definition}[Completed composition proof]
  \label{def:completed-composition-proof}
  Let $\operatorname{Valid}_{\rhlInterp}$ be the least predicate on finite
  terms generated by the rules
  \[
    \frac{
      \operatorname{Evidence}_{\rhlInterp}(e,K)
    }{
      \operatorname{Valid}_{\rhlInterp}(\operatorname{evidence}(K,e))
    }
  \]
  and
  \[
    \frac{
      \begin{gathered}
        \operatorname{Inst}_{\rho}
        (\operatorname{root}(t_1),\ldots,\operatorname{root}(t_n);K)
        \\
        \operatorname{Valid}_{\rhlInterp}(t_i)\quad(1\leq i\leq n)
      \end{gathered}
    }{
      \operatorname{Valid}_{\rhlInterp}
      (\operatorname{step}(K,\rho;t_1,\ldots,t_n))
    }.
  \]
  A skeleton $\pi$ is a \emph{completed composition proof} of a
  composition judgment $K$ when $\operatorname{root}(\pi)=K$ and
  $\operatorname{Valid}_{\rhlInterp}(\pi)$.
\end{definition}

There is no validation rule for $\square_K$, so a completed proof has no
open claims. A rule with no premises, such as \textsc{Assumption}, is
validated by the second rule with $n=0$. Matching an \textsc{Embed}
instance requires the composition judgment to carry the local assumption
context, while matching an \textsc{Oracle} instance requires the stated
oracle check. Local subderivations may use assumptions that are
subsequently discharged or carried to the root.
These conditions depend on the finished tree, so the model may construct
its steps and supply their justifications in any order.

A valid term is a derivation in the following sense. By induction on
$\operatorname{Valid}_{\rhlInterp}$, the root of every valid term is
established: a composition, local RHL, or entailment judgment is derivable
by the rules of $\mathcal R$, and a check holds. In particular, a valid
term whose root is a local judgment $\Gamma\rhlDerives_{\mathrm{RHL}}J$
yields a contextual derivation of that judgment in the sense of
Appendix~\ref{app:local-obligations}, whether it is an evidence leaf or is
built from steps.

\subsection{Composition Soundness}
\label{app:composition-soundness}

\begin{proof}[Proof of Proposition~\ref{prop:composition-soundness}]
  Fix an instantiation satisfying the stated requirements. Assertions use
  the fixed interpretation $\rhlInterp$, and commands use the store
  semantics $\rhlSemObj{\cdot}$. By the definition of semantic validity
  (Appendix~\ref{app:relational-judgments}), the conclusion of the
  proposition states that
  $\validJudgment{\prog}{\progprime}{\rhlPhi}{\rhlPhiPrime}$. We prove the
  following claim by induction on valid terms
  (Definition~\ref{def:completed-composition-proof}): if $t$ is a valid
  term with root
  $\proofSysContextJudgment{\proofSysAssumptions}{\prog}{\progprime}{\rhlPhi}{\rhlPhiPrime}$
  and every judgment in {\proofSysAssumptions} is semantically valid, then
  $\validJudgment{\prog}{\progprime}{\rhlPhi}{\rhlPhiPrime}$. The
  proposition is the case $t=\pi$.

  An evidence leaf cannot justify a composition judgment, so $t$ is a step
  by \textsc{Embed}, \textsc{Conjunctive Composition}, or
  \textsc{Sequential Composition}, the only rules that conclude one. In
  each of these rules, the context of the conclusion is the union of the
  contexts of its relational premises. Every premise context therefore
  contains only valid judgments, and the induction hypothesis applies to
  every composition premise. Each entailment premise is concluded by
  \textsc{Oracle} and thus denotes an inclusion of relations
  (Appendix~\ref{app:entailment-validation}). We also use that lifting is
  monotone: $R\subseteq R'$ implies $\rhlLift{R}\subseteq\rhlLift{R'}$.

  \emph{Embed.} The premise is a valid term with root
  $\rhlContextJudgment{\proofSysAssumptions}{\prog}{\progprime}{\rhlPhi}{\rhlPhiPrime}$,
  hence a contextual derivation of this judgment
  (Appendix~\ref{app:proof-skeletons}). Every judgment in
  {\proofSysAssumptions} is valid, so Lemma~\ref{lem:contextual-soundness}
  gives $\validJudgment{\prog}{\progprime}{\rhlPhi}{\rhlPhiPrime}$; Embed
  leaves the programs and assertions unchanged.

  \emph{Conjunctive Composition.} Let $(s_1,s_2)\in\rhlSem{\rhlPhi}$ and
  $r=(\rhlSemObj{\prog}(s_1),\rhlSemObj{\progprime}(s_2))$. The first
  entailment gives $(s_1,s_2)\in\rhlSem{\rhlPhiIndexed{i}}$ for every $i$,
  so the induction hypothesis gives
  $r\in\rhlLift{\rhlSem{\rhlPhiIndexedPrime{i}}}$ for every $i$. Because
  command denotations are functions, every premise constrains the same
  pair $r$. Since $n\geq1$, either $r=(\rhlBottom,\rhlBottom)$, which
  belongs to $\rhlLift{\rhlSem{\rhlPhiPrime}}$, or $r$ is a pair of stores.
  In the latter case $r\in\rhlSem{\rhlPhiIndexedPrime{i}}$ for every $i$,
  so $r$ satisfies their conjunction, and the second entailment gives
  $r\in\rhlSem{\rhlPhiPrime}$.

  \emph{Sequential Composition.} Here
  $\prog\equiv\subprogone\rhlSeq\subprogtwo$ and
  $\progprime\equiv\subprogoneprime\rhlSeq\subprogtwoprime$. Let
  $(s_1,s_2)\in\rhlSem{\rhlPhi}$. The first entailment gives
  $(s_1,s_2)\in\rhlSem{\rhlPhiOne}$, so by the induction hypothesis
  $(\rhlSemObj{\subprogone}(s_1),\rhlSemObj{\subprogoneprime}(s_2))
  \in\rhlLift{\rhlSem{\rhlPhiOnePrime}}$.
  If this pair is $(\rhlBottom,\rhlBottom)$, then~\eqref{eq:sequencing}
  gives $\rhlSemObj{\prog}(s_1)=\rhlBottom=\rhlSemObj{\progprime}(s_2)$.
  Otherwise it is a pair of stores
  $(s'_1,s'_2)\in\rhlSem{\rhlPhiOnePrime}$, and the middle entailment gives
  $(s'_1,s'_2)\in\rhlSem{\rhlPhiTwo}$. The induction hypothesis for the
  second premise, the final entailment, and the monotonicity of lifting
  give
  \[
    \bigl(\rhlSemObj{\subprogtwo}(s'_1),
    \rhlSemObj{\subprogtwoprime}(s'_2)\bigr)
    \in\rhlLift{\rhlSem{\rhlPhiTwoPrime}}
    \subseteq\rhlLift{\rhlSem{\rhlPhiPrime}},
  \]
  and by~\eqref{eq:sequencing} this pair is
  $(\rhlSemObj{\prog}(s_1),\rhlSemObj{\progprime}(s_2))$.

  \textsc{Assumption} and \textsc{Obligation Discharge} conclude local
  judgments and are covered by Lemma~\ref{lem:contextual-soundness}. A
  local assumption that is not discharged passes through \textsc{Embed}
  into the context and is therefore covered by the hypothesis on
  {\proofSysAssumptions}.
\end{proof}

\subsection{Encoding the Equivalence Task}
\label{app:task-encoding}

For a task $\tau_r=(p_1,p_2,I)$ in language $\mathcal L$, we take
${\prog}=p_1$ and ${\progprime}=p_2$ and use the program and state
representations of Appendix~\ref{sec:object-language}, so that
$I\subseteq\rhlStates$, and require~\eqref{eq:store-output-adequacy}.
Choose an initial assertion $\Phi_{\mathrm{in}}$ and an output-equality
criterion $\Phi_{\mathrm{eq}}$ satisfying
\begin{equation}
  \label{eq:proof-task-boundary}
  \begin{aligned}
    \{(s,s)\mid s\in I\}
    &\subseteq\rhlSem{\Phi_{\mathrm{in}}},\\
    \rhlSem{\Phi_{\mathrm{eq}}}
    &\subseteq\{(s_1,s_2)\mid
      \operatorname{obs}_{\mathcal L}(s_1)
    =\operatorname{obs}_{\mathcal L}(s_2)\}.
  \end{aligned}
\end{equation}
The first inclusion ensures input coverage; the second ensures that the
final-store relation implies equal task outputs. For example, if the task
output is the value of a designated variable $y$, then
$\operatorname{obs}_{\mathcal L}(s)=s(y)$, and we may take
$\Phi_{\mathrm{eq}}\equiv(\rhlTagOne{y}=\rhlTagTwo{y})$; other final
variables may differ. This assertion expresses the output agreement needed
for equivalence without requiring equality of the entire final stores.
For {\proclangobs} and {\proclangalt}, the output consists of the final
values of the variables in scope
(Section~\ref{sec:familiarity-controls}), and $\Phi_{\mathrm{eq}}$ may be
taken to equate each of these variables across the two stores.

Neither membership in an arbitrary input set nor equality of observed
outputs need be exactly expressible in the assertion language. The
boundary conditions require only the stated inclusions:
$\Phi_{\mathrm{in}}$ may include additional initial state pairs, and
$\Phi_{\mathrm{eq}}$ may impose a stronger condition than equal
observations. Such choices strengthen what the proof must establish while
remaining sufficient for the task. The task encoding must provide these
inclusions, or they must be checked before the evidence is accepted.

\begin{proposition}[Task-level soundness]
  \label{prop:task-soundness}
  Assume the hypotheses of Proposition~\ref{prop:composition-soundness},
  the store/output correspondence~\eqref{eq:store-output-adequacy}, and the
  boundary conditions~\eqref{eq:proof-task-boundary}. Let $\pi$ be a
  completed composition proof with root
  \[
    \proofSysContextJudgment
    {\proofSysAssumptions}
    {\prog}
    {\progprime}
    {\Phi_{\mathrm{in}}}
    {\Phi_{\mathrm{out}}}
  \]
  for some postcondition $\Phi_{\mathrm{out}}$, where every judgment in
  {\proofSysAssumptions} is semantically valid and
  $\rhlEntailmentDerivation{\Phi_{\mathrm{out}}}{\Phi_{\mathrm{eq}}}$ is
  established through \textsc{Oracle}. Then $p_1\equiv_{\mathcal L,I}p_2$.
\end{proposition}

\begin{proof}
  Let $s\in I$. The first inclusion in~\eqref{eq:proof-task-boundary} gives
  $(s,s)\in\rhlSem{\Phi_{\mathrm{in}}}$, so
  Proposition~\ref{prop:composition-soundness} and the soundness of the
  final entailment give
  \[
    \bigl(\rhlSemObj{p_1}(s),\rhlSemObj{p_2}(s)\bigr)
    \in\rhlLift{\rhlSem{\Phi_{\mathrm{out}}}}
    \subseteq\rhlLift{\rhlSem{\Phi_{\mathrm{eq}}}}.
  \]
  If both results are $\rhlBottom$, then
  by~\eqref{eq:store-output-adequacy} both $\llbracket p_1\rrbracket(s)$
  and $\llbracket p_2\rrbracket(s)$ are undefined. Otherwise the results
  are stores $(s'_1,s'_2)\in\rhlSem{\Phi_{\mathrm{eq}}}$. The second
  inclusion in~\eqref{eq:proof-task-boundary} gives
  $\operatorname{obs}_{\mathcal L}(s'_1)=\operatorname{obs}_{\mathcal L}(s'_2)$,
  and~\eqref{eq:store-output-adequacy} gives
  $\llbracket p_1\rrbracket(s)=\operatorname{obs}_{\mathcal L}(s'_1)
  =\operatorname{obs}_{\mathcal L}(s'_2)=\llbracket p_2\rrbracket(s)$.
  In both cases the two programs agree on $s$, as the equivalence
  criterion of Section~\ref{sec:equivalence-task} requires.
\end{proof}

The completed proof, the validity of its remaining assumptions, and the
validated boundary checks therefore justify the task-level judgment
\[
  R_r\!\left(p_1,p_2,
    \{(s,s)\mid s\in I\},
  \{(o,o)\mid o\in O\}\right),
\]
whose precondition and postcondition are the input and output relations
required by Section~\ref{sec:reasoning-tasks}.

\section{Implementation Details}
\label{app:implementation}

This appendix describes how the benchmark is generated, the proof-generation
prompts and feedback, the proof representation and processing operations used
in the experiments, the concrete checking procedures, and the experimental
setup, models, and cost. \footnote{The code is available upon request.
Please contact the authors at \href{mailto:haozhi5@ualberta.ca}{haozhi5@ualberta.ca}.}

\subsection{Base-Program Generation}
\label{app:base-programs}

Each benchmark pair starts from a base program. Gemini-3.8-Flash writes a set of
\emph{program templates}, each a {\proclangobs} program containing zero, one, or two
placeholder statements. A sampler then fills each placeholder with the more
involved implementation from the transformation library described below.
Programs are generated in {\proclangobs} only;
the {\proclangalt} versions are obtained afterwards by rewriting control-flow
constructs (Section~\ref{sec:data-generation}).

\paperparagraph{Transformation library.}
To vary the local equivalence reasoning required, we use a hand-written library
of \emph{local transformations}. Each entry pairs a direct implementation of a
computation with a more involved alternative and declares its read, write,
and scratch variables. The sampler inserts the more involved implementation
into the base program and uses the direct alternative in the initial candidate
partner (Appendix~\ref{app:equivalent-programs}). The surrounding control flow
and its rewrite are generated separately; the completed program pair must pass
the equivalence check described below. We call the number of library entries
instantiated in a program pair its \emph{insertion count} and use it as a
heuristic control on proof difficulty. The model chooses its own decomposition
of the resulting programs into proof fragments.

\paperparagraph{Generation prompt.}
Each request asks for a single template. The prompt gives the {\proclangobs}
grammar and its big-step semantic rules. It asks for a program of at least ten
lines that uses only these constructs and whose loops terminate within an
evident bound. To vary difficulty, it asks for different control-flow shapes,
levels of nesting, and mixes of arithmetic and control flow.

For a template with placeholders, the prompt adds a placeholder
section. It requires the program to contain the placeholder markers, which are
complete statements. When there are two placeholders, they must sit at different nesting depths. Placeholder $i$ has a
fixed interface: it reads \texttt{p$i$} and \texttt{q$i$} and may modify
\texttt{p$i$}, \texttt{q$i$}, \texttt{r$i$}, and \texttt{s$i$}. The template must
use the modified variables after the placeholder without first overwriting
them, so that each inserted computation can influence the result. The prompt
specifies only the placeholder interfaces and does not expose any library
entry, so the generated context does not depend on which library entry
later fills each placeholder.

\paperparagraph{Acceptance.}
The sampler strips code fences and labels from the response and parses it with
the {\proclangobs} parser. A template is rejected if it does not parse or if its
markers are not exactly what the prompt asks for. It is also rejected if it repeats an earlier
template, either verbatim or after every variable name and numeral is replaced
by a placeholder token. After a rejection, the request is sent again.

An accepted template with placeholders is then filled in, in five steps:
\begin{enumerate}
  \item \emph{Inputs.} The first literal assignments to two distinct variables
    in the straight-line prefix, before any placeholder, are removed, so that
    these variables become inputs. A template without two such assignments is
    discarded.
  \item \emph{Insertion.} For a template with $k$ placeholders, the sampler draws
    $k$ distinct entries uniformly at random from the transformation library. It
    renames each entry's read and written variables to its placeholder's
    interface variables, renames its scratch variables to fresh names, and
    replaces each marker with the more involved implementation, enclosed in
    braces.
  \item \emph{Liveness.} The sampler searches for concrete inputs under which
    every inserted computation is live, meaning that replacing it with
    \texttt{skip} changes the final state (ignoring scratch variables). The
    interface variables a computation reads are assigned values in $\{1,2,3\}$ and
    the remaining inputs are set to $2$. If none of the first 24 assignments
    works, the library entries are redrawn; after four draws, the template is
    discarded.
  \item \emph{Execution.} The program is run with the reference interpreter on
    these inputs to record the final state and the execution trace. A program
    whose execution fails is discarded.
  \item \emph{Input sensitivity.} Each of the two lifted inputs must change some
    output when its value is perturbed with the other inputs held fixed;
    otherwise the program is discarded.
\end{enumerate}
A template without placeholders is executed as written. Its symbolic inputs
are chosen when the pair is built: among the first four literal assignments in
its straight-line prefix, the first two whose values both affect an output.
Discarded templates are replaced by new requests.

A base program enters the benchmark only if an equivalent program is accepted
for it (Appendix~\ref{app:equivalent-programs}). Base programs without an
accepted partner are replaced by new templates until each insertion count
reaches its quota. A program whose parsed form matches one already in the
benchmark is excluded.

\subsection{Equivalent-Program Generation}
\label{app:equivalent-programs}

Gemini-3.8-Flash rewrites each base program's template, and the direct
implementation from each selected library entry is spliced into the rewrite.
A pair is accepted only after cvc5 validates its relational goal.

\paperparagraph{Relational goal.}
The relational goal is derived from the base program. Its outputs are the
variables in the base program's final state. Its symbolic inputs are the
variables the base program may read before assigning
them. The relational precondition equates each symbolic input across the two
programs, and the postcondition equates each output. The concrete values
chosen in Appendix~\ref{app:base-programs} serve only as execution inputs and
do not appear in the precondition.

\paperparagraph{Transformation prompt.}
The prompt gives the {\proclangobs} grammar and semantic rules, followed by
the template, with its markers still in place, and the lists of input and
returned variables. For a base program without placeholders, the template is
the base program itself. The prompt asks for ten rewrites that preserve the
program's semantics, use only the constructs of the grammar, and differ
structurally from the template. Each rewrite must change the control flow and
apply at least two of the following techniques: statement reordering,
constant propagation, constant folding, common-subexpression elimination,
introduction of intermediate variables, and modification of branch
conditions. The prompt repeats the syntactic restrictions of
Appendix~\ref{app:base-programs} and fixes the response format: each rewrite
lists the program, its input variables, and its returned variables.

For a template with placeholders, the prompt adds a section on the markers.
A marker resembles a dead assignment, so the prompt states that each marker
stands for an unseen block that may change its interface variables
to unpredictable values. The model must
keep each marker exactly as written and must not propagate, fold, or reuse
the value of an interface variable across a marker. It must also not rename
or eliminate those variables, and must not rename any other existing
variable; new variables need fresh names. Because the model initially rewrites the
template rather than the base program, it does not see the library
implementations during rewriting.

\paperparagraph{Candidate selection.}
Before any solver call, a candidate is rejected if it does not parse, if its
markers differ from the template's, if it omits an interface variable or
output that the template mentions, or if it parses to the same program as the
template. The remaining candidates are ranked by how much they change the
control flow. The most different come first. In this order, the sampler splices
the direct implementation from each selected library entry into the
candidate's markers, using the same interface variables as
in the base program and fresh scratch names, and validates the resulting
pair. cvc5 checks each output equality separately. It encodes values as
fixed-width bit-vectors and restricts the symbolic inputs to a bounded range.

\paperparagraph{Candidate repair.}
After an unsuccessful equivalence check, a further request can provide the
base program, the failed candidate, and checker feedback to the model. The
model may repair the complete candidate, including the inserted implementations.
The repair must preserve the required outputs for symbolic inputs; it may add
case splits but may not impose a new precondition or specialize to the concrete
execution input. Each repaired candidate is checked against the same relational
goal. The first validated candidate is accepted.

\paperparagraph{{\proclangalt} pairs.}
Each accepted pair is translated into {\proclangalt} by negating the
condition of every conditional and loop and using {\proclangalt}'s
inverted-guard constructs. The translation preserves the behavior of both
programs, so the validation of the {\proclangobs} pair carries over to the
{\proclangalt} pair.

\paperparagraph{Equivalent program differences.}
To give a sense of how far an equivalent program departs from its base program, we
compute the tree edit distance between their abstract syntax trees: the
fewest node insertions, deletions, and relabelings that turn one tree into
the other. Base programs have 91.5 nodes on average ($\sigma=24.4$), and the
two programs of a pair are 52.7 edits apart ($\sigma=31.6$), or 0.49
($\sigma=0.18$) relative to the larger tree.

\subsection{Proof-Generation Prompts}
\label{app:proof-generation-prompts}

The two prompt stages expose different information. Proof-construction
prompts provide the language grammar, RHL and composition rules, program
fragments, and relational precondition and postcondition; they omit the
language-specific semantic rules. Semantic-judgment prompts provide those
semantic rules and an extracted assumed judgment or nontrivial connecting
entailment. They ask the same model for a correctness verdict
(\texttt{true}, \texttt{false}, or \texttt{null} if undetermined) and a
free-form justification. These are separate calls, rather than requests
to revise the supplied proof. The reported proof-generation result without
feedback scores the composed proof from the first iteration, after json
repair and but before any correction, refinements, or diagnostics are returned for
whole-proof revision. The recorded model judgments analyzed in the evaluation
are also obtained before those diagnostics are returned.
Figure~\ref{fig:prompt-templates} gives abbreviated templates.

\paperparagraph{Feedback diagnostics.}
Table~\ref{tab:feedback-outcomes} summarizes the diagnostics delivered during
assisted generation and the final outcomes of runs receiving each diagnostic.
Only feedback followed by a recorded model response counts as received, and a
run may receive several categories. The categories therefore overlap and
describe the complete workflow rather than the effect of any individual
feedback type. For a refuted assumption or connecting entailment, the revision
prompt reports the failed condition but omits the concrete counterexample
state retained in the checker report.

\begin{table}[!htbp]
  \centering
  \small
  \caption{Feedback supplied during assisted generation and the resulting
    selected-final proof correctness. Runs are counted once per category among
    the 1,000 model--task evaluations; categories overlap. Percentages use
    the receiving runs in that row as the denominator. Response-format
  feedback includes parsing repair in recursive refinement calls.}
  \label{tab:feedback-outcomes}
  \label{tab:retained-work}
  \begin{tabular}{@{}>{\raggedright\arraybackslash}p{0.19\linewidth}
    >{\raggedright\arraybackslash}p{0.46\linewidth}rr@{}}
    \toprule
    Feedback type & Information returned & Runs & Final passes \\
    \midrule
    Response format & JSON or response-schema error; requests a corrected
    submission. & 393 & 231 (58.8\%) \\
    \addlinespace[2pt]
    Judgment syntax & Identifies a proof judgment that cannot be parsed.
    & 179 & 76 (42.5\%) \\
    \addlinespace[2pt]
    Rule application & Explains mismatches in rule applications or parsing
    errors in side conditions. & 669 & 473 (70.7\%) \\
    \addlinespace[2pt]
    Premise linkage & Identifies a premise with no matching earlier conclusion.
    & 194 & 120 (61.9\%) \\
    \addlinespace[2pt]
    Task goal & Identifies a mismatch between the final judgment and the
    required whole-program goal. & 40 & 17 (42.5\%) \\
    \addlinespace[2pt]
    Assumed judgment & Identifies a refuted assumed judgment and the
    postcondition conjuncts that fail. & 210 & 98 (46.7\%) \\
    \addlinespace[2pt]
    Connecting entailment & Identifies a refuted side condition, including
    failing conjuncts when available. & 132 & 64 (48.5\%) \\
    \bottomrule
  \end{tabular}
\end{table}

\begin{figure}[p]
  \centering
  \lstset{
    basicstyle=\ttfamily\scriptsize,
    columns=fullflexible,
    keepspaces=true,
    showstringspaces=false,
    breaklines=true,
    breakindent=1em,
    aboveskip=1pt,
    belowskip=1pt,
    escapeinside={(*}{*)},
    literate={⊢}{{$\vdash$}}1 {∼}{{$\sim$}}1 {Φ}{{$\Phi$}}1
    {⇒}{{$\Rightarrow$}}1 {⊨}{{$\models$}}1 {Γ}{{$\Gamma$}}1
  }%
  \def\fld#1{\textrm{\itshape\{#1\}}}%
  \def\elide#1{\textrm{\itshape[#1]}}%
  \def\part#1{\par\smallskip\hrule\smallskip{\small\bfseries #1}\par}%
  \begin{minipage}[t]{0.485\linewidth}
    \raggedright
    \part{(a) Proof construction: system prompt}
\begin{lstlisting}
You are an expert in formal program semantics.
Below is the ProcLang language definition, plus
the rules of Relational Hoare Logic (RHL) and the
Composition Proof System (C), which work together.
Use these to construct the requested
program-equivalence proof.
(*\elide{ProcLang grammar, without semantic rules; relational notation; RHL judgments and the 36 RHL rules; the four C rules; how \texttt{[Embed]} connects RHL and C}*)
## Proof Construction Strategy
- Only apply [Assumption] when no RHL derivation
  exists for the code fragment under the required
  Φ ⇒ Φ' (*\ldots*); never use [Assumption] as a
  shortcut when a derivation is possible.
- Use [Embed] to move an available RHL judgment
  into the C proof. If it depends on an
  undischarged assumption set, record
  EXTERNAL OBLIGATION: ⊨ Γ_name as a side
  condition.
## Task
Given two ProcLang programs P and P', a
relational precondition Φ_in, and a relational
postcondition Φ_eq (the equivalence criterion),
construct a proof (*\ldots*) The proof must conclude
the C judgment for the whole programs
  ⊢_C P ∼ P' : Φ_pre ⇒ Φ_post
with (*\ldots*) Φ_pre exactly Φ_in and (*\ldots*)
Φ_post exactly Φ_eq. (*\ldots*) Clearly
distinguish derived RHL judgments from assumed RHL
judgments, and do not present an assumed judgment
as derived.
## Required Output
(*\elide{one JSON object in the step schema of Appendix~\ref{app:proof-processing}; every rule fully instantiated; premises matching earlier conclusions exactly}*)
\end{lstlisting}
    \part{(a) Proof construction: user message}
\begin{lstlisting}
## Program 1
(*\fld{first program}*)
## Program 2
(*\fld{second program}*)
## Relational Precondition (Φ_in)
(*\fld{precondition}*)
## Relational Postcondition / Equivalence
   Criterion (Φ_eq)
(*\fld{postcondition}*)
## Valid Assumptions
None.
\end{lstlisting}
    \part{(b) Parsing repair: follow-up message}
\begin{lstlisting}
Your previous response has these problems:
(*\fld{parsing problems}*)
Produce the complete proof again, fixing these
problems. Follow the Required Output format
exactly: respond with one JSON object containing
the "steps" array and nothing else.
\end{lstlisting}
  \end{minipage}\hfill
  \begin{minipage}[t]{0.485\linewidth}
    \raggedright
    \part{(c) Refinement: note appended to (a)}
\begin{lstlisting}
## Further Derivation
An earlier proof recorded this judgment as an
[Assumption], so it is still a proof obligation.
Derive it with the rules of RHL and C, reducing it
to assumptions smaller than the judgment itself
where it cannot be derived outright. If no further
derivation is possible, so it can only stay a
single assumption, respond with {"steps": []}
instead of a proof.
\end{lstlisting}
    \part{(d) Revision}
\begin{lstlisting}
(*\elide{system prompt of (a), with this task:}*)
## Revision Task
A proof that two ProcLang programs P and P' are
equivalent was written with the proof system
above, and checking it found problems. Repair the
proof by editing it (*\ldots*)
(*\elide{pasted subproofs end in \texttt{[Sub-proof]} steps}*)
Repair only what the problems require. (*\ldots*)
(*\elide{user message of (a), followed by:}*)
## Previous Proof
Steps with problems: (*\fld{step numbers}*)
(*\fld{composed proof, one numbered step at a time, each followed by its problems}*)
## Your Response
Respond with edits to the previous proof that fix
its problems: one JSON object whose only key is
"edits", a list of edits. (*\ldots*)
(*\elide{replace, insert, or delete numbered steps; uncovered steps are kept}*)
\end{lstlisting}
    \part{(e) Semantic judgment of an assumption}
\begin{lstlisting}
Determine whether the provided relational
judgment is correct under the ProcLang grammar
and semantics below. State your answer and give a
free-form justification.
(*\elide{the judgment form and its partial-correctness reading}*)
Assess the supplied judgment itself. Its inclusion
in Γ does not establish its correctness and must
not be used to justify it.
(*\elide{ProcLang grammar and big-step semantic rules}*)
## Response Format
Return only one valid JSON object with exactly
these keys:
- "correct": true if the judgment is correct,
  false if incorrect, or null if you cannot
  determine correctness.
- "justification": a string containing your
  free-form explanation.
(*\elide{user message:}*)
Is the following judgment correct? Respond in the
required JSON format.
(*\fld{assumed judgment}*)
\end{lstlisting}
  \end{minipage}
  \caption{Abbreviated prompt templates for {\proclangobs}. Braced italics
    mark substituted fields, bracketed italics summarize omitted material,
    and ellipses mark omitted sentences. (a)~starts every proof; with the
    note in (c), filled with an assumed judgment's code and assertions, it
    refines that judgment. (b)~follows a response that contains a malformed proof, and
    (d)~starts each revision. The entailment prompt matches (e), with an
    entailment in place of the judgment. Only (e) includes semantic rules.
    The {\proclangalt} templates differ in the grammar and in the guards of
  the conditional and loop rules.}
  \Description{Five abbreviated prompt templates: the proof-construction
    system prompt and user message, the parsing-repair follow-up, the
    refinement note, the revision prompt with its edit-based response, and
  the semantic-judgment prompt for an assumption.}
  \label{fig:prompt-templates}
\end{figure}

\subsection{Proof Representation and Processing}
\label{app:proof-processing}

\paperparagraph{Representation.}
Models submit JSON representations of proof skeletons, recording
proposed rule applications and references to relational assertions.
A skeleton may contain assumed local judgments as well as explicit
derivations. Structural validation checks this representation before
rule applications and semantic premises are assessed. Without feedback, malformed JSON is repaired deterministically; with verifier feedback, parsing errors are returned to the model for at most two corrections (three attempts including the initial response), separately from the revision rounds. A premise cites an earlier step by repeating that step's conclusion. When a subproof is integrated, it is followed by a \texttt{[Sub-proof]} step that
concludes the assumed judgment, as the parent wrote it, from the subproof's
final {\proofSysAbbrev} judgment with the same code and assertions. The
parent's premises therefore still cite a matching conclusion, and none of
them is rewritten.
A premise can remain open in two ways. A premise that matches no earlier
conclusion it may cite is reported as unlinked and, under the assisted
protocol, returned to the model. A remaining \texttt{[Assumption]} step is
an open obligation, recorded with its origin and the reason it was not
refined, such as the model declining or the depth limit. Its assumption set
must appear in every RHL conclusion that depends on it and be listed when
\texttt{[Embed]} lifts the judgment into {\proofSysAbbrev}. Remaining
assumptions and nontrivial entailment side conditions go to the semantic
checks (Appendix~\ref{app:checking-procedures}).

\paperparagraph{Proof schema.}
The JSON object has a single key, \texttt{steps}, which lists the rule
applications in derivation order, each step after the steps it cites:
{\small
\begin{verbatim}
{"steps": [
  {"rule": "[Sequential Composition]",
   "premises": ["<judgment>", "<judgment>"],
   "side_conditions": ["<condition>", "<condition>"],
   "conclusion": "<judgment>"},
  ...]}
\end{verbatim}
}
\noindent The \texttt{rule} field names the applied rule. The
\texttt{premises} field lists the step's derivational premises, one RHL or
{\proofSysAbbrev} judgment per element, each the exact conclusion of an
earlier step. The \texttt{side\_conditions} field lists the step's
entailments $\rhlEntailmentClaim{\rhlPhi}{\rhlPhiPrime}$, its
partial-equivalence and termination conditions, and its assumption-set
conditions $\rhlModels \proofSysAssumptions$, which are marked as external
obligations when undischarged. The \texttt{conclusion} field is the
judgment the step derives. Judgments and conditions are strings in the
prompt's relational notation, with every component instantiated by
concrete code and assertions.

\paperparagraph{Subproblem extraction.}
To refine an assumption, we treat the assumed judgment
\rhlJudgment{\rhlC}{\rhlD}{\rhlPhi}{\rhlPhiPrime} as a new goal to prove. The
new problem has the same form as the whole task: the programs are $\rhlC$ and
$\rhlD$, the precondition is $\rhlPhi$, and the target postcondition is
$\rhlPhiPrime$. The model receives it through the same proof-construction
prompt and is asked to derive the judgment. The new problem starts with no
assumptions, so the parent's assumptions {\proofSysAssumptions} are not carried
over. Section~\ref{sec:proof-generation} specifies
which assumptions are refined and to what depth.

\paperparagraph{Subproof integration.}
Generated subproofs are integrated into the parent argument, updating
assertion references and affected conclusions and recording the origins
of composed steps. When a subproof is integrated, the names $\Phi_{\mathrm{in}}$ and
$\Phi_{\mathrm{eq}}$ in its steps are replaced by the subproblem's
precondition and postcondition, so that in the composed proof they refer
only to the whole task. Each subproof targets the same instantiated judgment
as the assumption it refines. Assumptions remaining in a subproof
remain open obligations in the composed argument. Refinement can thus
expose more of a proposed derivation without completing all its local
justifications.

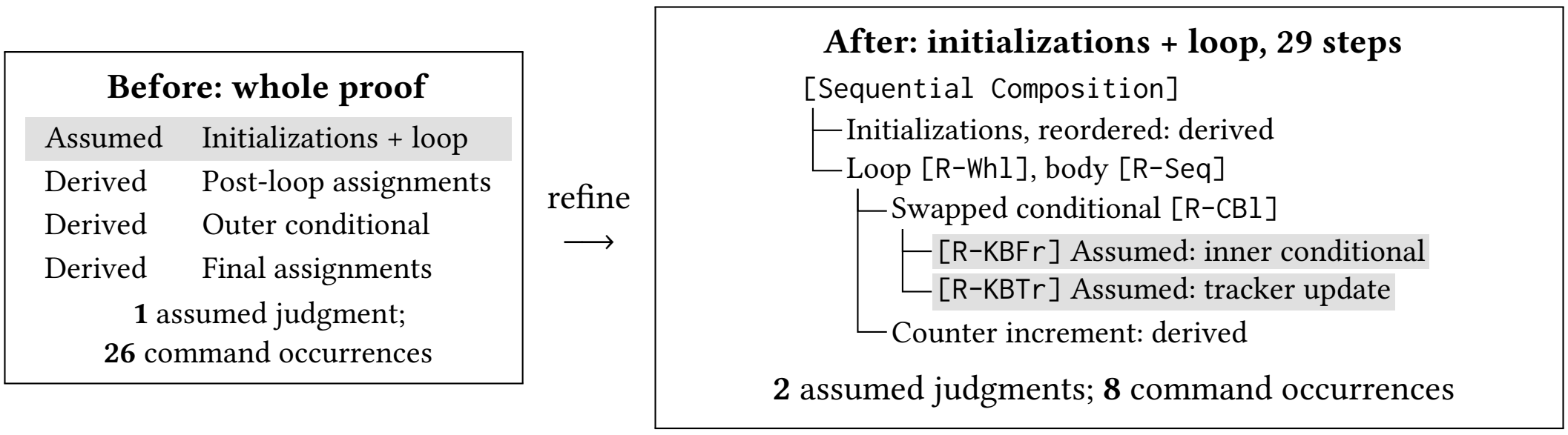
\begin{figure}[htbp]
  \centering
  \small
  \setlength{\fboxsep}{5pt}
  \fbox{
    \begin{minipage}[c]{0.30\linewidth}
      \centering
      \textbf{Before: whole proof}\par\smallskip
      \footnotesize
      \renewcommand{\arraystretch}{1.1}
      \begin{tabular}{@{}ll@{}}
        \rowcolor{black!12}Assumed & Initializations + loop \\
        Derived & Post-loop assignments \\
        Derived & Outer conditional \\
        Derived & Final assignments
      \end{tabular}\par\smallskip
      \textbf{1} assumed judgment;\\
      \textbf{26} command occurrences
  \end{minipage}}
  \hfill
  \begin{minipage}[c]{0.07\linewidth}
    \centering
    refine\par
    $\longrightarrow$
  \end{minipage}
  \hfill
  \fbox{
    \begin{minipage}[c]{0.55\linewidth}
      \centering
      \textbf{After: initializations + loop, 29 steps}\par\smallskip
      \footnotesize
      \begin{tikzpicture}[
          x=4mm, y=-3.6mm,
          row/.style={anchor=west, inner sep=1pt, minimum height=3.1mm},
          asm/.style={row, fill=black!12},
          link/.style={line width=0.4pt},
        ]
        \node[row] (r0) at (0,0) {\texttt{[Sequential Composition]}};
        \node[row] (r1) at (1,1) {Initializations, reordered: derived};
        \node[row] (r2) at (1,2) {Loop \texttt{[R-Whl]}, body \texttt{[R-Seq]}};
        \node[row] (r3) at (2,3) {Swapped conditional \texttt{[R-CBl]}};
        \node[asm] (r4) at (3,4) {\texttt{[R-KBFr]} Assumed: inner conditional};
        \node[asm] (r5) at (3,5) {\texttt{[R-KBTr]} Assumed: tracker update};
        \node[row] (r6) at (2,6) {Counter increment: derived};
        \draw[link] (0.35,0.45) |- (r1.west);
        \draw[link] (0.35,0.45) |- (r2.west);
        \draw[link] (1.35,2.45) |- (r3.west);
        \draw[link] (1.35,2.45) |- (r6.west);
        \draw[link] (2.35,3.45) |- (r4.west);
        \draw[link] (2.35,3.45) |- (r5.west);
      \end{tikzpicture}\par\smallskip
      \small
      \textbf{2} assumed judgments; \textbf{8} command occurrences
  \end{minipage}}
  \caption{Refinement reduces the code covered by assumptions in a GLM
    proof on {\proclangalt}. Before refinement, the proof joins four pieces
    of the programs by sequential composition, and one assumption covers the
    initializations and the loop (26 command occurrences). A 29-step
    sub-proof replaces it; its tree, shown with each judgment above its
    premises, derives everything except two smaller assumptions covering eight
  occurrences. Shading marks assumptions; counts include both programs.}
  \Description{Left: the proof before refinement, four pieces of the programs
    of which the initializations and loop are a single assumption covering 26
    command occurrences. Right: the 29-step sub-proof that replaces it, drawn
    as an indented tree. It derives the reordered initializations, the loop
    and its body, the conditional with swapped branches, and the counter
    increment; the inner conditional and the tracker update remain assumed,
  covering eight command occurrences.}
  \label{fig:rq3-refinement}
\end{figure}

\paperparagraph{Refinement example.}
In 180 of 1,000 runs, a refinement replaces one broad assumption with
several that together cover fewer command occurrences.
Figure~\ref{fig:rq3-refinement} shows one GLM proof on {\proclangalt}: GLM
initially assumes the relation between the initializations and loops, then
derives the initialization, outer-branch correspondence, and loop
structure; only the inner-conditional and tracker-update relations remain
assumed, covering eight command occurrences instead of 26.

\subsection{Checking Procedures}
\label{app:checking-procedures}

\paperparagraph{Rule checking.}
Our verifier checks model-generated relational derivations against the
local RHL and composition rules, including the program fragments,
assertions, and premises of each application. Assumed judgments remain
explicit dependencies. Replacing an assumption with a completed,
assumption-free RHL subproof implements the \textsc{Obligation Discharge}
rule; validating a rule application does not itself discharge its
assumed premises. Assumptions that remain after refinement are carried to
the root context.

\paperparagraph{Semantic checks.}
We extract relational obligations and connecting entailments from each
submitted argument. Entailments use the same checking interface: an
entailment from $\rhlPhi$ to $\rhlPhiPrime$ is encoded as a relational
judgment over $\rhlSkip$ on both sides, with precondition $\rhlPhi$ and
postcondition $\rhlPhiPrime$. Semantic-judgment
calls supply the evaluated model with the appropriate language semantics
and an extracted assumption or nontrivial entailment to assess. This is
the semantic-judgment stage of Section~\ref{sec:proof-generation}. We retain
the model and solver results separately. Model verdicts support the
validator-agreement analysis; they do not determine proof acceptance or
provide proof-revision feedback. The assisted protocol supplies revision
feedback from the structural verifier and cvc5 and uses cvc5 to choose which assumptions to refine.

\paperparagraph{Solver settings.}
During proof construction, cvc5 checks each assumption and each nontrivial
entailment in 16-bit signed bit-vector arithmetic, unrolling loops 10 to 80
times. When a check cannot be decided, it is repeated under a smaller bound,
with every initial input in $[-7, 7]$. Each check has a 600-second
wall-clock limit. Because cvc5 searches in 16-bit arithmetic, every counterexample it returns
is replayed by running both programs with the reference interpreter over
unbounded integers. A check fails only if the replayed initial states
satisfy the precondition, both runs terminate, and the final states violate
the postcondition. When the replay does not confirm a counterexample, the
unrolling bound is doubled if a loop may have exceeded it.

\paperparagraph{Extensive checking undecided results.}
We rerun the checks that remained unknown, or that failed only through
16-bit overflow, on a computing cluster (Appendix~\ref{app:setup}). Each
check first runs over unbounded integers with cvc5, and then
over input ranges that widen from $[-7, 7]$ up to $[-32767, 32767]$.
Each solver query has a 300-second limit, and each stage a one-hour limit, raised to at most six hours.

The scope of solver results is determined by the bit-vector encoding
and loop-unrolling bounds. Applying these results to the integer
semantics requires arithmetic preservation and coverage of the relevant
loop executions. Validating the assumptions that remain in the root
context establishes the hypothesis of
Proposition~\ref{prop:composition-soundness}.

\paperparagraph{Results and reports.}
The verifier returns validity results and failed-premise reports.
Rule violations identify invalid derivation steps; missing
justifications leave premises open, and inconclusive checks leave
validation unresolved. Independently validated subderivations can
remain useful when the overall argument is unfinished, following
Section~\ref{sec:evidence-assessment}.

\subsection{Experimental Setup and Cost}
\label{app:setup}

\paperparagraph{Implementation and hardware.}
We implement the framework in Python~3.12, with about 31K lines of code
excluding tests, and use cvc5~1.4.2 and z3~4.15.4 as solvers. The
experiments run on an Apple M5 Max machine (18-core CPU, 64\,GB memory,
macOS~27.0), which sends the API model calls and runs the cvc5 checks
during proof construction with four worker processes; for Qwen, these
checks run on the cluster that serves the model. The rechecks of
Appendix~\ref{app:checking-procedures} run on the Vulcan cluster of the
Digital Research Alliance of Canada. Each Vulcan node has two Intel Xeon
Gold 6448Y CPUs (64 cores), 512\,GB memory, and four NVIDIA L40S 48\,GB
GPUs, and runs Ubuntu~24.04.5 LTS. Each recheck job uses 4 to 8 cores with
5\,GB of memory per core and a 24-hour limit.

\paperparagraph{Models.}
We query GPT-6.1-Sol, GPT-6-Luna, and Opus-5.5 through their providers'
APIs, and GLM-5.3 through Tinker. For proof construction, semantic
judgment, and the execution task, Qwen3.8-27B runs on Vulcan in BF16 with
vLLM~0.21.0, tensor-parallel across four L40S GPUs, with 16 CPU cores and
128\,GB memory per job. For the free-form task, it runs in FP8 on a
Together AI dedicated endpoint with two H100 80\,GB GPUs.

\paperparagraph{Token usage.}
The experiments make 35,287 model calls, with 245M input and 393M output
tokens in total, or 7.0K input and 11.1K output tokens per call on
average. Proof construction, including semantic-judgment calls, averages
8.3K input and 12.5K output tokens per call; the initial proof prompt is
about 10.3K tokens, or 16.4K with Opus's tokenizer. The execution task
averages 2.9K input and 9.9K output tokens per call, and the free-form task
3.1K input and 5.1K output tokens.

\paperparagraph{Cost.}
At the providers' list prices, with cached input billed at the reported
cached rates, the experiments cost about \$3.9K
(Table~\ref{tab:cost}). The Qwen figure is an estimate, because we served
that model ourselves.

\begin{table}[!htbp]
  \centering
  \small
  \caption{Cost of the experiments by model, at list prices.}
  \label{tab:cost}
  \begin{tabular}{@{}l>{\raggedright\arraybackslash}p{0.5\linewidth}r@{}}
    \toprule
    Model & Price basis & Cost \\
    \midrule
    GPT-6.1-Sol & OpenAI list price & \$239 \\
    GPT-6-Luna & OpenAI list price & \$19 \\
    Opus-5.5 & Anthropic list price & \$1,321 \\
    GLM-5.3 & Tinker's 256K-context price & \$1,854 \\
    Qwen3.8-27B & Alibaba Cloud's API rate (\$0.50 input, \$3.00 output
    per 1M tokens) & \$449 \\
    \midrule
    Total & & \$3,882 \\
    \bottomrule
  \end{tabular}
\end{table}

\section{Additional Evaluation Examples}
\label{app:evaluation-examples}

Figure~\ref{fig:rq3-diagnostics} shows the two failure locations discussed in
Section~\ref{sec:reasoning-results}: (a) Opus's claims hold, but its invariant
permits $\texttt{step}=3$ where a branch requires $2$; (b) GLM's loop invariant
omits $\texttt{factor}=2$, so its assumed claim about the loop body is false.

\begin{figure}[htbp]
  \vspace{-4pt}
  \centering
  \footnotesize
  \setlength{\fboxsep}{5pt}
  \fbox{
    \begin{minipage}[t]{0.455\linewidth}
      \textbf{(a) Opus: a failed connection}\par\smallskip
      $s$ abbreviates \texttt{step}.\par
      $A\equiv s_1=s_2\land 0<s_1\land 0<s_2$\par
      \hspace*{1.8em}$\land\neg(s_1=1\land s_1\ne2)$
      \par\smallskip
      \textbf{Required branch condition}\par\smallskip
      \hfill $A\rhlEntails s_2=2$\hfill\mbox{}\par\smallskip
      \textbf{Integer counterexample}\par\smallskip
      Both stores: $\texttt{step}=3$;\par
      $A$ holds, but $s_2=2$ is false.
  \end{minipage}}\hfill
  \fbox{
    \begin{minipage}[t]{0.455\linewidth}
      \textbf{(b) GLM: a false assumed claim}\par\smallskip
      $c$, $f$, $w$ abbreviate \texttt{counter}, \texttt{factor}, \texttt{weight}.\par
      Loop bodies $B_1$, $B_2$ set $w\rhlAssign c\cdot f$ and $w\rhlAssign c\cdot 2$.\par
      $I\equiv c_1=c_2\land f_1=f_2\land w_1=w_2\land\cdots$
      \par\smallskip
      \textbf{Assumed claim}\par\smallskip
      \hfill $\rhlJudgmentForm{B_1}{B_2}{I}{I}$\hfill\mbox{}\par\smallskip
      \textbf{Integer counterexample}\par\smallskip
      Both stores: $(c,f)=(-3,-1)$, so $w_1=3$ but $w_2=-6$.\par
      $I$ omits $f_1=2$, so the bodies break it.
  \end{minipage}}
  \caption{Two failure locations: (a) Opus's connecting entailment on
    {\proclangobs}; (b) GLM's assumed claim on {\proclangalt}.
  Notation is normalized, and only decisive conjuncts are shown.}
  \Description{Two failure locations. Opus's invariant permits step=3,
    falsifying a connecting entailment requiring step=2. GLM's assumed
    loop invariant omits factor=2; with counter=-3 and factor=-1, the
    left body sets weight to 3 and the right body sets it to -6.}
  \label{fig:rq3-diagnostics}
  \vspace{-6pt}
\end{figure}

Figure~\ref{fig:rq3-matched-proofs} shows one of the 146 tasks for which both
Sol and Luna have accepted selected-final proofs
(Section~\ref{sec:reasoning-results}). Both proofs are
accepted, but only Sol derives the swapped conditionals in the loop bodies;
Luna assumes the complete loop-body relation, which cvc5 validates.

\begin{figure}[htbp]
  \centering
  \small
  \lstset{
    basicstyle=\ttfamily\footnotesize,
    columns=fullflexible,
    keepspaces=true,
    showstringspaces=false,
    breaklines=true,
    breakindent=0pt,
    aboveskip=2pt,
    belowskip=2pt,
    literate={⊢}{{$\vdash$}}1 {Φ}{{$\Phi$}}1 {Γ}{{$\Gamma$}}1
    {∧}{{$\land$}}1 {⇒}{{$\Rightarrow$}}1 {∼}{{$\sim$}}1
    {◇}{{$\diamond$}}1 {□}{{$\square$}}1 {△}{{$\triangle$}}1
    {…}{{$\ldots$}}1
  }
  \begin{minipage}[t]{0.485\linewidth}
    \raggedright
    \hrule\smallskip
    \textbf{Sol: derives both conditionals}\par\smallskip
    \textbf{Step 31:} \texttt{[R-CBl]}
    \begin{lstlisting}
⊢_RHL
if aux ◇ 0 then { skip }
else { cursor := cursor □ 1 }
∼
if # (aux ◇ 0) then
  { cursor := cursor □ 1 }
else { skip }
: (Φ_eq ∧ var(factor, 1) = 2)
⇒ (Φ_eq ∧ var(factor, 1) = 2)
    \end{lstlisting}
    Step 23 \texttt{[R-CBl]} derives the other swapped conditional.\par
    \textbf{Retained assumptions:} none.
  \end{minipage}\hfill
  \begin{minipage}[t]{0.485\linewidth}
    \raggedright
    \hrule\smallskip
    \textbf{Luna: assumes the loop bodies}\par\smallskip
    \textbf{Step 5:} \texttt{[Assumption]}
    \begin{lstlisting}
Γ_loop_body ⊢_RHL
{ step := cursor △ factor;
  …; cursor := cursor □ 1 }
∼
{ step := cursor △ 2;
  …; cursor := cursor □ 1 }
: … ⇒ …
    \end{lstlisting}
    Step 6 \texttt{[R-Whl]} uses this relation as its body premise.\par
    \textbf{Retained assumption:} complete loop bodies (solver-verified).
  \end{minipage}
  \caption{Representative records from two accepted proofs of the same
    {\proclangobs} task. Sol derives both swapped-conditional relations and
    retains no assumptions; Luna retains the complete loop-body relation as
    an assumption and uses it in \texttt{[R-Whl]}. Ellipses omit intervening
  commands and assertions from Luna's judgment.}
  \Description{Side-by-side records from two accepted proofs. Sol derives
    the swapped conditional relations and retains no assumptions. Luna
    assumes the complete loop-body relation and uses it in the relational
  while rule.}
  \label{fig:rq3-matched-proofs}
\end{figure}